\documentclass[12pt,preprint]{aastex}
\newcommand{\manq}{manqu\'{e}}
\newcommand{\msun}{M_\sun}
\newcommand{\teff}{T_{\rm eff}}

\begin{document}
\title{A Re-Evaluation of the Evolved Stars in the Globular Cluster
  M13\footnote{Based on observations made with the NASA/ESA Hubble Space
    Telescope, obtained from the data archive at the Space Telescope Science
    Institute. STScI is operated by the Association of Universities for
    Research in Astronomy, Inc. under NASA contract NAS5-26555.}}

\author{Eric L. Sandquist, Mark Gordon, Daniel Levine} 
\affil{Department of Astronomy, San Diego State University,
  San Diego, CA 92182} 
\author{Michael Bolte}
 \affil{UCO/Lick Observatory, University of California, Santa Cruz, CA 95064}

\begin{abstract}
  We have analyzed photometry from space- and ground-based cameras to identify
  {\it all} bright red giant branch (RGB), horizontal branch (HB), and
  asymptotic giant branch (AGB) stars within 10\arcmin ~ of the center of the
  globular cluster M13. 

    We identify a modest (7\%) population of HB stars redder than the primary
  peak (including RR Lyrae variables at the blue end of the instability strip)
  that is somewhat more concentrated to the cluster core than the rest of the
  evolved stars. We find support for the idea that they are noticeably evolved
  and in the late stages of depleting helium in their cores. This resolves a
  disagreement between distance moduli derived from the tip of the red giant
  branch and from stars in or near the RR Lyrae instability strip. We
  identified disagreements between HB model sets on whether stars with $\teff
  \la 10000$ K (near the ``knee'' of the horizontal branch in optical CMDs)
  should evolve redward or blueward, and the differences may depend on the
  inclusion of diffusion in the stellar interior.  The sharp cut at the red
  end of M13's HB provides strong evidence that stars from the dominant HB
  group must still be undergoing blue loops, which implies that diffusion is
  being inhibited. We argue that M13's HB is a somewhat pathological case ---
  the dominant HB population occurs very near the ``knee'' in optical CMDs,
  and evolved stars exclusively appear redward of that peak, leading to the
  incorrect appearance of a continuation of the unevolved HB.

We identify two stars as ``blue hook'' star candidates --- the faintest stars
in optical bands that remain significantly subluminous in the shortest
ultraviolet wavelength photometry available. M13 also has a distinct group of
stars previously identified with the ``second $U$ jump''. Based on far UV
photometry, we find that these stars have genuinely high temperatures
(probably 26000 K $\la \teff \la 31000$ K), and are not produced by a jump in
brightness at lower temperature ($\teff \approx 22000$ K) as previously
suggested. These stars are brighter than other stars of similar color (either
redder or bluer), and may be examples of ``early hot flashers'' that ignite
core helium fusion shortly after leaving the red giant branch.
We used ultraviolet photometry to identify hot post-HB stars, and based on
their numbers (relative to canonical AGB stars) we estimate the position on
the HB where the morphology of the post-HB tracks change to $I \sim 17.3$,
between the two peaks in the HB distribution.

Concerning the possibility of helium enrichment in M13, we revisited
the helium-sensitive $R$ ratio, applying a new method for correcting
star counts for larger lifetimes of hot horizontal branch stars. We
find that M13's $R$ ratio is in agreement with theoretical values for
primordial helium abundance $Y_P = 0.245$ and inconsistent with a
helium enhancement $\Delta Y = 0.04$.  The brightness of the
horizontal branch (both in comparison to the end of the canonical HB
and to the tip of the red giant branch) also appears to rule out the
idea that the envelopes of the reddest HB stars have been
significantly enriched in helium.  The absolute colors of the turnoffs
of M3 and M13 may potentially be used to look for differences in their
mean helium abundances, but there are inconsistencies in current datasets
between colors using different filters that prevent a solid conclusion.




The numbers of stars on the lower red giant branch and in the red
giant bump agree very well with recent theoretical models, although
there are slight indications of a deficit of red giant stars above the
bump. There is not convincing evidence that a large fraction of stars 
leave the RGB before undergoing a core helium flash.

\end{abstract}

\section{Introduction}




M13 was one of the first globular clusters identified as having unusually blue
horizontal branch (HB) stars, and it remains one of the prototypes of the
``long blue tail'' with stars approaching the main sequence for helium stars.
As \citet{smith} notes, differences
between stars on both giant branches and stars on the horizontal branch can be
discerned from data in papers as early as \citet{barn09,barn14}. Along with
the nearby, massive, and little reddened cluster M3, M13 forms half of the
best known ``second parameter'' pair. The chemical composition of the hydrogen
envelopes of HB stars affects their observable properties via opacity and mean
molecular weight. The ``first parameter'' is heavy element content, where
higher metallicity produces higher envelope opacity and generally redder
stars. Because M3 and M13 have nearly the same iron abundances
($\langle$[Fe/H]$\rangle = -1.53$ for M13 versus $\langle$[Fe/H]$\rangle =
-1.45$ for M3; \citealt{sned}) but M3 has a much redder HB including a huge
number of RR Lyrae variable stars, a second parameter is needed. In addition
to the color shift between the HBs of M3 and M13, the HB stars in M13 show a
bimodal distribution that is not present in M3. On its own, this fact implies
that the HB stars were produced by at least two different populations of
cluster stars or involve different methods of producing HB stars.  M3 and M13
share a number of similarities beyond iron abundance, and we tabulate some of
their characteristics in Table \ref{m13m3}.

Many theories have been proposed to explain the HB differences, and two of the
main goals of this paper are to 1) assemble a large and complete set of
photometric data for M13, and 2) use the photometric data to examine questions
bearing on the production of M13's horizontal branch stars. Because of the
complexity of the HB, it is very doubtful that one explanation can cover all
of its aspects. Before we describe our results, we briefly
summarize the main hypotheses we will examining, and the primary reasons they
are viable. We emphasize that they are not mutually exclusive.

\underline{The $\Delta t$ Hypothesis.} Early models showed that as the
mass of the hydrogen-rich envelope of an HB star is decreased, the
surface temperature increases with relatively little change in
luminosity. In this hypothesis, age differences between populations of
stars lead to differences in mass between stars leaving the main
sequence and between stars reaching the HB. While it is natural to
expect that clusters in the Milky Way were born at different times,
age differences are hard to prove except for clusters that are much
younger than the average. \citet{rey} compared M13 with M3 and found
that turnoff-to-giant branch color differences and changes in HB
morphology were consistent with an age difference $\Delta t = 1.7 \pm
0.7$ Gyr (with M13 older). \citet{catrev} finds that the age
differences implied by differences in the cluster CMDs near the
turnoff can explain the HB morphology as long as M3 is younger than
about 12 Gyr and the EHB stars in M13 are presumed to arise from a
process that is unrelated to the age. However, there are aspects of M13's
population that cannot be explained in this hypothesis. For example, 
neither of the studies above could reproduce the bluest HB stars in M13 
in synthetic HB simulations using the same chemical
composition and dispersion in stellar mass used for M3. 

\underline{The $\Delta Y$ Hypothesis.}  Variations in helium content
($Y$) result in differences in position on the HB, largely because
greater helium abundance allows lower mass stars to leave the main
sequence at the present day \citep{dant08b}. \citet{jb} proposed that
a helium abundance difference $\Delta Y \sim 0.05$ (with M13 the more
helium rich) could be responsible for many of the unusual features of
the color-magnitude diagram (CMD), including an interesting difference
in the slopes of the subgiant branch. \citet{cda} also examined data
for M3 and M13, finding the luminosity of the red giant bump and RR
Lyrae stars relative to the cluster turnoff are consistent with an
enhancement $\Delta Y \sim 0.04$. This picture has been taken very
seriously with the discovery of multiple stellar populations in
some clusters. For example, NGC 2808 was
found to have at least three identifiable main sequences
\citep{pio07}, while $\omega$ Cen has a blue main sequence
\citep{bed04} that appears to be helium enriched \citep{pio05}.
$\omega$ Cen and NGC 2808 are among the most massive clusters known in
the Milky Way, which may enable them to retain gas that has been
processed and released by a first generation of stars.

While the helium abundance is very difficult to measure except in limited
circumstances, spectroscopic observations of other heavy element species lead
to the belief that helium was probably enriched in some clusters.  Stars in
M13 are well-known to have star-to-star abundance differences in O and Na that
can be traced from the giant branch \citep{sbcn,yong,jkp,sbh,sned} to the main
sequence turnoff \citep{cm,briley}. O depletion and Na enrichment can
only be accomplished in hydrogen-fusion regions where significant production
of helium is accomplished \citep{dandd,lang}, and star-to-star variations on the main
sequence require that they must have been present in the gas forming the
stars.

The $\Delta Y$ hypothesis is attractive for M13 because it may explain the
blueward shift of the main body of HB stars compared to M3's population, and
the bimodality of the HB (as an additional population within
the cluster).  \citet{dant08b} conducted a fit to M13's HB using
helium-enriched models, and found that a fit required 70\% of the population
to be enriched to $0.27 < Y < 0.35$ (a mean $\Delta Y \approx 0.04$), and the
remaining 30\% to be enriched to $Y \sim 0.38$.  There are some difficulties
with this picture though. In the \citeauthor{dant08b} models of M13, they
still needed to assume a rather large (but constant) total mass loss on the
RGB ($0.18 \msun$).  Because M13 has few HB stars in the instability strip or
redward (where M3's HB is heavily populated), {\it virtually all} of M13's HB
stars must also be more helium rich than M3's.  This is in striking contrast
to more massive clusters that show strongly bimodal HB star distributions in
which the redder HB population is interpreted as a first generation of stars
formed from primordial material, while subsequent generations have varying
degrees of enrichment and are bluer. The massive clusters that are inferred to
have such large spreads in $Y$ generally also have multiple main
sequence or subgiant branch populations, whereas M13 has shown no sign of
multiple populations to date.  There is also not a clear bimodality in the
spectroscopic abundances and a definite connection has never been made between
the abundances and HB morphology \citep[although see][for indications that the
maximum $\teff$ extent of the HB is correlated with the extent of observed
Na-O anticorrelations]{carrb}. \citet{rey} attribute some of \citeauthor{jb}'s
conclusions to slight missteps in the implementation of their relative age
comparison.

\underline{The $\dot{M}$ Hypothesis.} From early models, it was recognized
that a significant amount of mass loss is needed, probably on the red giant
branch (RGB), to produce the colors of the majority HB populations in
most clusters. Dispersion in colors was then taken to mean
that there are star-to-star differences in mass loss, although a mechanism
to produce these differences has not been identified.

Independent of the majority of HB stars, there is a population that seems to
{\it require} strong mass loss: the ``blue hook'' stars
\citep{cast93,dcruzbhk,cast06,millb}. In ultraviolet color-magnitude diagrams
of some of the most massive clusters \citep{dcruzomega,brown}, stars are found
fainter and redder than the zero-age HB (ZAHB) at its blue end, meaning that
they must have almost no hydrogen envelope. If a star loses virtually all of
its hydrogen envelope before reaching the tip of the RGB, it can leave the RGB
without igniting core helium fusion. As the star contracts onto the He white
dwarf cooling curve, a late He flash can be ignited that drives a convection
zone that reaches hydrogen rich layers \citep{brown}.


\underline{The Evolution Hypothesis.} As relatively cool HB stars convert He
into carbon and oxygen, they are expected to eventually evolve brightward and
redward toward the asymptotic giant branch (AGB). Depending on the
distribution of stars on the HB, evolving HB stars could be mistaken for
fainter, more slowly evolving HB stars, thereby misrepresenting the brightness
of the HB. Clusters with large blue HB populations (like M13) are most
susceptible to this effect because evolutionary tracks may nearly parallel the
ZAHB at the blue end of the HB distribution where the relative number of
unevolved stars drops rapidly. Because the HB is a frequently used standard
candle in astronomy, it is worth studying the degree to which this affects
stellar populations.

Evolutionary effects have been inferred from the pulsation properties of RR
Lyrae stars. For example, \citet{jurc} used magnitudes and periods to identify
RR Lyraes in different stages of their HB evolution.  \citet{caccm3}
identified mean lines in the period-amplitude diagram for different subsets of
M3 variables, and labeled them as regular or ``well-evolved''. They showed
that in at least some other clusters, the majority of variables could be
identified with one group or the other.



For the purposes of this paper we focused on post
main-sequence evolution. Using datasets from telescopes and
instruments having a wide range in spatial resolution and field size, 
we attempted to completely survey the evolved stars from the
center far into the outskirts of the cluster. We discuss the
observational material and the analysis of the photometry in \S 2. In
\S 3, we describe the steps used to identify the evolutionary status
of the evolved cluster stars. We examine the red giant branch and
horizontal branch populations in greater detail in \S 4 and \S 5,
respectively. In \S \ref{ratios} we look at population ratios and their
relationship to the evolution timescales for stars in different
stages. Finally in \S \ref{disc} we discuss the body of evidence
involving second parameter effects (cluster to cluster variations) and
intracluster differences between stars.

\section{Datasets}

\subsection{Archival Hubble Space Telescope Imaging}

We used WFPC2 images taken in three different studies and ACS images from one
additional study for photometric measurements.  The principal investigators
and filters used are listed in Table \ref{tbl-1}.  The images from proposal
8174 were composed of three partially overlapping fields that were reduced
separately and later combined.

\begin{figure}
\epsscale{.90}
\plotone{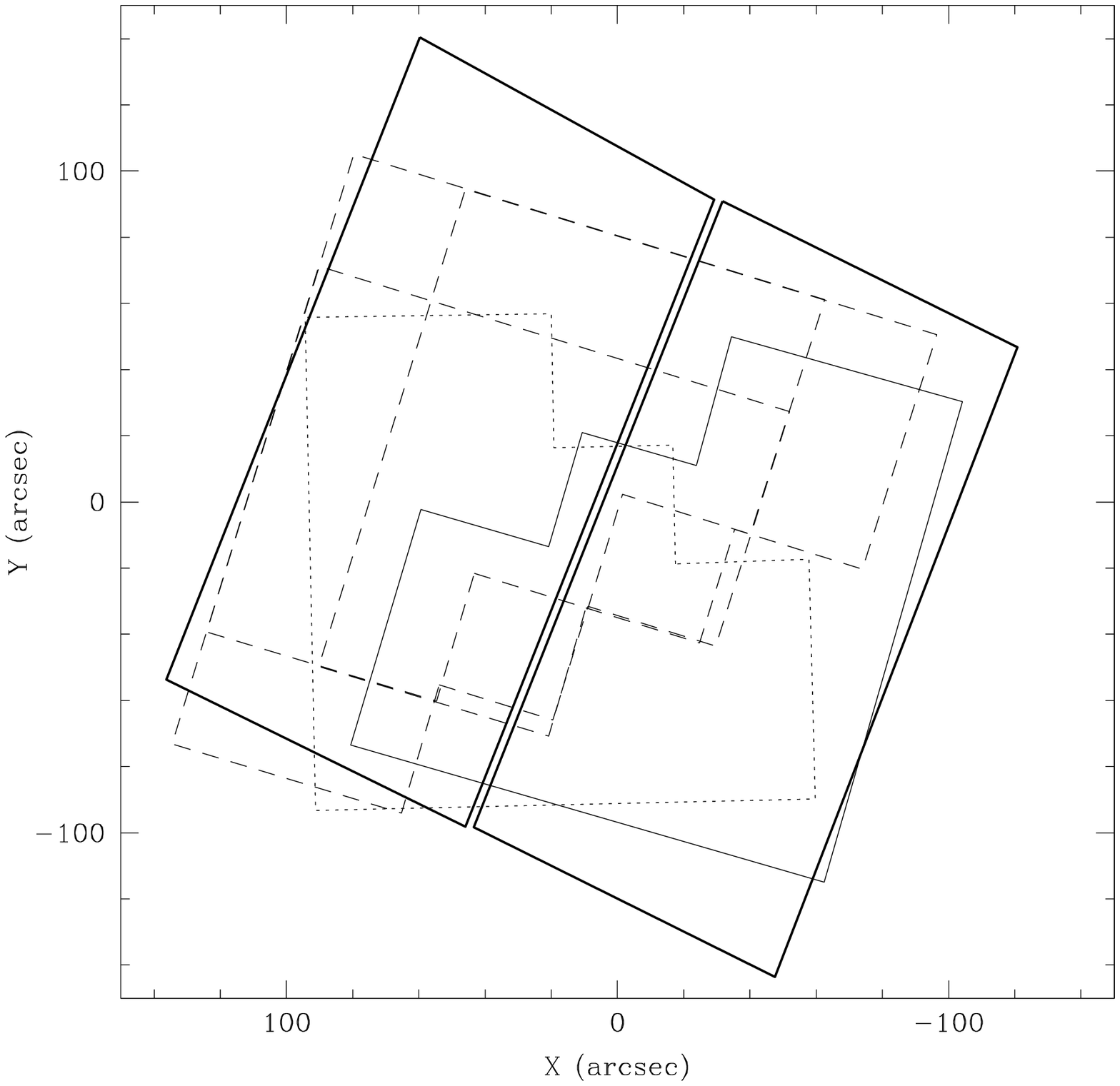}
\caption{Outlines of the observed HST 
fields for M13. 
The HST proposal 5903 field with a dotted border, the proposal 8278
field with a solid border, and the proposal 8174 fields with a dashed
border.  (Three overlapping fields were observed as part of proposal
8174.) The ACS fields of the two WFC chips are shown with a dark solid
line (WFC1 on the left).
\label{fig1}}
\end{figure}

All of the WFPC2 images were processed using the HSTPhot\footnote{\tt
  http://purcell.as.arizona.edu/hstphot/} package (version 1.1.7b),
which was described by \citet{hstphot}.  Star positions for each set
of processed data were derived using the IRAF task METRIC, which we
used to convert the HSTPhot pixel coordinates to RA and DEC. METRIC
positions often have absolute errors of up to 0\farcs5, but absolute
errors are not important for our purposes.  The RA and DEC coordinates
were converted to a system relative to the cluster center
(Fig. \ref{fig1}).  The data for each star, including coordinates,
flight system magnitudes, and errors, were extracted from each dataset
and merged into one master file.  In cases where a star had multiple
measurements in the same filter, the values were averaged using
weights derived from the measurement errors.  This master dataset was
used to construct CMDs that we used to identify post-MS stars.  
The F160BW images were reduced separately because the
filter distorts star positions relative to the rest of the images,
complicating the matching of stars within HSTPhot.  After completing
the photometry, we were then able to match all of the stars with
previously identified HB and AGB stars.

The ACS Wide Field Camera (WFC) images were processed using the
DOLPHOT\footnote{\tt http://purcell.as.arizona.edu/dolphot/} package
with its module for ACS data. DOLPHOT tasks mask bad pixels, correct
for effective pixel area, and do point-spread function photometry.
Individual frames (prior to drizzling in the ACS pipeline) were
obtained from the HST Archive in order to get photometry on stars
covering the widest possible dynamical range (including stars that
were saturated on long exposures) and to avoid the pixel resampling
that goes along with the drizzling process.

Because the WFC sits far from the optical axis of {\it HST}, there is
significant geometric distortion of these images. For the purposes of
the photometry, the differences in effective pixel area were corrected
within DOLPHOT through the use of a pixel area map provided on the ACS
website\footnote{\tt
  http://www.stsci.edu/hst/acs/analysis/PAMS}. DOLPHOT, however, does
not currently correct pixel positions to sky coordinates. For the
purposes of {\it relative} astrometry, we used the fourth-order
polynomial corrections provided in the most recent IDCTAB file for the
dataset (qbu1641sj\_idc.fits).  These corrections reduce distortions
to about 0.1 pixels, which is more than adequate for our purposes. The
star positions were later put on a common coordinate system with
photometry from WFPC2 images as described below.

\subsection{Archival CFHT Images}

The Hubble Space Telescope is very effective at resolving the core of
the cluster, but its approximately 3\arcmin ~ by 3\arcmin ~ field of
view is insufficient to capture most of the cluster stars.  We thus turned to
the archive of the Canada-France-Hawaii Telescope (CFHT) CFH12K camera
for data covering the majority of the globular cluster
(Fig. \ref{fig1}), with data collected by J.-H. Park, Y.-J. Sohn, and
S. J. Oh in Feb. 2001 using $B$, $V$, and Cousins $I$ filters.  This
camera is composed of a mosaic of 12 CCD's, of which we only used
numbers 11 and 12 because these two chips covered the most heavily
populated parts of the cluster.  Each chip covers a roughly 7\arcmin ~
by 14\arcmin ~ area of sky.  The images we obtained from the CFHT
archive were already bias-subtracted and flat-fielded. For the
photometry, we used the DAOPHOT II/ALLSTAR packages \citep{daophot}.
Star positions in the CFHT images, like the HST images, needed to be
transformed from pixel coordinates into angular offsets from the
cluster center.  We cross-identified stars from the USNO A2.0 catalog,
covering an approximately 18.3\arcmin ~ by 16.7\arcmin ~ area of sky
around the cluster that encompassed the CFHT field.  These catalog
stars were used to convert the CFHT pixel positions into RA and DEC
coordinates relative to the core of M13.  In regions of overlap, the
HST datasets were placed on the same coordinate system.

\subsection{KPNO 0.9m Imaging}

We also reduced imaging the KPNO 0.9 m telescope taken in $B$,
$V$, and Cousins $I$ filters by Bolte and Sandquist on two nights in
May 1995. The camera used a single 2048 $\times$ 2048 CCD with a scale
of $0\farcs68$ per pixel, for a field of view $23\farcm2$ on a
side. The photometry was done similarly to the CFHT dataset, and
coordinates were transformed to the CFHT system.

\subsection{Proper Motion (PM) Studies}

We cross identified all stars from the proper motion studies of \citet{cmpm}
and \citet{cud}. The field covered by these studies is similar in size to that
of the KPNO images. For the brightest stars, membership probabilities were
available to within about $1\arcmin$ of the cluster center, while for fainter
stars the innermost $2-3\arcmin$ was effectively excluded. Their photometry
had a faint limit at $V \sim 15.6$, which is about 0.5 mag fainter than the
red end of the HB. As a result, the proper motion membership probabilities
were useful primarily for identifying bright stars that would otherwise
contaminate post-HB and giant star samples.

\subsection{Archival UIT Imaging}

The archive of the Ultraviolet Imaging Telescope contains aperture and
point-spread function photometry of M13 produced following the procedures
described in \citet{stech}, and the images involved have previously
been discussed by \citet{par}.  However, in examining the UIT images,
we found a large number of sources that were not recorded by
\citeauthor{par} or in the UIT archive file, and these match up with
horizontal branch stars identified with optical photometry. It
appears that the earlier photometry had an unnecessarily
high faint limit.

We therefore rereduced the two longest exposure far UV images (using
the B5 filter with central wavelength 1620 \AA) taken on the Astro 2
mission (fuv2418, 192.5 s; fuv2419, 953.5 s). We downloaded archived IDL
reduction routines called
MOUSSE\footnote{http://archive.stsci.edu.uit/analysis.html} for use on
linearized, flat-fielded digitizations of the original photographic
images. The {\it uit\_find} routine identifies significant sources
using an algorithm based on DAOPHOT, and determines centroided
positions for each. The {\it uit\_aper} routine conducts aperture photometry.
We used a 3 pixel radius aperture recommended by
\citet{par}
and verified that the measurements matched the archived aperture
photometry to within 0.01 mag for sources in common. The shorter fuv2418
exposure was analyzed to allow the brightest sources to be
incorporated, as they were saturated on the longer exposure. 

Many sources could be identified in the core of the cluster because
crowding is relatively mild in the far UV. However, crowding does
significantly modify the photometry, so we will generally restrict our
discussion to stars more than 120\arcsec ~ from the cluster center, as
did \citet{par}. This partly resulted from the pixel scale (pixels in
the digitized photographs have a scale of about $1\farcs14$) and
tracking inaccuracy in the longer fuv2419 exposure. We identified a
few instances of blending in the UIT observations outside this radius
when stars on the blue tail of the HB were close enough together on
the sky. We rated UIT detections by hand based on the apparent degree
of contamination, and will only plot stars with the best ratings.

We note that only in one case (ID 72) did we find a star that was very
likely to be a proper motion nonmember, and in two other cases (UIT
source 568; ID 3904) did we find stars that could not be identified as
HB or AGB stars. ID 72 has optical photometry that places it only
about 0.2 mag brighter than the HB level. We found no clear
counterpart to UIT source 568 ($m_{1620} = 15.98$), and hypothesize
that it might be related to the faint UV sources found in HST
photometry by \citet{faintuv}. Source (ID 3904) corresponds to a
bright blue straggler star in optical photometry.

\subsection{Photometry Calibration\label{photcal}}

\subsubsection{CFHT\label{cfhtcal}}

We calibrated the $BVI$ PSF photometry from CFHT to the standard system using
standard stars from M13 described in \citet{stet}.  These standard stars are
rigorously matched to the same photometric system as the earlier \citet{lan92}
study. We obtained the 2 May 2005 update star list from the CADC
website\footnote{\tt http://cadcwww.dao.nrc.ca/standards/}. The conditions for
the CFHT observations did not appear to be photometric, so the magnitude
zeropoints for each image were free parameters in the calibration.

\begin{figure}
\epsscale{.90}
\plotone{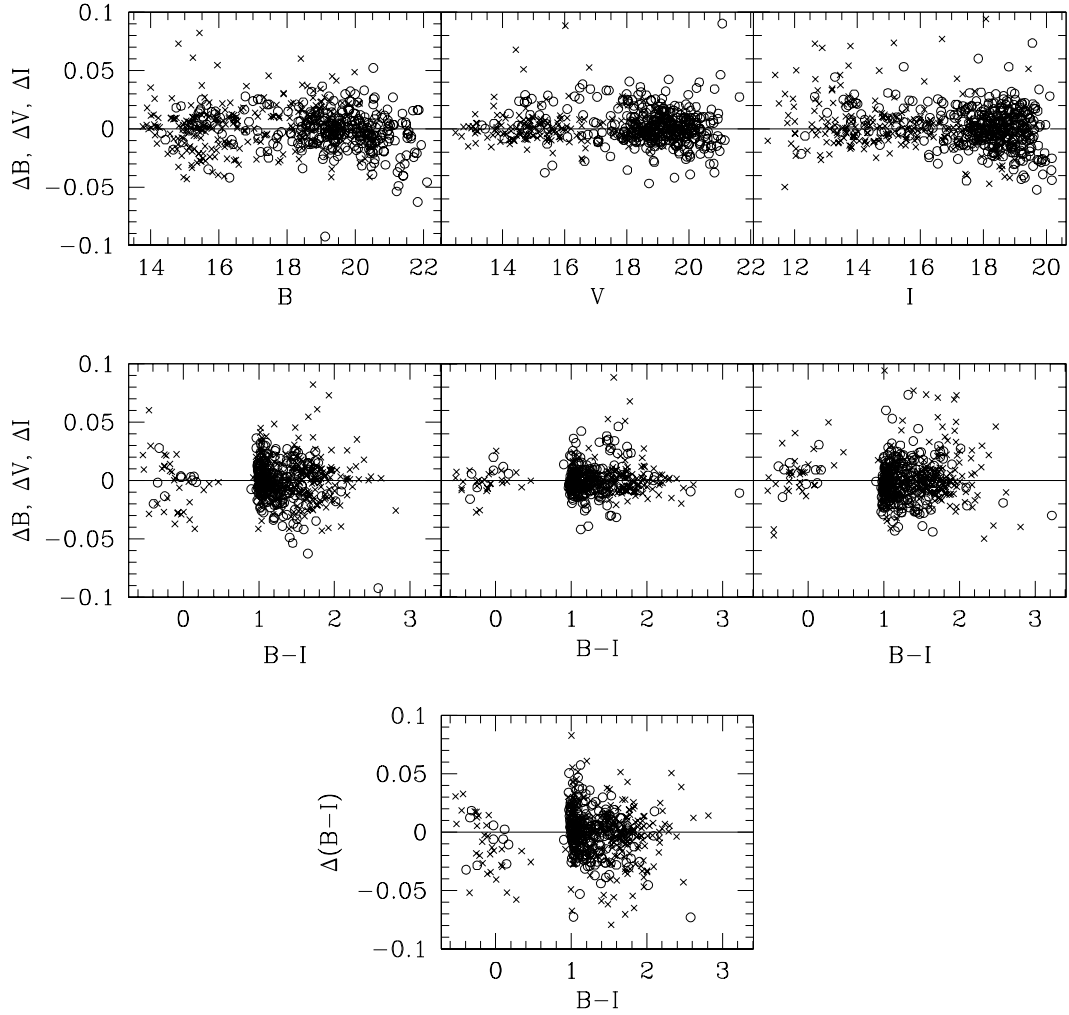}
\caption{Photometric residuals between our calibrated photometry and the
  standard magnitudes of \citep{stet} in the sense of this study minus
  Stetson's.  Stars measured on chip 11 of the CFHT field are shown with open
  circles, and stars from chip 12 are shown with crosses.\label{rsdstet}}
\end{figure}

Standard stars in the Stetson field cover almost the entire range of colors for
evolved stars: from stars at the faint end of the HB distribution to a giant
star within 0.5 $V$ mag of the RGB tip. Three redder stars (probably field)
were included to extend the color calibration closer to the color of the tip
($B-I \approx 3.1$). We found it
necessary to calibrate the two CFHT12k CCD samples separately because of
significant differences in color terms. (This is consistent with the results
of other studies --- for an example, see \citealt{kali}.)  Using the CCDSTD
\citep[e.g.,][]{stet92}
program, the standard star transformation equations were found to be:
\[ b = B + a_{i}+(-0.025\pm0.006)(B-I)+(0.0084\pm0.0029)(B-I)^2\]
\[ v = V + b_{i}+(-0.0038\pm0.0026)(B-I)\]
\[ i = I + c_{i}+(-0.0079\pm0.0025)(B-I) \]
for chip 11, and 
\[ b = B + a_{i}+(-0.030\pm0.004)(B-I)+(-0.0173\pm0.0018)(B-I)^2\]
\[ v = V + b_{i}+(-0.0084\pm0.0013)(B-I)\]
\[ i = I + c_{i}+(-0.0011\pm0.0012)(B-I) \]
for chip 12, where $b$, $v$, and $i$ are instrumental magnitudes, $B$, $V$, and
$I$ are standard magnitudes, and $a_{i}$, $b_{i}$, and $c_{i}$ are the zero
points for individual frames. The $(B-I)$ color was chosen primarily for its
wide wavelength spacing, which helps minimize the importance of photometric
errors in $B$ or $I$. We compared the final
calibrated measurements with the Stetson standard values, as shown in Fig.
\ref{rsdstet}. The comparisons include 417 stars from chip 11 and 286 stars
from chip 12. The median residuals $\Delta B$, $\Delta V$, $\Delta I$,
and $\Delta (B-I)$ were all less than 0.003 mag for the samples from both
CCDs, and consistent with zero. There do not seem to be significant systematic
trends at the extremes of the color range either.

\subsubsection{HST}

For most purposes, we preferred to leave the HST datasets in flight
system magnitudes using the VEGAMAG zeropoints in order to maintain
the relative precision of the original photometry and to make it
possible for others to use improved transformations to a standard
system in the future. However, to produce uniform samples of stars
from photometric datasets in different regions of the cluster, we did
need to derive some transformations.

\begin{figure}
\epsscale{.90}
\plotone{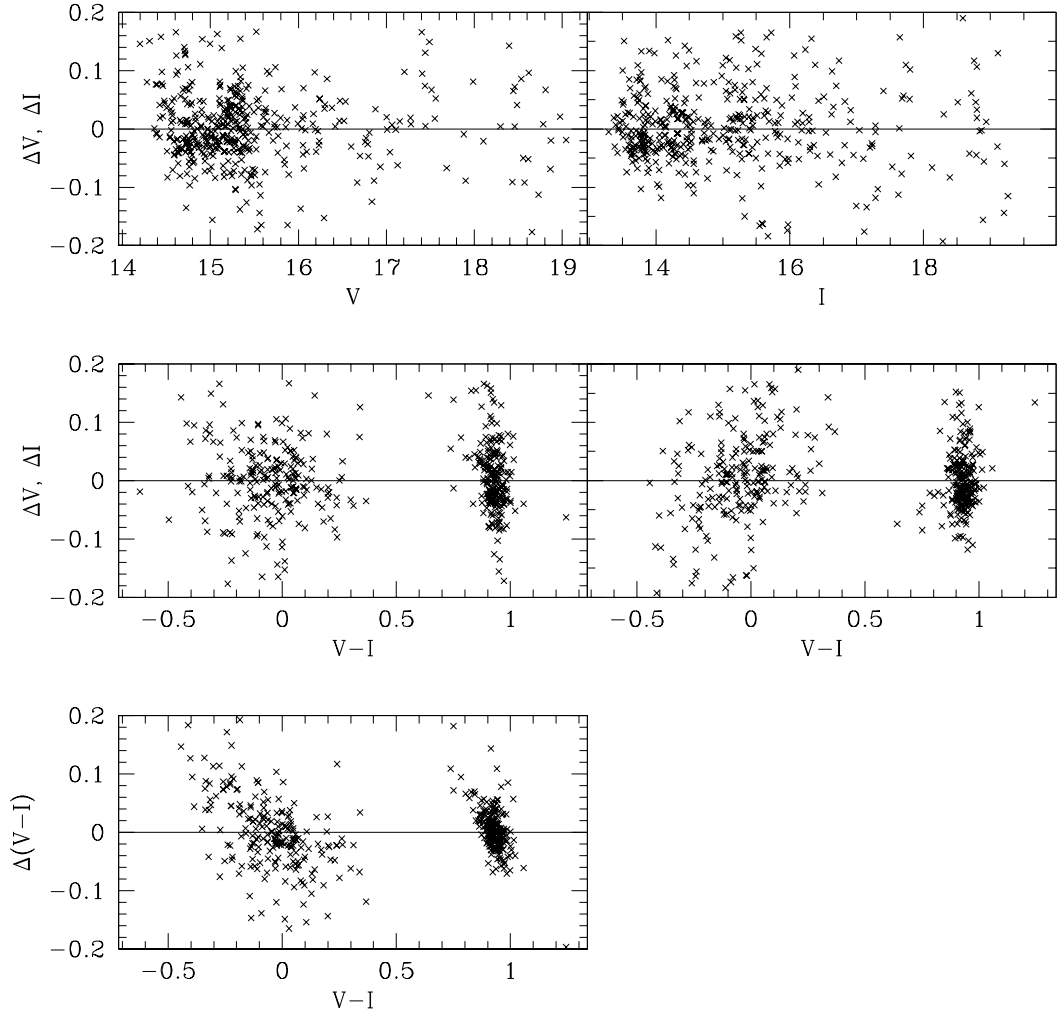}
\caption{Photometric residuals between our calibrated CFHT photometry and the
  ACS photometry calibrated to the standard system (in the sense of CFHT minus
  ACS). \label{rsdacs}}
\end{figure}

Because the CFHT photometry appears to be accurately matched to the
standard system and because the CFHT images had good spatial
resolution in the cluster core, we use this dataset to examine the
calibration of the ACS WFC photometry in $V$ and $I$. The comparison
involves HB and faint RGB stars because stars
high on the RGB were heavily saturated even on the shortest
exposures. In comparing the ACS flight system magnitudes to the CFHT
photometry, it was clear that second-order color terms in color were needed to
transform the F606W observations, but a first order term seemed sufficient
to transform the F814W observations. This is in agreement with the
transformations derived by \citet{sir}. However, we found that when we used
the \citeauthor{sir} transforms, the RGB stars were systematically too faint
by about 0.04 mag. We therefore derived our own transformations. It was
necessary to iterate the process, eliminating stars with large positive
residuals (most likely due to blending in the CFHT dataset).
The following transformations were derived from 437 stars:
\[ m_{606} = V + (0.010\pm0.004)+(-0.0592\pm0.0174)(V-I)+(-0.1553\pm0.0207)(V-I)^2\]
\[ m_{814} = I + (-0.005\pm0.004)+(0.0290\pm0.0053)(V-I) \]
Fig. \ref{rsdacs} shows the comparison between the CFHT photometry and the ACS
photometry calibrated to $VI$.

\begin{figure}
\epsscale{.90}
\plotone{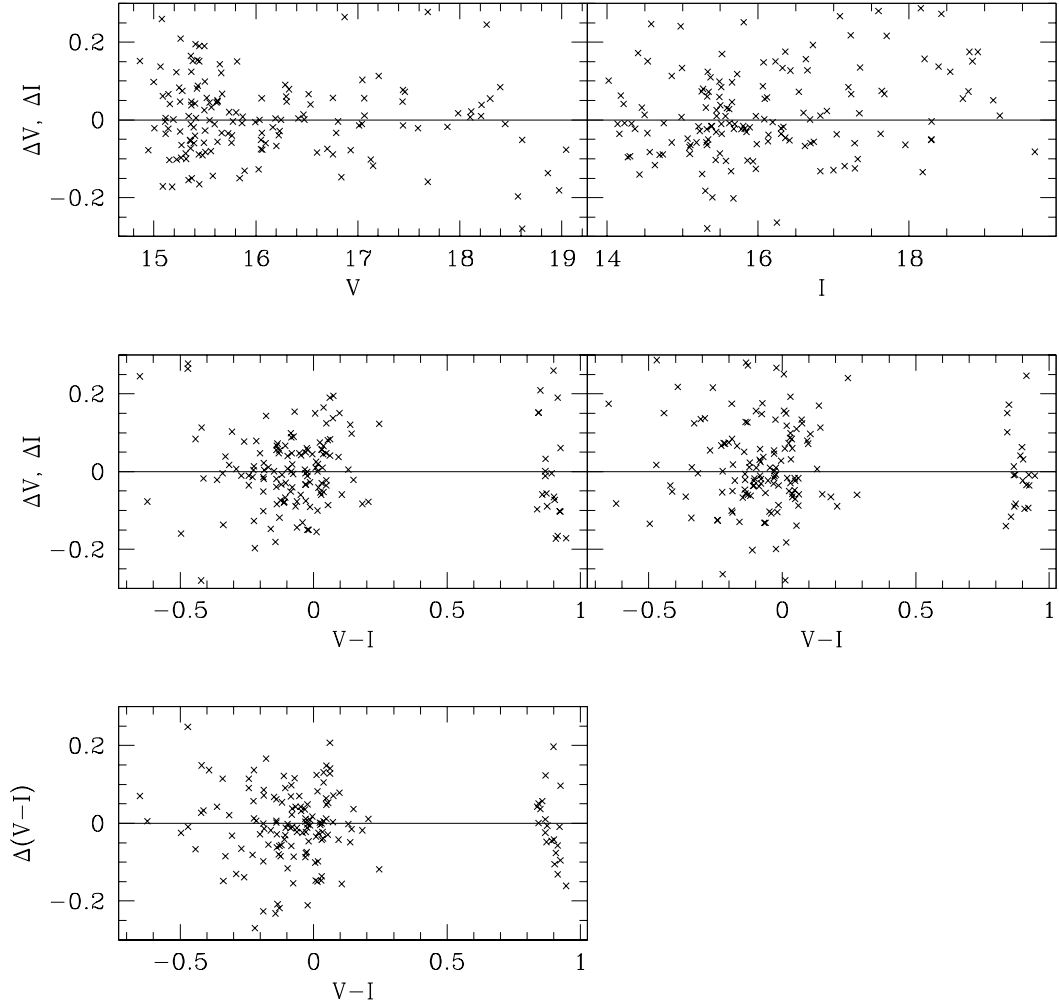}
\caption{Photometric residuals between our calibrated CFHT photometry
  and the WFPC2 photometry (F555W and F814W filters) calibrated to the
  standard system (in the sense of CFHT minus WFPC2). \label{rsdwfpc}}
\end{figure}

For a small number of stars, WFPC2 images were the primary
source for photometry, thanks to crowding and/or gaps in sky coverage.
We used WFPC2 photometry in the VEGAMAG system, but have
calibrated measurements in the F555W and F785LP bands to $V$ and $I$. We first
used the standard calibration of \citet{hstcal}, but found that there were
large color-dependent residuals. As a result, we determined new transformation
equations using 153 RGB and HB stars in common with the CFHT data (eliminating
stars that were most likely to be blended, as we did with the ACS data):
\[ m_{555} = V + (0.065\pm0.009)+(0.0860\pm0.0364)(V-I)+(0.0207\pm0.0498)(V-I)^2\]
\[ m_{785} = I + (0.051\pm0.010)+(0.1346\pm0.0413)(V-I)+(-0.2869\pm0.0557)(V-I)^2 \]
The transformed WFPC2 and CFHT datasets are compared in Fig. \ref{rsdwfpc}.
Although the formal errors on the quadratic term for the $m_{555}$ equation imply it has low significance, it was our qualitative judgement that it improved the transformation at the red end of the color range.

For a small number of stars at the faint end of the RGB in the cluster core,
the only reliable near-infrared measurements were in the WFPC2 field of
proposal 8278, and so this was also calibrated to CFHT measurements using a
sample of 87 RGB and HB stars:
\[ m_{555} = V + (0.052\pm0.011)+(0.0538\pm0.0428)(V-I)+(-0.0378\pm0.0629)(V-I)^2\]
\[ m_{814} = I + (0.021\pm0.012)+(0.3233\pm0.0481)(V-I)+(-0.3328\pm0.0688)(V-I)^2 \]
In the color range covered by the stars with WFPC2 photometry, the two
calibrations typically agree to better than 0.05 mag.

\subsubsection{KPNO}

We observed Landolt standard fields on one photometric night from the KPNO
run, and a standard star calibration using these data was used in calibrating
cluster data from previous studies \citep{har,pol,fek}. In this paper we have
opted to calibrate the KPNO photometry directly against the well-observed
Stetson (2000) standards in the M13 field used to calibrate the CFHT data.  To
ensure the calibration of the evolved populations, we restricted the standard
star sample to 343 stars brighter than the approximate turnoff of the cluster
($V < 19$).  The transformation equations are
\[ b = B + a_{i}+(-0.0279\pm0.0018)(B-I)\]
\[ v = V + b_{i}+(-0.0002\pm0.0017)(B-I)\]
\[ i = I + c_{i}+(-0.0130\pm0.0020)(B-I)\]

\begin{figure}
\epsscale{.90}
\plotone{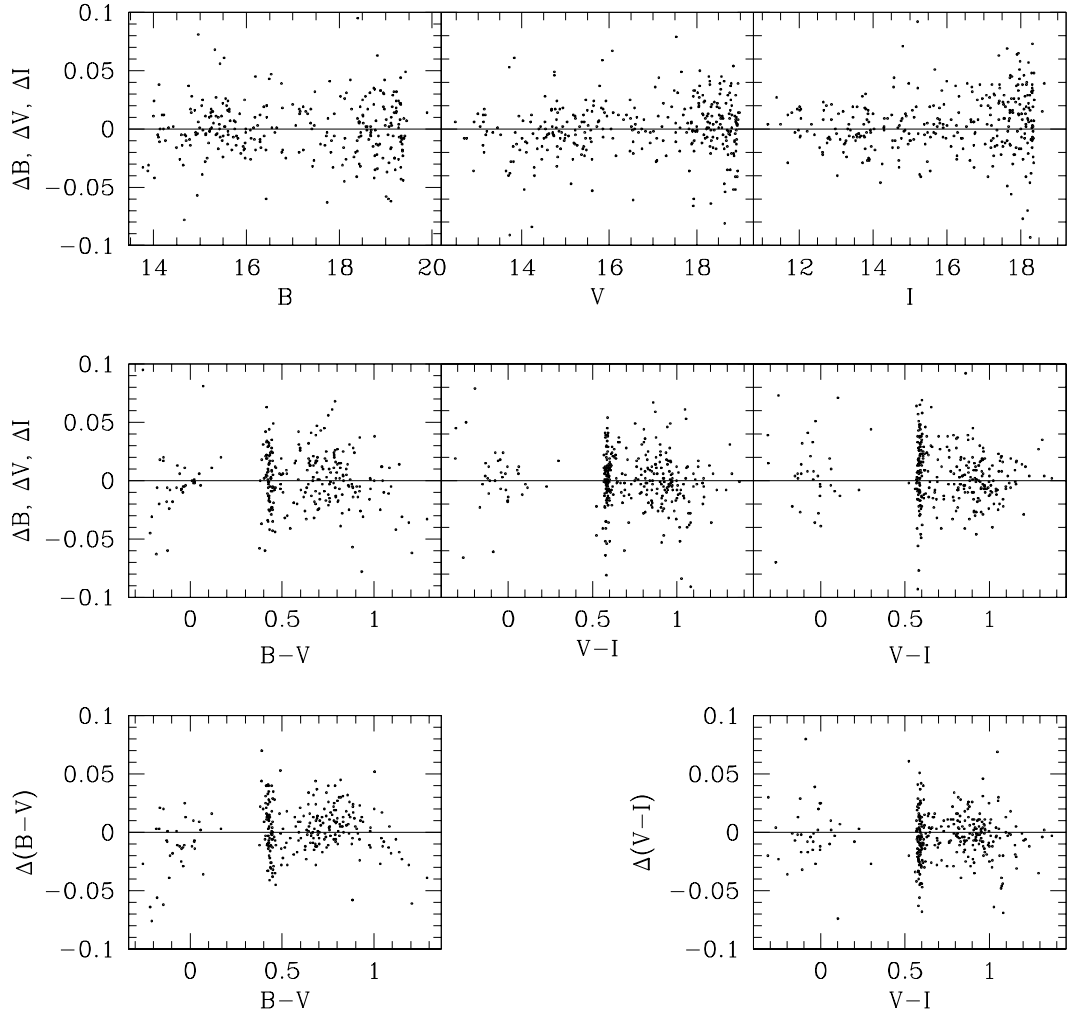}
\caption{Photometric residuals between our calibrated KPNO photometry and the
  standard magnitudes of \citet{stet}. \label{rsdkpno}}
\end{figure}

The residuals of the calibration are shown in Fig. \ref{rsdkpno}.  The
median residuals in all cases ($B$, $V$, $I$, $B-V$, and $V-I$) are
less than 0.005 mag, and well within the uncertainty in the fit.
There may be some small systematic trends at the extreme ends of the
color range for the standard stars: $(B-V) \la -0.1$ and $(B-V) >
1.2$. After additional experiments though, we found that second-order color
terms or changes to the standard star sample (fewer main sequence
turnoff standard stars, so that the giant and horizontal branch stars
carried more weight) did not significantly improve the fits.

\begin{figure}
\epsscale{.90}
\plotone{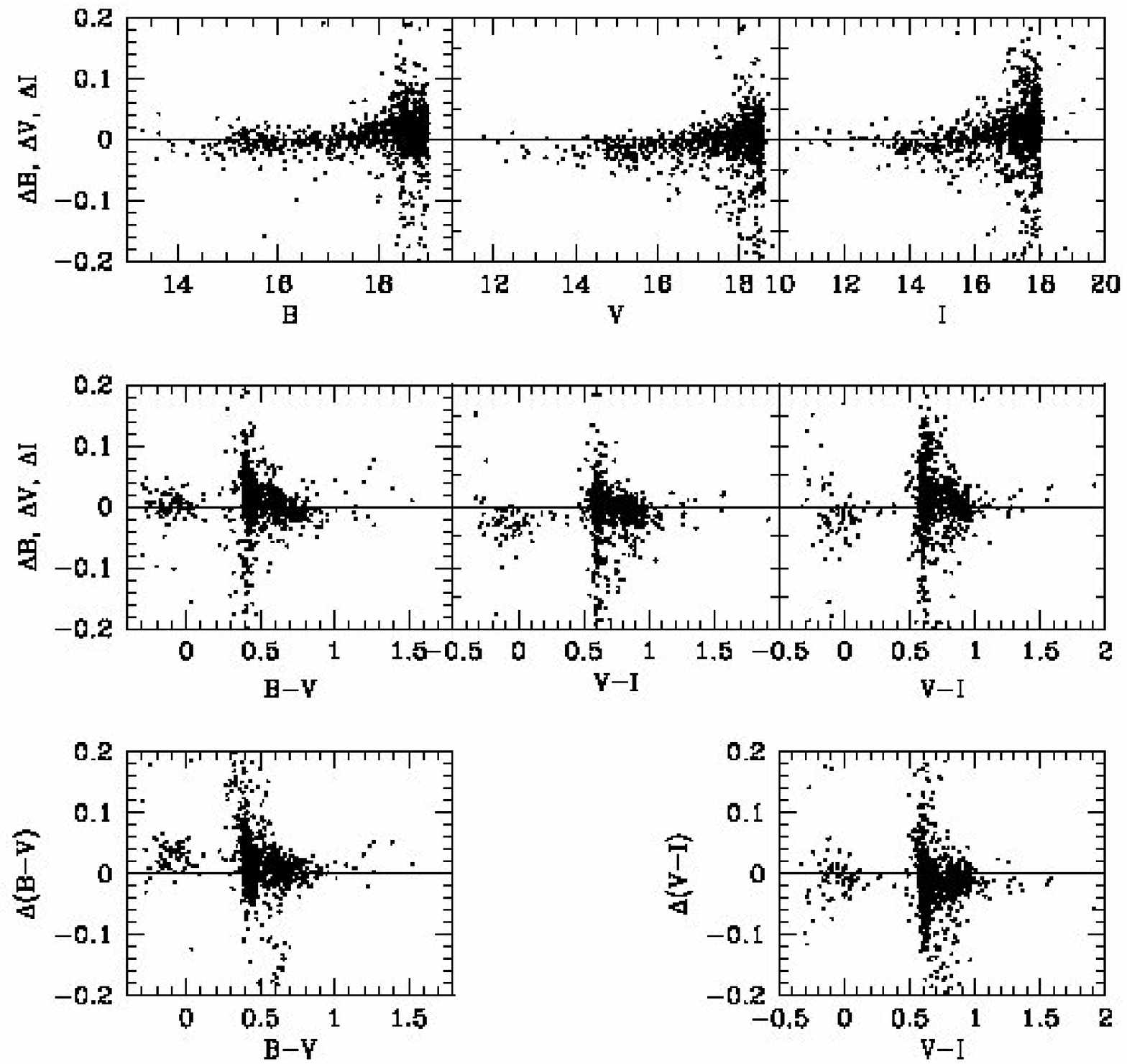}
\caption{Photometric residuals between our calibrated KPNO and CFHT photometry 
  (chip 11) in the sense of KPNO minus CFHT. The sample has been restricted
  to stars brighter than the main sequence turnoff (with the exception of
  faint HB stars). \label{rsdkc11}}
\end{figure}

\begin{figure}
\epsscale{.90}
\plotone{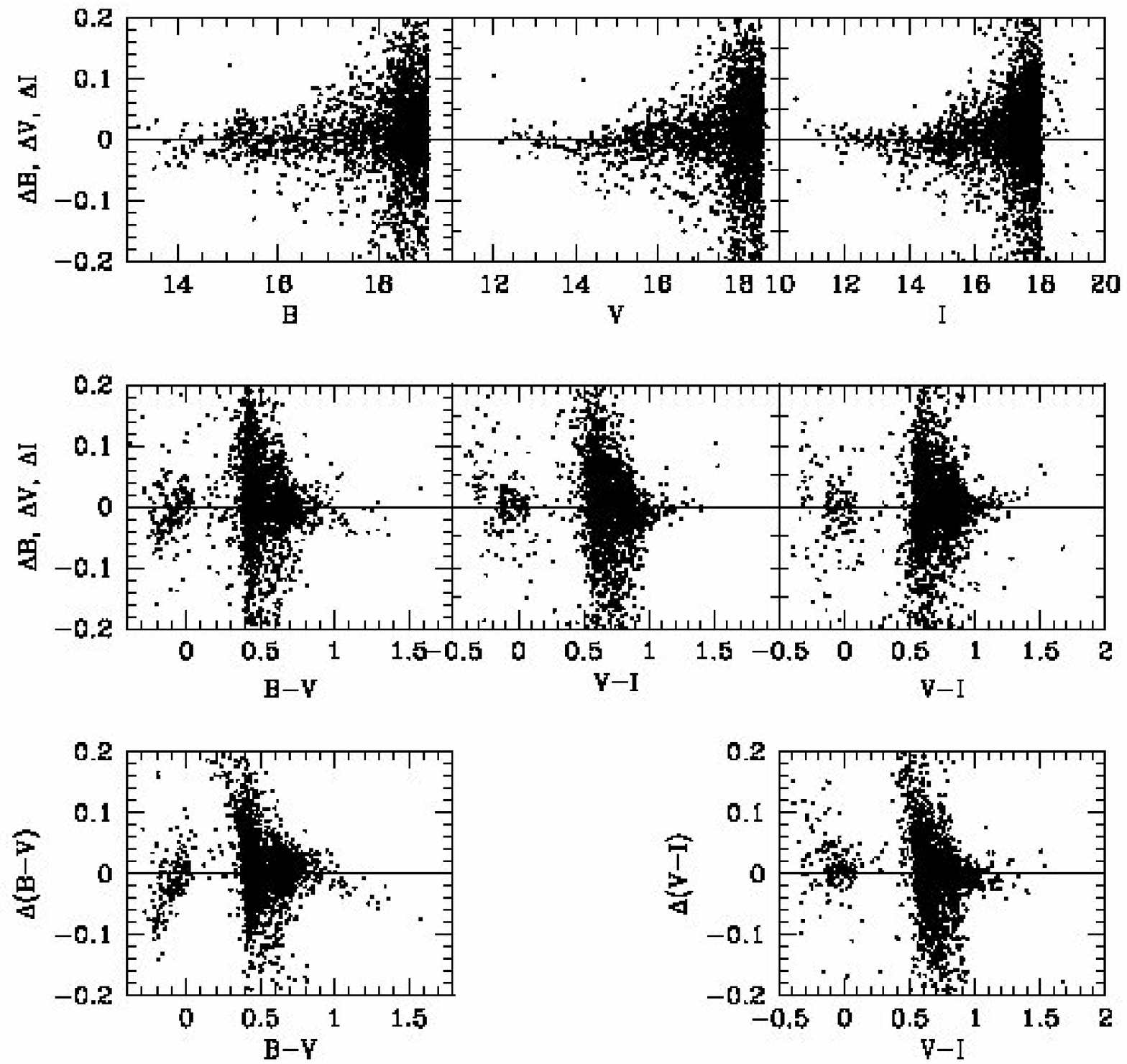}
\caption{Photometric residuals between our calibrated KPNO and CFHT photometry 
  (chip 12) in the sense of KPNO minus CFHT. The sample has been restricted
  to stars brighter than the main sequence turnoff (with the exception of
  faint HB stars). \label{rsdkc12}}
\end{figure}

For most of the scientific applications later in the paper, the relative
precision of photometric measurements within one dataset is most important.
However, for the discussion of HB and RGB luminosity functions, it is
important that there are not large systematic errors between the datasets we
have merged together.  The luminosity functions will employ the $I$-band
measurements, so particular attention should be focused on that filter.
Figures \ref{rsdkc11} and \ref{rsdkc12} show comparisons between the KPNO
photometry and the CFHT photometry from chip 11 and chip 12.  There appear to
be signs of systematic residuals in various places [notably some small offsets
in $V$ and $I$ in chip 11, and a trend in $\Delta (B-V)$ versus $(B-V)$ for chip
12].  However, for the great majority of stars, the residuals are quite small
($\la 0.04$ mag). In the KPNO-CFHT comparisons, the $I$-band measurements seem
to be well calibrated.

\section{Identification of Evolutionary Status\label{sel}}

Our first step was to first identify all known variable stars (RR
Lyrae, BL Her, bright giants) from \citet{kopacki02} in order to
eliminate stars that might confuse classifications.  Our subsequent
procedure for separating stars by evolutionary stage involved all of
the datasets described above (from widest area coverage to smallest:
PM, KPNO, CFHT, HST ACS, and HST WFPC2). We relied primarily on the
observations with the highest available spatial resolution for each portion of
the field. 

\begin{figure}
\epsscale{0.9}
\plotone{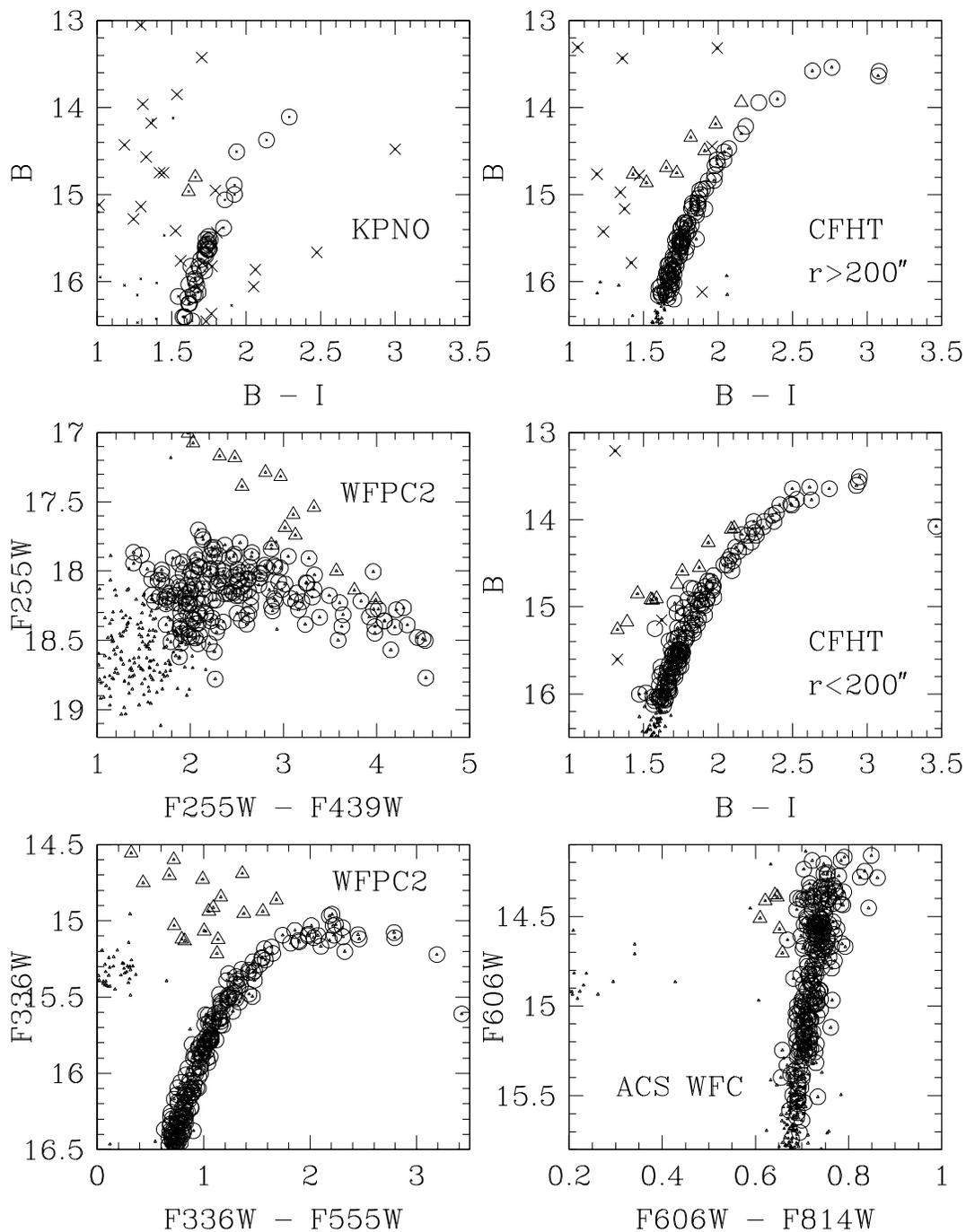}
\caption{Color-magnitude diagrams used in the identification of bright RGB
  stars. RGB stars have $\bigcirc$, AGB stars have $\bigtriangleup$, and known
  field stars have $\times$ symbols.\label{rgbsel}}
\end{figure}

The main CMDs used to identify giant stars (RGB and AGB) are shown in
Fig. \ref{rgbsel}.  For the outermost portion of the field, the KPNO dataset
was used along with proper motion information. The CFHT dataset was used for
the bright RGB and AGB samples even within the core, driven by the fact that
the brightest cluster stars were saturated in the HST images (with the
exception of some exposures in the ultraviolet). Crowding effects were minor
for the CFHT images, which were taken with good seeing and with a high
resolution camera (0\farcs2 pixels). Because giant stars (RGB and AGB) were
all observed in $BVI$ filters and because the $B-I$ color covers the widest
wavelength range, we used $B-I$ as their primary discriminant.  For fainter
RGB stars ($I > 14.5$), HST photometry was available in the core, and so ACS
WFC photometry was used when available, and WFPC2 photometry for all other
stars falling between WFC chips or outside the WFC field.


The HB and AGB manqu\'{e} are comprised of hotter stars, so we relied
on the shortest wavelength filters to identify them. For the portions
of the core observed in the ultraviolet, the F160BW, F255W, and F336W
filters on HST are particularly good discriminants (see
Fig. \ref{hbuv}), with the F160BW observations effectively selecting
{\it only} the hottest HB stars and a handful of manqu\'{e} stars.
Otherwise the HST ACS observations were the primary source for
identifications for the bluest horizontal branch stars in the core. HB
candidates in the core were selected based on CMD position, but
derived $\chi^2$, sharpness, and crowding values were used to flag
candidates that might be spurious.  Questionable candidates were also
examined by eye, and stars that clearly fell on chip defects or very
close to brighter stars were rejected. The crowding parameter in
DOLPHOT (the change in measured brightness if close neighbors had not
been simultaneously fit with point-spread functions) was very
effective at flagging spurious stars. We found that a change of 0.4
mag indicated a spurious star in nearly all cases. The optical CMDs
used in the HB star selections are shown in Fig. \ref{hbopt}.

Just outside the observed HST fields, there is noticeable scatter in
the photometry measured from the CFHT images, as can be seen in the
lower right panel of Fig. \ref{hbopt}. To reliably identify blue HB
stars in this region, we made use of UIT observations.  UIT
observations cover almost the entire field discussed in this paper,
and even though blending sometimes made accurate photometry a problem
in the core, the UIT observations allowed us to identify blue HB stars
even there. As a result, we believe we have virtually all of the faint
HB stars in our sample, all the way from the cluster center out to the
edge of our large ground-based fields.

AGB \manq ~ stars also deserve discussion. In optical CMDs these stars
(post-HB stars that will not reach the traditional AGB) are predicted
to evolve almost parallel to the zero-age HB (ZAHB), and the hottest
ones can sometimes be hard to distinguish from normal HB
stars. Ultraviolet-optical CMDs can flatten out the blue HB and make
it possible to identify AGB manqu\'{e} stars by their relative
brightness. For the majority of the field we identified candidates
from the HB sample in the highest resolution optical CMD first. We
subsequently used observations in F160BW and F255W for the core, and
the wide-field UIT observations elsewhere to validate most of the
candidates. We found that nearly all of the bluest candidates selected
from the optical CMDs were verified with the ultraviolet observations,
but 5 additional manqu\'{e} stars were identified first in the UV.
For the UV observations, we defined stars to be manqu\'{e} if they were more
than about 0.8 mag brighter than the faintest HB stars at the same
color. (This criterion is based on where theoretical tracks predict evolution
slows again after core He exhaustion.) The UIT observations were subject to
blending in the central regions and some candidates were rejected based on
high-resolution optical photometry, but in all cases it was possible to
identify the stars that were the primary UV sources based on other photometry.
In addition, some of the \citet{whit} sources did not appear to have cluster
counterparts at the tabulated positions. We do not have an explanation for
this. Two stars (HB 72 and 295) present reasonably strong cases in optical
CMDs for being \manq, but are not discernibly unusual in the UV, so we have
left them as HB stars. Finally, two \manq ~ stars (AGB 12 and 26) were
identified using only optical photometry because they fell outside the HST
ultraviolet observations, but in a region of the cluster where the UIT images
were too crowded to yield reliable measurements. All of our cross
identifications are given in Table \ref{idtab}. The photometry for the
identified stars can be found in Tables \ref{hbtab}-\ref{nmtab}. With the use
of the UV observations, we believe we have isolated nearly all AGB manqu\'{e}
stars from the HB sample.

\section{The Red Giant Branch}\label{rgb}


We assembled a comprehensive tabulation of the bright RGB stars in M13 from
the tip of the RGB to the base of the RGB, where it meets the subgiant branch.
As described in \S \ref{sel}, we eliminated AGB stars from the sample using
the best photometry available (see Fig. \ref{rgbsel}). This is a relatively
easy task in M13 because the AGB is well-separated in color from the RGB
except near the bright end.

In a study of the cluster core using the HST Planetary Camera,
\citet{cohm13} previously found evidence that there was an excess of
bright giants compared to models.  Though their photometry was not
precise enough to separate AGB and RGB stars, the excess appeared to
be too large to be explained by the AGB star population. To examine
the evolution rates of the bright giants, we followed the procedure
described by \citet{sm}. In this method theoretical cumulative
luminosity functions in $I$ band are shifted in magnitude so that the
tip of the giant branch matches the brightest giant, and the
luminosity function is then normalized to the total number of stars
found brighter than the RGB bump. \citeauthor{sm} found that this has
the benefit of making the comparison extremely insensitive to chemical
composition and age, which makes it possible to test the physics of
the upper giant branch in clusters with large giant populations. In
that study, \citeauthor{sm} found that the giant population in NGC
2808 was significantly depleted relative to the model predictions, and
hypothesized that this could be due to underestimated neutrino
emission rates or loss of the giant envelope (causing some giants to
leave the RGB before undergoing a normal helium flash). We have
therefore tested whether M13's very blue HB morphology is
reflected in the character of the giant population. \citet{sh}
recently examined a large sample in the core of the distant cluster
NGC 2419 (another cluster with a long blue HB tail), which showed only
weak signs of a depleted bright giant population.

To conduct the test for M13, we shifted Victoria-Regina \citep{vr} and
Teramo \citep{ter} models to the $I$ magnitude of the brightest giant. We need
to be careful because almost all of the brightest giants in M13 are
known to vary irregularly on timescales of more than 40 d \citep{kopacki02}.
The variability itself should be a relatively small effect based on the
variation amplitudes observed by \citet{kopacki02}: $0.07 < \Delta V < 0.38$
for the 12 stars studied, with most having amplitudes toward the low end of
that range. We identified all of the known RGB variables, and find that V11
appears to be the brightest ($I \approx 10.3$) in our study and most others.
(Our $V$-band measurements appear to have been made in brighter periods in the
star's variation. According to \citeauthor{kopacki02}, the star's mean $V$
magnitude is only a few hundredths of a magnitude different from other bright
giants.)

Fig. \ref{fig9} shows the comparison for a model having [Fe/H] $= -1.41$ and
[$\alpha$/Fe] $= +0.3$. We tested models with different metallicity, but as
described earlier, the exact value does not significantly affect the results.
In addition, the Victoria-Regina and Teramo models agree well with
each other.  The plotted Victoria-Regina model shows the best agreement with
the observed position of the RGB bump. We conducted Kolmogorov-Smirnov (K-S)
tests with the observed and theoretical luminosity functions using different
faint cutoffs for the RGB sample. (Because the RGB sample increases
exponentially with increasing magnitude, this was necessary to ensure that
deviations on the less-populated bright RGB are not washed out.) The test
results are presented in Table \ref{tbl-2}. The probability $P$ that the
observed stars are consistent with being drawn from the theoretical
distribution reaches a minimum of about 5\% when the cutoff is at $I = 12.76$
in response to a clump of stars at $I \approx 12$ (see below). Otherwise,
the probabilities remain greater than 15\%, and the overall impression from
the test is that there is at most a slight deficit of stars compared to the
model predictions, in contrast to the \citet{cohm13} results.

As discussed by \citet{sm}, this method of comparing with models is
insensitive to chemical composition for [Fe/H] $\la -1$, which means that it
should be possible to combine samples from different clusters to improve the
statistical significance of the comparisons. M5 has a readily available sample
of RGB stars \citep{sb04}, and issues related to combining the two samples are
minimal.  For example, M5 and M13 have very similar distance moduli
[$(m-M)_{0,M13} - (m-M)_{0,M5} \approx 0.06$, \citealt{ferr99}; $-0.02$,
  \citealt{reb}; $-0.08$, \citealt{car}] and reddening [E$(B-V)_{M13} -
  $E$(B-V)_{M5} = -0.01$, \citealt{harris}; -0.02, \citealt{reb}] values, and
their metallicities are similar (with M13 universally identified as more metal
poor by about 0.2 dex).  The M5 sample is similar in size to the M13 sample as
well. In the comparisons below we use a model with a compromise [Fe/H] $=
-1.31$, although once again the composition choice is not critical.  In light
of the similar and relatively well-determined distance moduli, we have chosen
to combine the samples after correcting for the small difference in distance
modulus. This is probably superior to shifting the cluster samples so that the
brightest giants in each cluster match because i) many of the brightest giants
in M13 are variable, and ii) statistically speaking, the brightest giants may
fall at different brightness levels fainter than the tip of the red giant
branch (TRGB).

The combined sample is shown in Fig. \ref{clfcomb}.  K-S test probabilities
only go below 10\% in small ranges of magnitude (for example, $P = 0.05$ for a
cutoff at $I - I_{TRGB} = 3.00$).  Overall, we regard the agreement with the
models as good, and a confirmation of the most recent plasmon neutrino
emission rates \citep{hrw,itoh}, which influence the evolution rates near the
TRGB. There are variations in counts around the theoretical predictions, but
are not significant at more than a 2$\sigma$ level. With these samples, we see
no reason to consider particles beyond the standard model with very large mean
free paths (such as axions and WIMPS) that could affect cooling of the
degenerate core \citep[e.g.][]{cat96}.

The evolution rates for fainter giants can also be examined using star
numbers.  Using a sample covering a smaller portion of the cluster,
\citet{cho} found that M13 appeared to have ``extra'' stars in and
slightly fainter than the RGB bump. On the other hand, in their study
of the bumps of a large sample of clusters, \citet{riello} found that
M13 appeared to have a lower than expected number of bump stars (stars
with $V$ within $\pm 0.4$ mag of the bump center) compared to giant
stars between 0.5 and 1.5 $V$ magnitudes fainter than the bump. Both
\citeauthor{riello} and \citet{bono} found M13 had a lower value than
the mean for their whole sample of clusters, and in the case of
\citeauthor{riello} the value was almost $3 \sigma$ lower. Using
$V_{bump} = 14.75$, our M13 samples are more than twice as large as
the ones tabulated by \citeauthor{riello} We find $R_{bump} = 297 /
609 = 0.487 \pm 0.034$, which is quite consistent with the mean value
found by \citeauthor{riello} and with theoretical predictions
\citep{riello,bjork}. The slope of the lower giant branch in the
cumulative luminosity function (which is related to the evolution
timescale) for M13 is also in very good agreement with the predictions
of the Victoria-Regina and Teramo models. Based on our large sample,
the M13 giants are evolving at the rate predicted
by models.


As mentioned above, the CLF of M13 also shows a slight enhancement in the
number of RGB stars at $I \approx 12.0$ or $V \approx 13.15$. This feature can
be seen by eye in some of the CMDs (for example, see Fig. \ref{ivi}). This is
reminiscent of the ``heap'' feature seen in NGC 2808 by \citet{heap}, which
was identified about 1.4 $V$ magnitudes brighter than the RGB bump. This
feature falls approximately where the RGB and AGB begin to overlap. While
there have been quite a number of spectroscopic studies of M13, there is not
an unambiguous spectroscopic signature to separate AGB stars from RGB stars.
For example, \citet{sbcn} find that photometrically-identified AGB stars had
uniformly weak CN bands, indicative of O$\longrightarrow$N conversion.
However, there is a significant group of CN-weak RGB stars even fainter than
the level where they could reasonably be confused for AGB stars.


We present the differential luminosity function for the bright end of the RGB
in Fig. \ref{dlf}. As \citeauthor{heap} found for NGC 2808, the heap is not
statistically significant. Its appearance in the CMD comes largely from the
contrast with slightly fainter RGB stars, which appear to be less common than
predicted theoretically. In our M13 sample (which is close to as large as it
will get for this cluster), the heap appears to be significant at about the
$2.5\sigma$ level based on K-S tests on the cumulative luminosity function.




To summarize, after gathering the largest possible sample of giant stars in
the cluster, we see at most low-significance deviations from theoretical
luminosity functions. We emphasize that our luminosity function comparisons
have been done in a relative sense, by forcing agreement between the
brightness of the observed and theoretical TRGB. In contrast to the case for
NGC 2808 (but in agreement with NGC 2419), we do not see evidence of large
numbers of stars leaving the RGB early that could account for the
bluest HB stars in the cluster.

\section{The Horizontal Branch}

\subsection{Brightness and Radial Distributions}

M13 possesses a horizontal branch comprised mainly of stars bluer than the
instability strip (Fig. \ref{hbopt}).  Before we turn to comparisons of the HB
population with other post-main-sequence samples, we will first examine the
distribution of HB stars in the CMD and as a function of radius in the
cluster.

Our dataset has been derived from a wide-range of source material, but $V$
photometry is common to all of the sources, and $I$ photometry is available
for all but one of the sources (the proper motion data of \citealt{cmpm}). We
use $I$ photometry to describe the relative positions of stars on the HB
because it is an {\it observable} quantity, and HB stars grow monotonically
fainter in $I$ with decreasing mass. The relationship between $I$ magnitude
and mass results from competition between decreasing radius and increasing
surface temperature as the mass of the star's envelope decreases, and from the
rapidly increasing bolometric corrections that result. The use of a directly
observable coordinate for HB position also reduces the difficulties in
comparing models and observations \citep{vargas}.

Because of the potential influence of blending, we drew the $I$ magnitude of a
star from the photometric dataset with the highest spatial resolution: ACS
WFC, CFHT, and KPNO. Five stars fell in the gap between the two ACS WFC chips,
and their photometry was derived from WFPC2 observations in F785LP (\S
\ref{photcal}). In Fig. \ref{ihist}, we show a histogram of the HB star
distribution in $I$, clearly showing the well-known bimodality. For conceptual
purposes, we have broken the HB into three parts: $I < 16.25$ (the bright
peak, hereafter P1), $16.25 < I < 18$ (intermediate stars), and $I > 18$ (the
faint peak, hereafter P2).  The break at $I = 18$ corresponds to gap G3
identified by \citet{gaps}, while the break at $I = 16.25$ is slightly fainter
than their gap G1 and at approximately the same position as the ``$u$ jump''
identified by \citet{ujumpm13}. \citet{ujumps} tentatively identified this as
the place on the HB where radiative levitation of heavy elements becomes
important, and spectroscopic studies \citep{behr03,fabb05,pace07} confirmed a
change in the atmospheric abundances of heavy elements at similar positions in
a number of other clusters. The gap G1 appears to correspond to a local
minimum in the HB distribution at $I \approx 15.7$.  There is no doubt that
\citet{gaps} saw clear evidence of gap G1 in their dataset, but the gap in
their data was fairly wide and it is difficult to assign a precise location
using their smaller dataset.

Because the well-studied cluster M3 shows a clear change in HB morphology with
distance from the center \citep{catm3}, we examined the distribution for signs
of radial variations. In the top panel of Fig.  \ref{ihist}, we see that the
HB and RGB distributions are essentially identical. In the lower panel,
we plot the
HB distributions for samples with $r < r_h$ and $r > r_h$ (380 and 404 stars,
respectively). There appear to be slight differences between the sample
distributions, with the second peak at the end of the blue tail becoming less
prominent in the outer portion of the cluster. As shown in the cumulative
radial distributions in the bottom panel of Fig. \ref{crd}, there are modest
differences in the innermost 140\arcsec ~ of the cluster, with the intermediate
HB stars being the least centrally concentrated and the faint peak stars being
the most centrally concentrated. However, for this global comparison, a
Kolmogorov-Smirnov test indicates that there is still a 21\% chance that those
two distributions could be drawn from the same parent population.  When the
comparison is restricted to $r < r_c$ or $r_h$, the probability does not
change significantly.



Table \ref{tbl-hb} shows the number counts for different radial samples. The
counts show a tendency for stars in the faint peak to be more
centrally concentrated. They are almost as abundant as bright peak stars in
the very center of the cluster ($r < r_c / 2$), but the fraction drops quickly
so that for $r \ga r_c$, they are consistently about as abundant as the
intermediate group. There is a marginal trend in the fraction of HB stars in
the faint blue peak ($f_{P2}$) as a function of distance, but a significant
change in the difference in the fractions in the bright red and faint blue
peaks ($f_{P1} - f_{P2}$).

\citet{cohm13} found that the blue HB stars in M13 appear to be centrally
depleted relative to other types of stars for their sample with $r <
60\arcsec$.  We believe the difference is probably due to incompleteness at
the faint end of their horizontal branch sample.  The appearance of the
horizontal branch in their Fig. 4 clearly shows the bright peak, but there is
little or no sign of the faint peak (most stars in $18 < V < 19$), which falls
near the completeness limit of their dataset.


\subsection{Notable Stars}\label{notable}

Only nine RR Lyrae stars are known in the entire cluster, but there are stars
redder than the instability strip. We find a conspicuous group of eight stars
just bluer than the RGB but approximately 0.5 mag fainter than the AGB clump
(seven are seen in Fig. \ref{rhb}). A star at the red end of the theoretical
ZAHB locus plotted in Fig. \ref{rhb} has a mass of $0.8 \msun$, so HB stars in
this part of the CMD would be more massive than stars at the cluster turnoff
(the HB is theoretically expected to curve back toward higher temperature and
luminosity for more massive stars). We believe that these stars are probably
evolved counterparts to blue straggler stars, as originally proposed by
\citet{fusi}. This possibility could be tested by comparing the number to the
size of the sample of cluster stragglers (and thereby comparing their relative
lifetimes), but this is beyond the scope of the current work. Although field
stars are relatively common in this part of the CMD, five of the eight stars
have proper motions indicating that they are high probability members
\citep{cud}. The remaining three stars are all within $45\arcsec$ of the
cluster center. Only one of the stars (HB 532) has UV observations --- in the
F255W passband it is significantly brighter than the RGB but fainter than
hotter HB stars, which rules out an optical blend of a hot HB and an RGB star.

From careful examination of the photometry, we identified three additional
stars that we believe may be red HB stars. 
HB 314 has proper motion information that identifies it
as a high probability member and photometry that places it at $V$
magnitude nearly identical to stars blueward of the instability strip.  The
two remaining stars do not have proper motion information, and their CMD
positions give them greater probabilities of being field stars. HB 793 has a
$V$ magnitude placing it at the cluster HB level, and it is fairly blue
compared to the known field stars of similar brightness, but it is outside the
proper motion field ($r = 625\arcsec$). HB 133 is closer to the cluster center
($r = 103\arcsec$), but has a $V$ magnitude placing it about 0.3 mag fainter
than the cluster's HB.  However, there are no other stars near it in the CMD,
and there is a small possibility that it is an undetected variable star,
although it was within the field studied by \citet{kopacki02} using a
sensitive image subtraction method.



Although quite a few globular clusters have long blue horizontal branch tails
(NGC 6752, M2, and M80 are examples), M13 is somewhat unusual in having a
secondary peak in its distribution of stars that falls near the
theoretically-predicted blue end of the HB. Clusters that have blue hook stars
(such as $\omega$ Cen, NGC 2808, and NGC 2419) have blue HB tails in optical
CMDs that extend to fainter magnitudes. As a result, we have looked at the
faintest HB stars in the optical to gauge whether there are any
blue hook stars present. 

\begin{figure}
\epsscale{0.9}
\plotone{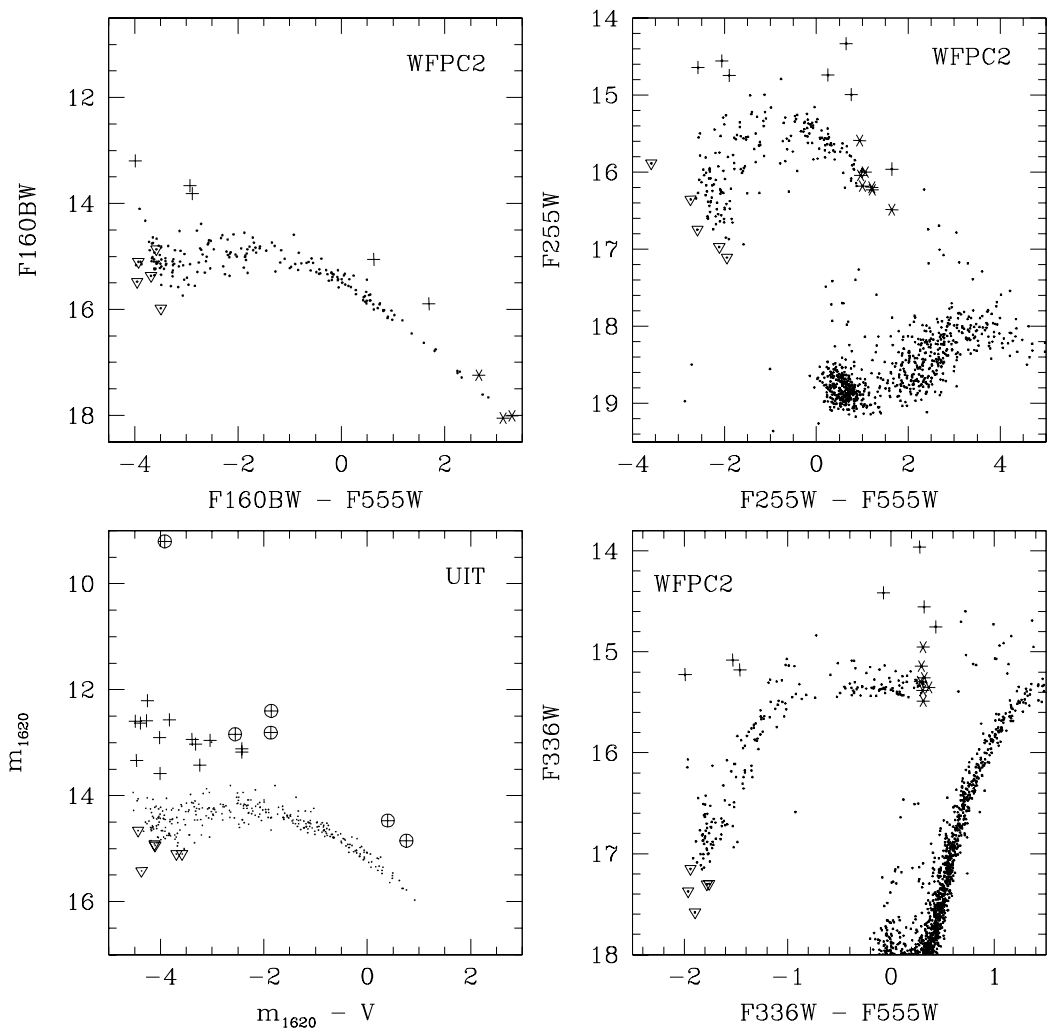}
\caption{Ultraviolet color-magnitude diagrams used in the identification of
  HB stars. Variable HB stars have $\ast$ symbols, AGB manqu\'{e} and supra-HB
  stars are $+$ symbols, ``UV-bright'' stars have $\bigcirc$ symbols, and the
  hottest HB stars have $\bigtriangledown$ symbols. In the UIT panel, the $V$
  magnitude comes from the images with the highest available resolution (ACS,
  CFHT, or KPNO).\label{hbuv}}
\end{figure}

\begin{figure}
\epsscale{0.9}
\plotone{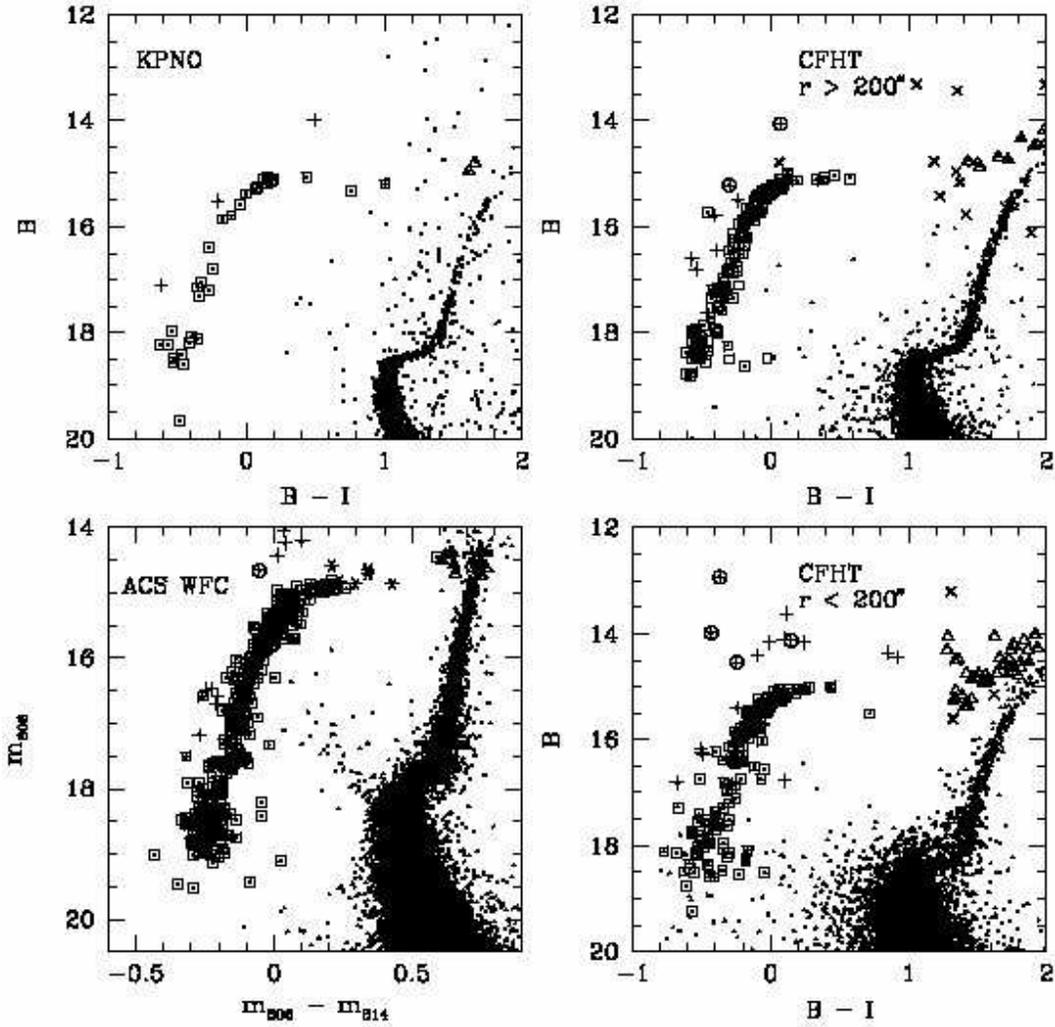}
\caption{Optical color-magnitude diagrams used in the identification of
  HB stars. Symbols are the same as in the right panels of Fig.  \ref{hbuv},
  with the exceptions that AGB stars are shown with $\bigtriangleup$ symbols,
  and unevolved non-variable HB stars are shown with $\sq$ symbols.\label{hbopt}}
\end{figure}

\begin{figure}
\epsscale{.90}
\plotone{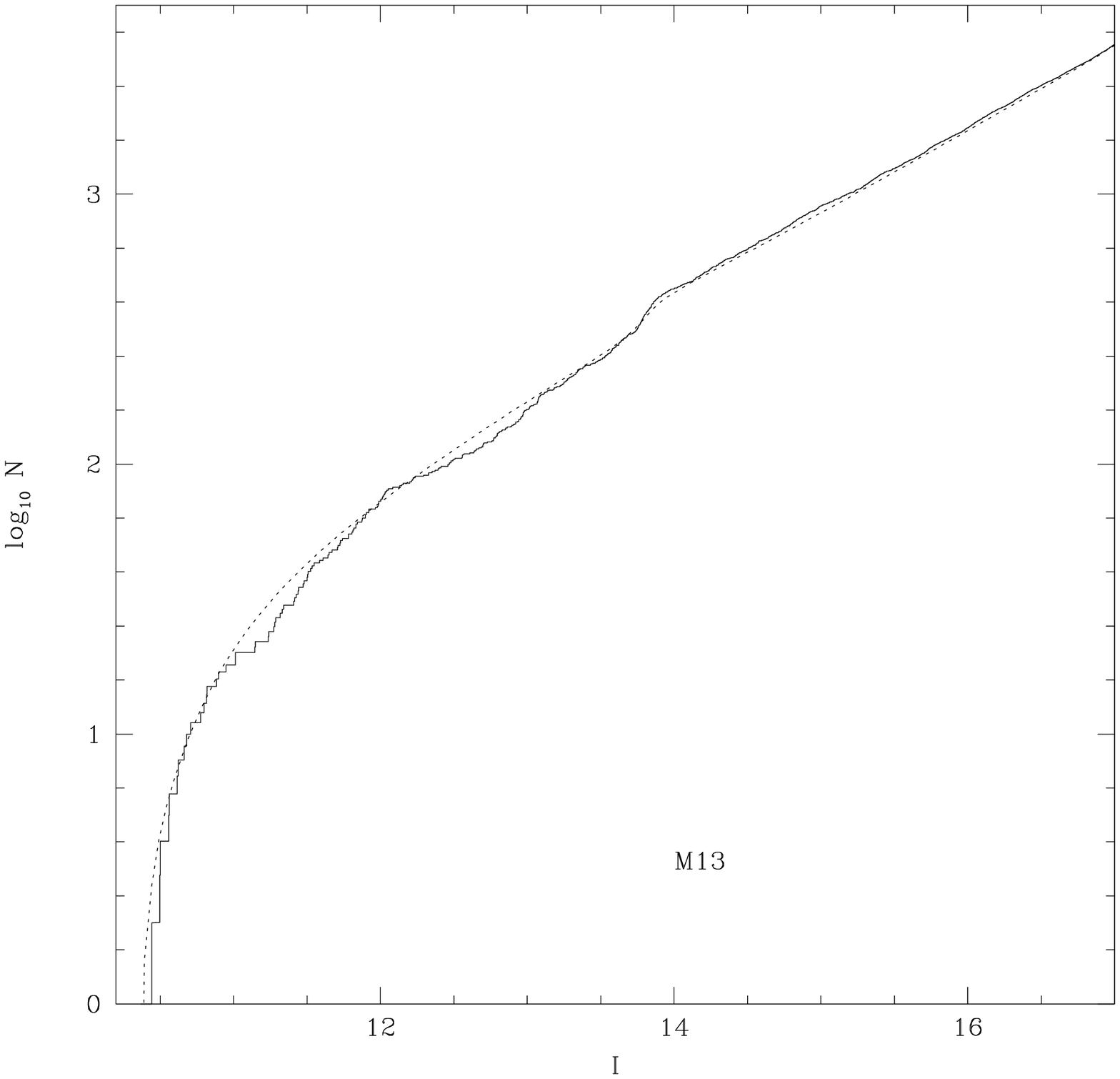}
\caption{Comparison of the cumulative luminosity function of 
  M13 RGB stars ({\it solid line}) with theoretical predictions from a
  Victoria-Regina model ({\it dotted line}; \citealt{vr}) with [Fe/H]$=-1.41$
  and [$\alpha$/Fe] $= +0.3$. The theoretical model has been shifted
  horizontally to match at the tip of the RGB, and vertically normalized to
  the RGB sample at $I = 13.7$, just brighter than the RGB bump \label{fig9}}
\end{figure}

\begin{figure}
\epsscale{.90}
\plotone{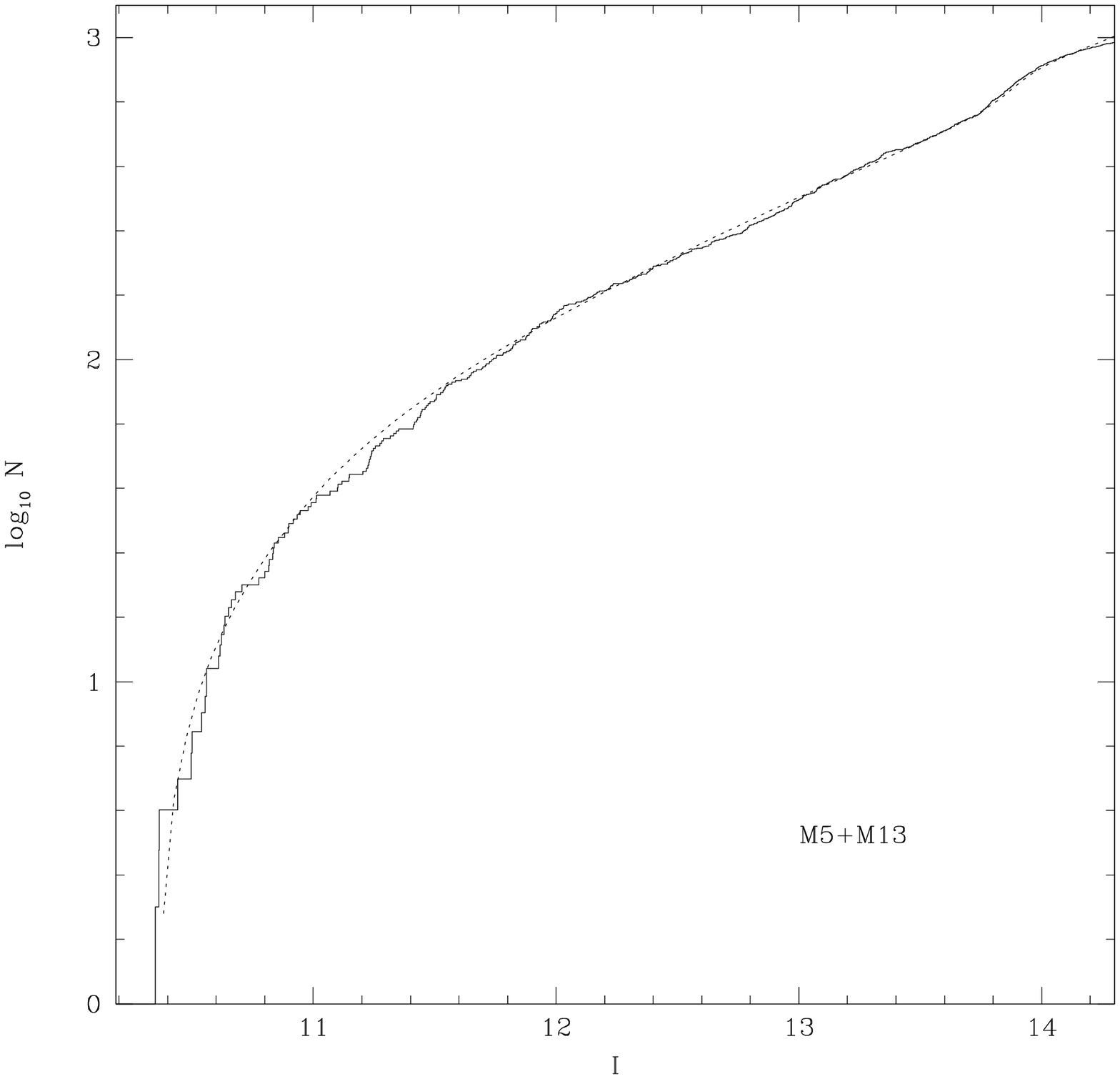}
\caption{Comparison of the combined cumulative luminosity function of 
  M13 and M5 RGB stars ({\it solid line}) with theoretical predictions from the
  Victoria-Regina model ({\it dotted line}; \citealt{vr}) with [Fe/H]$=-1.41$
  and [$\alpha$/Fe] $= +0.3$. The theoretical model has been shifted
  horizontally to match at the tip of the RGB, and vertically normalized to
  the RGB sample at $I = 13.7$, just brighter than the RGB bump \label{clfcomb}}
\end{figure}

\begin{figure}
\epsscale{.90}
\plotone{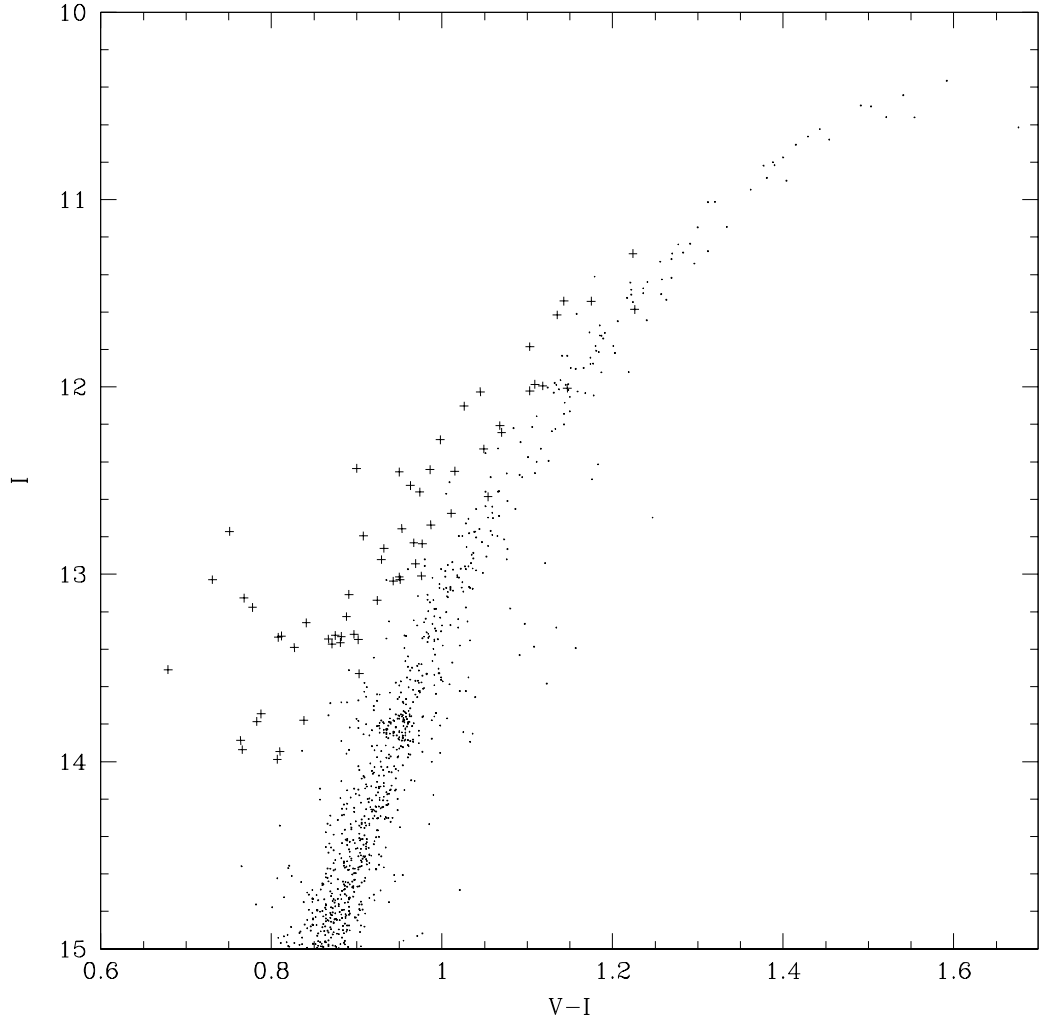}
\caption{Combined color-magnitude diagram for bright giant stars (AGB stars are
  identified with crosses) from all photometry sources.\label{ivi}}
\end{figure}

\begin{figure}
\epsscale{.90}
\plotone{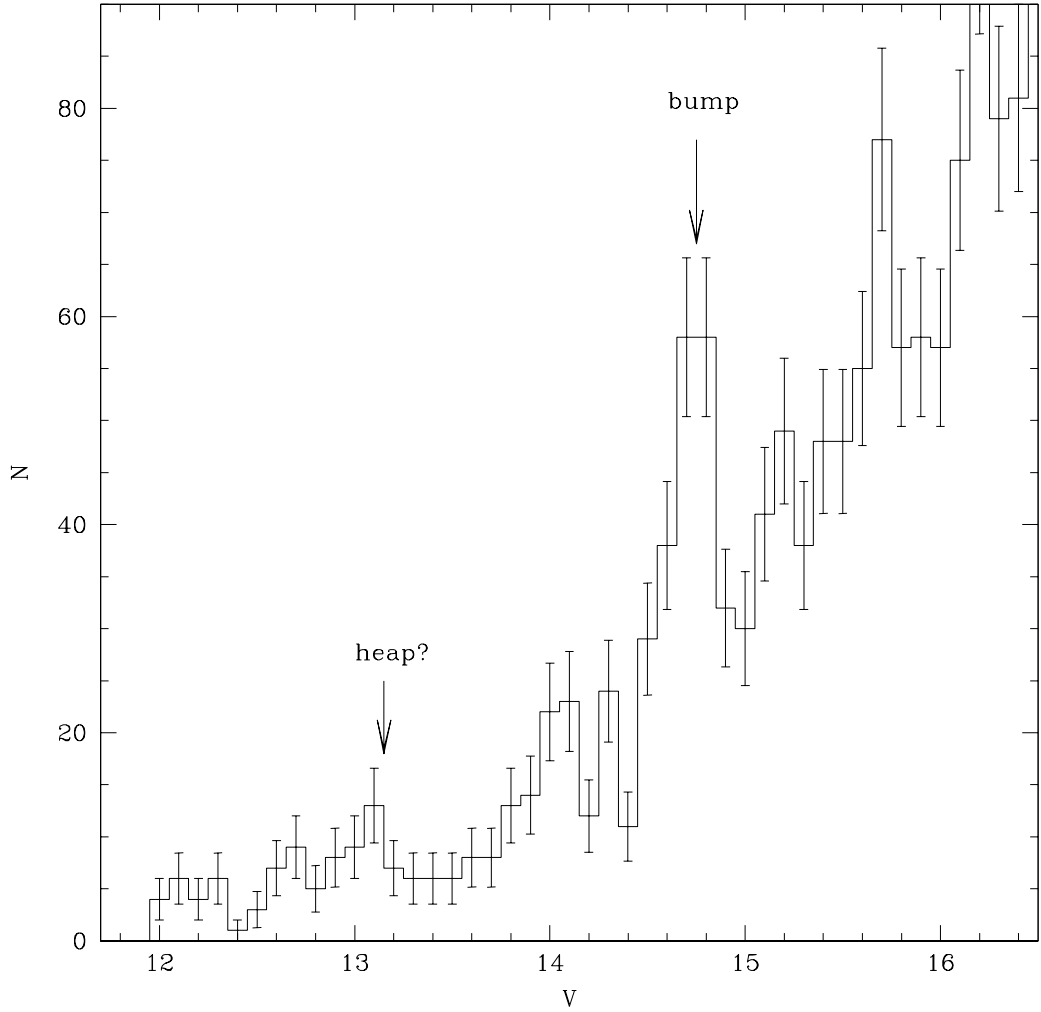}
\caption{The differential luminosity function of 
  M13 RGB stars with the bump and possible hump identified. Error bars
  are based on Poisson statistics.\label{dlf}}
\end{figure}

\begin{figure}
\epsscale{.90}
\plotone{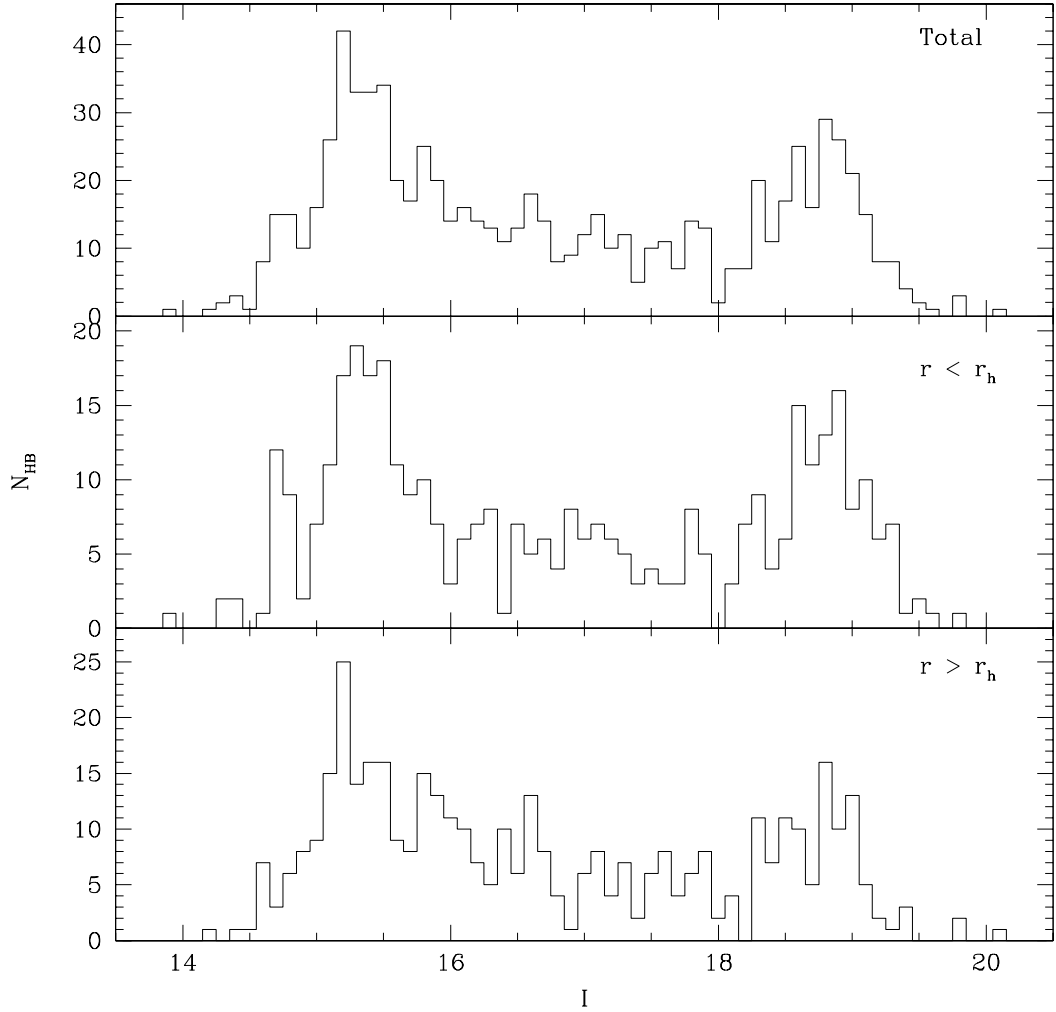}
\caption{$I$-band Distribution of horizontal branch stars in M13. {\it Top
    panel:} the total sample. {\it Middle panel:} stars within 1 half-light
  radius ($94\arcsec$). {\it Bottom panel:} stars outside 1 half-light radius.
\label{ihist}}
\end{figure}

\begin{figure}
\epsscale{.90}
\plotone{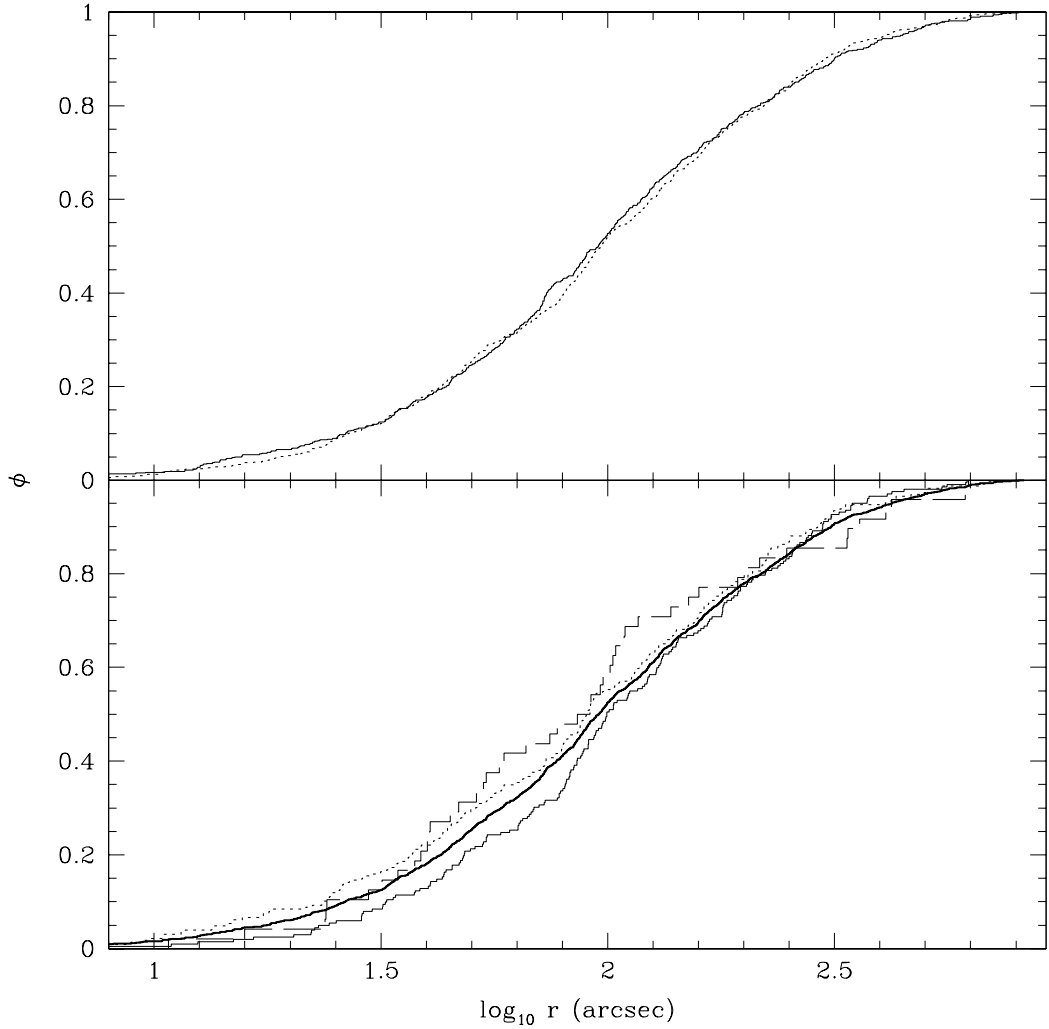}
\caption{Cumulative radial distribution of star populations. {\it Top panel:} 
  red giant stars ({\it solid line}), and horizontal branch stars ({\it dotted
    line}). {\it Bottom panel:} horizontal branch stars divided by magnitude
  into faint peak ($I < 18$; {\it dotted line}), intermediate stars ($16.25 <
    I < 18$; {\it solid line}), red HB stars ($I \la 14.9$; {\it dashed
    line}, and combined RGB/HB population ({\it dark solid line}).
\label{crd}}
\end{figure}

\begin{figure}
\epsscale{.90}
\plotone{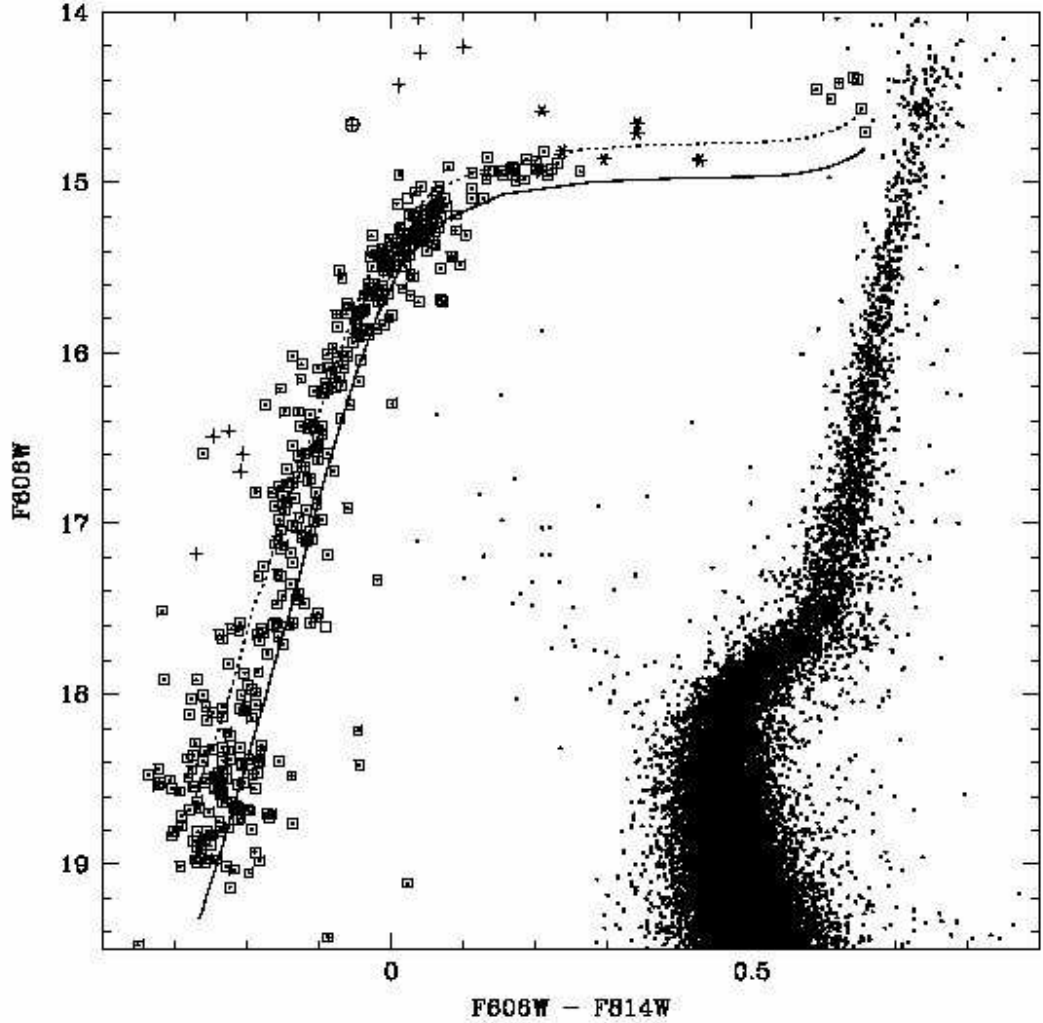}
\caption{Zoomed color-magnitude diagram from ACS data. Symbols are as in
  Fig. \ref{hbopt}. The ZAHB ({\it solid line}) and the $Y_c = 0.10$ locuses
  ({\it dotted line}) are shown for $\alpha$-enhanced Teramo HB models having
  $Z = 0.001$ and $Y = 0.246$, with $(m-M)_{606} = 14.55$ and E$(F606W-F814W)
  = 0.02$ assumed.
The AGB clump falls just above the
  bright end of the plot.\label{rhb}}
\end{figure}

\citet{sh} noted that the M13 HB largely terminates at the approximate
position of a gap in the EHB population of NGC 2419 when the HBs are aligned
at their brightest points in $B$.
In CMDs employing $B$, F439W, or F336W filters, we identified a handful of
candidates that were fainter than the faint HB peak P2. However, ultraviolet
CMDs must be used in conjunction with the optical CMDs in order to get a
full picture of the spectral output of the stars.  In Fig. \ref{hbuv}, we
present our UV CMDs for M13. As we progress from the near UV (F336W and F255W)
to the far UV (B5 and F160BW), bolometric corrections become less severe for
the hottest stars, and the hottest HB stars match the rest of the blue HB in
brightness.  Although the WFPC2 F160BW and UIT B5 filters cover similar
portions of the spectrum, and produce similar CMDs in Fig. \ref{hbuv}, the
F160BW filter reaches peak sensitivity more than 100 \AA ~ shorter than the B5
filter. We believe this is responsible for the slight upturn near the blue end
of the HB in F160BW.  In CMDs using even shorter wavelength observations, the
hottest HB stars would continue to change their configuration.

\citet{dcruzomega} found a large group of blue hook stars in $\omega$ Cen that
were subluminous in an (F160BW, F160BW-$V$) CMD compared to other stars at the
end of the HB.  A similar structure does not exist in Fig. \ref{hbuv} near
F160BW$-$F555W$ \approx -3.5$, but there are two stars that meet the typical
definition of the blue hook: fainter than the ZAHB in the far UV.  The optical
measurements for these two stars put them significantly fainter than the
predicted hot end of the HB. The UIT source (HB 2) is approximately 0.7 mag
fainter than the hot end of the HB in the B5 filter, and 1 magnitude fainter
in $B$. (It is the faintest HB source in the KPNO panel of Fig. \ref{hbopt}.)
HB 431 is among the faintest sources in the three WFPC2 UV filters, but is
faintest only in F160BW. (It is the second faintest HB source in the ACS WFC
panel of Fig. \ref{hbopt}.)

Nine other stars are fainter than the EHB in optical bands and sit at the
extreme blue/faint end of the HB in ultraviolet bands (although fairly close
to the extrapolated zero-age HB), and so we flag them as potentially
interesting stars. The UIT sources are HB 8, 56, 73, 86, and 757, 
while the WFPC2 sources are HB 373, 485, 611, and 636.


\subsection{Evolution of HB Stars}

With a large sample of HB stars, the sample of evolved HB stars also
grows, making it possible to trace post-HB evolution. The two most
densely populated portions of the HB contain 75\% of the stars, and so
to first order, post-HB stars will trace the evolution of these
two groups.

In Fig. \ref{hbter} and \ref{hbtracks}, we show a comparison between
theoretical models of \citet[hereafter, Teramo]{piet06} and \citet[hereafter,
  DSEP]{dsep} and our ground-based photometry in $(V,B-I)$ for $r >
200\arcsec$. To gauge the edges of the HB distribution, we used zero-age and
central helium depletion ($Y_c = 0.05$ or 0.10) CMD positions as a function of
mass.  Because there is some uncertainty about the precise chemical
composition of M13 \citep[e.g.,][]{jb}, we fitted the HB band to the
photometric data in order to better identify evolved stars in a differential
sense. We emphasize, however, that the fit does not completely validate the
models. In fact, there are mismatches between observation and theory that we
will examine below.

\subsubsection{The Reddest BHB Stars}\label{redhb}

Because M13 has a blue HB, traditional photometric indicators of distance
($M_V^{HB}$), age ($\Delta V_{TO}^{HB}$, for example) and helium ($R = N_{HB}
/ N_{RGB}$) that have the $V$ magnitude of the HB in their definitions cannot
be used without corrections of uncertain accuracy. As a result, redder stars
may better define the brightness of the HB if they can be proven to reside
near the ZAHB in the CMD.

As shown in Fig. \ref{vicmd}, we identified 55 stars redward of the most
heavily populated part of the HB (the bright peak P1; 308 stars in $16.25 > I
\ga 14.9$). The stars stretch to the instability strip ($0.10 < V-I < 0.22$),
and it is likely that at least some of the RR Lyrae stars belong to the same
group. These stars have largely been ignored in previous studies of M13
because many (but not all) are found in the cluster core.  The evidence of a
small gap at $V-I \approx 0.10$ between the reddest BHB stars and the bright
peak makes it worth considering whether they represent a separate population
and what their evolutionary status is. For the purpose of this section, we
restrict ourselves to comparisons with theoretical models, delaying additional
discussion of the luminosity of the HB to \S \ref{lhb}.

In examining recent stellar models from different research groups, we found
that there is still a great deal of variation in the morphology of evolution
tracks for stars that start their HB lives with $\teff \approx 10000$ K.  This
is an important question because it affects the interpretation of the reddest
BHB stars in M13. \citet{swei87} asserted that the relative importance of the
helium-fusion luminosity ($L_{He} / L$) is the main influence on whether the
evolution is primarily blueward or redward. Stars that have strong hydrogen
fusion shells significantly increase the helium core mass during the HB phase,
leading to higher $L_{He} / L$ and blueward evolution. As the helium becomes
depleted, the importance of core fusion is reduced and stars tend to evolve
redward, becoming more giant-like.  For similar compositions, some models
predict a modest blueward loop before the approach to the AGB
\citep[e.g.][]{swei87,dorman,yi} for HB evolution starting with $\teff \approx
10000$ K, while the most recent models \citep[e.g.  DSEP,][]{piet06} predict
stars evolve redward (see Fig.  \ref{hrcomp}). For the Teramo \citep{piet06}
models, slightly cooler HB stars (ones starting among the reddest blue HB
stars) have short blue loops.  However, in the case of DSEP models, stars with
$-2 \le \mbox{[Fe/H]} \le 0$ {\it all} evolve strongly redward from the start,
and remain close to the ZAHB during much of the HB phase. Consistent with
Sweigart's discussion, core helium burning remains significantly stronger than
the hydrogen fusion shell in DSEP models, even for stars starting quite close
to the giant branch.  If the DSEP models are good representations of actual
evolution, even the evolved HB stars in M13 could be reasonable indicators of
where the ZAHB is. Clearly there are significant differences between these
models that may help us to identify important physics.

Although there are many small differences between the physics inputs in the
DSEP and Teramo models, we believe the likely cause is diffusion. Diffusion
was consciously disabled in the Teramo models, while the DSEP models stop
diffusion in the outermost $0.05 \msun$ of stars and linearly ramp their
diffusion algorithm to full strength at $0.10 \msun$ below the surface. As
seen in more metal poor $0.73 \msun$ models with diffusion by \citet{mich07},
the introduction of diffusion produces redward evolution for a star that would
otherwise evolve blueward. There is clear evidence of diffusive processes in
the surface compositions of bluer HB stars in globular clusters
\citep[e.g.][]{moe03,moni09}, but not among stars like the reddest BHB
stars. Still, the surface layers are less likely to show chemical signatures
due to near-surface convection. The DSEP models may be demonstrating that
diffusion in the interior can be constrained by observations --- processes
such as rotationally-induced mixing could inhibit the action of diffusion.

The numbers of stars can also constrain the models --- because the early
stages of core helium fusion take the longest, the initial direction of the
evolution in color should have a big effect on how many stars are found to the
red of peak P1.  Fig. \ref{p1shb} shows a representative synthetic horizontal
branch generated from DSEP web
tools\footnote{http://stellar.dartmouth.edu/~models/shb.html We used
  ``empirical'' color transformations as these appear to do a better job than
  ``synthetic'' transformations in reproducing photometry in and around the
  subgiant branch \citep{dsep,sara}.} with a number of stars in the primary
peak comparable to observations.  We did not attempt to fully model the
fainter parts of the primary peak because most of those stars are expected to
exhaust their core helium before entering into the color range we are
considering, becoming too bright to be considered HB stars (see \S
\ref{supra}).  Even so, the number of stars actually observed is {\it
  considerably smaller} than predicted by the synthetic HB simulation. The
models predict that the red population ($I < 14.9$) should be approximately
28\% the size of the primary peak population, whereas the M13 population is
only about 19\%. Small variations in chemical composition (helium or heavy
elements) also do not affect this conclusion. Whatever the reason, the DSEP
evolutionary tracks do not seem to be accurately representing the evolution of
stars from the primary peak. A small amount of blueward evolution would
relieve the discrepancy.

Teramo tracks produce a somewhat better fit to the M13 data, but the
observations appear to need blueward evolution to continue at slightly higher
temperatures to explain the appearance of the rather sharp red edge at $(V-I)
= 0.07$ (pure redward evolution would tend to smear out such a feature in
color).  When the ZAHB and $Y_c = 0.05$ lines are fit to the magnitude extent
of the primary peak, the reddest BHB stars and RR Lyrae stars fall near the
$Y_c = 0.05$ line, implying they are significantly evolved (see
Fig. \ref{hbtervi}).  

While the Teramo models come close to explaining the appearance of the sharp
red edge of the HB distribution, the underlying question remains {\it why}
there should be such a sharp edge for M13's population.  Increased envelope
mass, as well as increased helium and metal abundance are known to encourage
blueward evolution. Helium enrichment has a significant effect on the HB
luminosity and some effect on the evolution of red HB stars.  For modestly
helium-enriched ($Y = 0.30$) Teramo models in Fig. \ref{hbterviy} (consistent
with the $\delta Y$ hypothesis), a large (and probably unrealistic) distance
modulus of about 14.75 is required, and the evolution of stars at the red end
of the bright peak P1 goes too bright to explain the reddest BHB stars. In
fact these stars can't be satisfactorily reproduced in the current
helium-enhanced models.

\citet{piet} recently computed HB models for ``extreme'' CNONa abundance
mixtures in which the sum of CNO elements is approximately a factor of two
higher at a given [Fe/H] than for a typical $\alpha$-enhanced mixture. The
mixture is intended to be representative of pollution resulting from
intermediate mass AGB stars, and N is by far the most abundant heavy element
due to nuclear processing.  An extreme CNONa mixture could realistically
produce blueward evolution for the dominant HB population.  However,
measurements of oxygen abundances among the cool HB stars in M13 all indicate
[O/Fe] is normal for globular clusters and super-solar \citep{pete}.
\citet{smith96} and \citet{cm} also found that [(C+N+O)/Fe] did not seem to
vary significantly within their samples of CN-strong and CN-weak giants, and
the CNO elements do not show ``extreme'' enhancement (the average was about
0.3 dex).  Similar results have been found for relatively unevolved stars in
other clusters \citep{car05},
although stars in NGC 1851 show a significant 0.57 dex spread \citep{yong09}.

From the theoretical comparisons above, our preferred explanation is that the
reddest BHB stars are stars with unenriched compositions that evolve somewhat
to the blue after reaching the ZAHB. However, it must be admitted that the
theoretical models disagree to a greater extent than we would like. We return
to the discussion of these stars in \S \ref{lhb}.

\subsubsection{Ultraviolet Bright Stars}\label{uvb}

We next identified hotter ultraviolet-bright stars (the brightest
cluster stars in $U$ band, but optically identifiable) from the study
of \citet{zng}.  ZNG 1 (Barnard 29) is a post-AGB star
\citep{con94,moe98,thom07}, and is the brightest star in the UV by
almost 3 mag (see the UIT panel of Fig. \ref{hbuv}).

In the lower left panel of Fig. \ref{hbuv}, ZNG 2, ZNG 6, and ZNG 7 sit at the
cool side of a group of stars we identify as AGB \manq ~ stars, which are hot
stars that are significantly brighter than the hot HB (see \S \ref{supra}).
ZNG 2 (G 43) was observed spectroscopically by \citet{moe03}, and its low
gravity ($\log g \approx 3$) clearly identifies it as being a post-HB star.
ZNG 7 falls slightly bluer than the ``knee'' of the HB in the upper right
panel of Fig. \ref{hbopt}, but is quite bright in the ultraviolet. ZNG 6 is
the only one of the UV bright stars that has HST ACS observations, which again
places it significantly brighter than the HB (see the lower left panel of Fig.
\ref{hbopt}).

In optical CMDs, ZNG3 and ZNG 4\footnote{For completeness, ZNG 5 is a
  non-member according to proper motions, and is not considered here.} are
about a magnitude brighter than the knee of the HB, and part of a small group
of stars seeming to parallel the HB. ZNG 3 and 4 are among the reddest objects
with reliable photometry detected using UIT (see Fig. \ref{hbuv}). ZNG4 was
observed spectroscopically by \citet{ambika}, who identified chemical
signatures of diffusion, which only shows up among blue HB stars with $T_{\rm
  eff} \ga 11000$ K. A likely explanation is that ZNG 4 ($T_{\rm eff} = 8500$
K) was once a hot blue HB star that evolved, and the chemical signature is
leftover from the earlier phase. A significant convective envelope is not
likely to appear until the star reaches lower $T_{\rm eff}$. We identify both
stars as supra-HB stars, which probably trace evolution from midway on M13's
HB, as discussed in the next subsection.

\subsubsection{Hot Post-HB Stars}\label{supra}

The origin and evolutionary behavior of the hottest HB stars has implications
beyond stellar evolution --- these stars contribute to, and perhaps dominate,
the ultraviolet light from old stellar populations in galaxies
\citep{dro}. Recently, \citet{brownm32} used a STIS UV CMD of the dwarf
elliptical M32 to study the main stellar contributors in the ultraviolet: the
hot HB, the AGB \manq, and the post-AGB.  The authors found that the CMD could
constrain the chemical evolution of the population, but that the post-HB
evolution did not seem to be in good agreement with models. Although galactic
populations provide better leverage on the shortest stages of post-HB
evolution, a massive globular cluster with a simpler population and precise
photometry should also constrain theoretical models.

\begin{figure}
\epsscale{.80}
\plotone{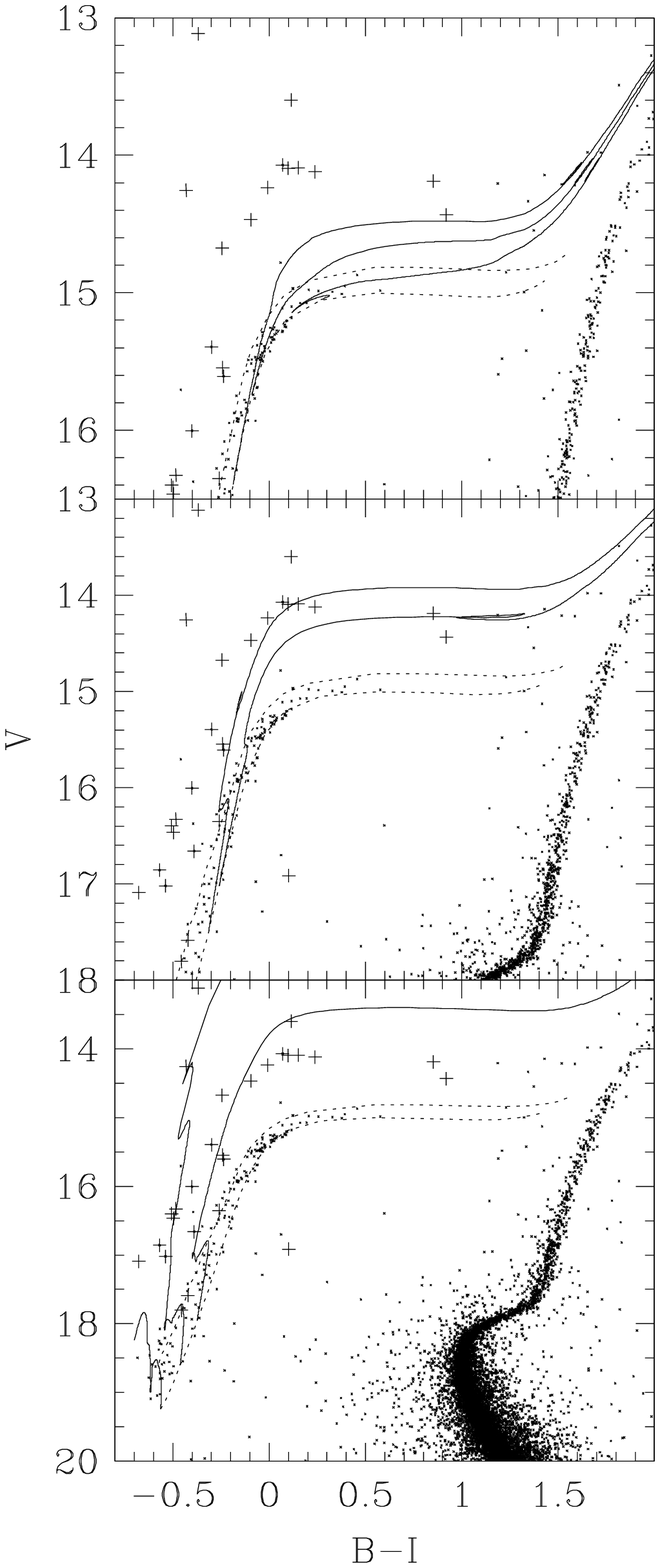}
\caption{Comparison of ground-based CFHT and KPNO photometry ($r >
  200\arcsec$) with Teramo \citep{piet06} HB models. In all panels theoretical
  values are shifted by 14.45 in $V$ and 0.02 in $B-I$ to fit the envelope of
  HB stars. {\it Left panels:} Comparison with evolution tracks having
  [Fe/H]$=-1.62$ and [$\alpha$/Fe] $\sim +0.4$ ([M/H] $= -1.27$, $Z = 0.001$).
  The ZAHB and the $Y_c = 0.10$ locuses are shown with dotted lines. The
  tracks are for 0.4912, 0.50, and 0.52 $\msun$ ({\it bottom panel}), 0.54 and
  0.56 $\msun$ ({\it middle panel}), and 0.58, 0.6, and 0.63 $\msun$ ({\it top
    panel}). \label{hbter}}
\end{figure}

\begin{figure}
\epsscale{.80}
\plotone{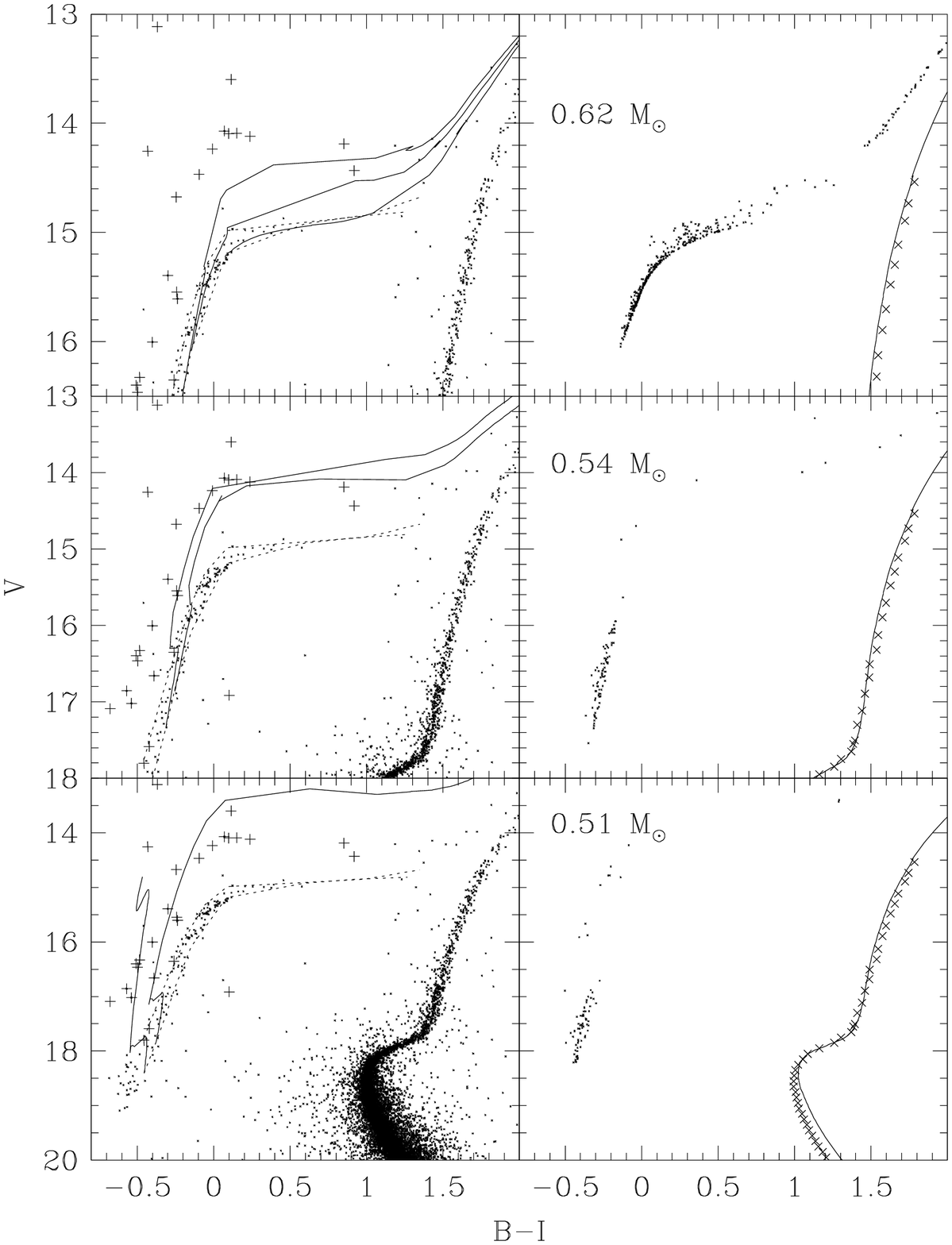}
\caption{Comparison of ground-based CFHT and KPNO photometry ($r >
  200\arcsec$) with DSEP HB 
  models. In all panels theoretical values are shifted by 14.28 in $V$ and 0.02
  in $B-I$ to fit the envelope of HB stars. {\it Left panels:} Comparison with
  evolution tracks having [Fe/H]$=-1.5$ and [$\alpha$/Fe] $= +0.4$. The ZAHB
  and the $Y_c = 0.05$ locuses are shown with dotted lines. The tracks are for
  0.50 and 0.52 $\msun$ ({\it bottom panel}), 0.54 and 0.56 $\msun$ ({\it
    middle panel}), and 0.58, 0.6, and 0.65 $\msun$ ({\it top panel}).  {\it
    Right panels:} Example synthetic horizontal branch populations having the
  same initial composition and a mass dispersion of 0.01 $\msun$, The number
  of synthetic HB stars roughly corresponds to the number in the
  total sample (see Fig.  \ref{ihist}).  Also included
  ({\it solid line}) is an isochrone for the same composition and an age of 14
  Gyr. Crosses show our mode fits to MS data, and individual stars that
  represent the mean HB line.  \label{hbtracks}}
\end{figure}

\begin{figure}
\epsscale{.90}
\plotone{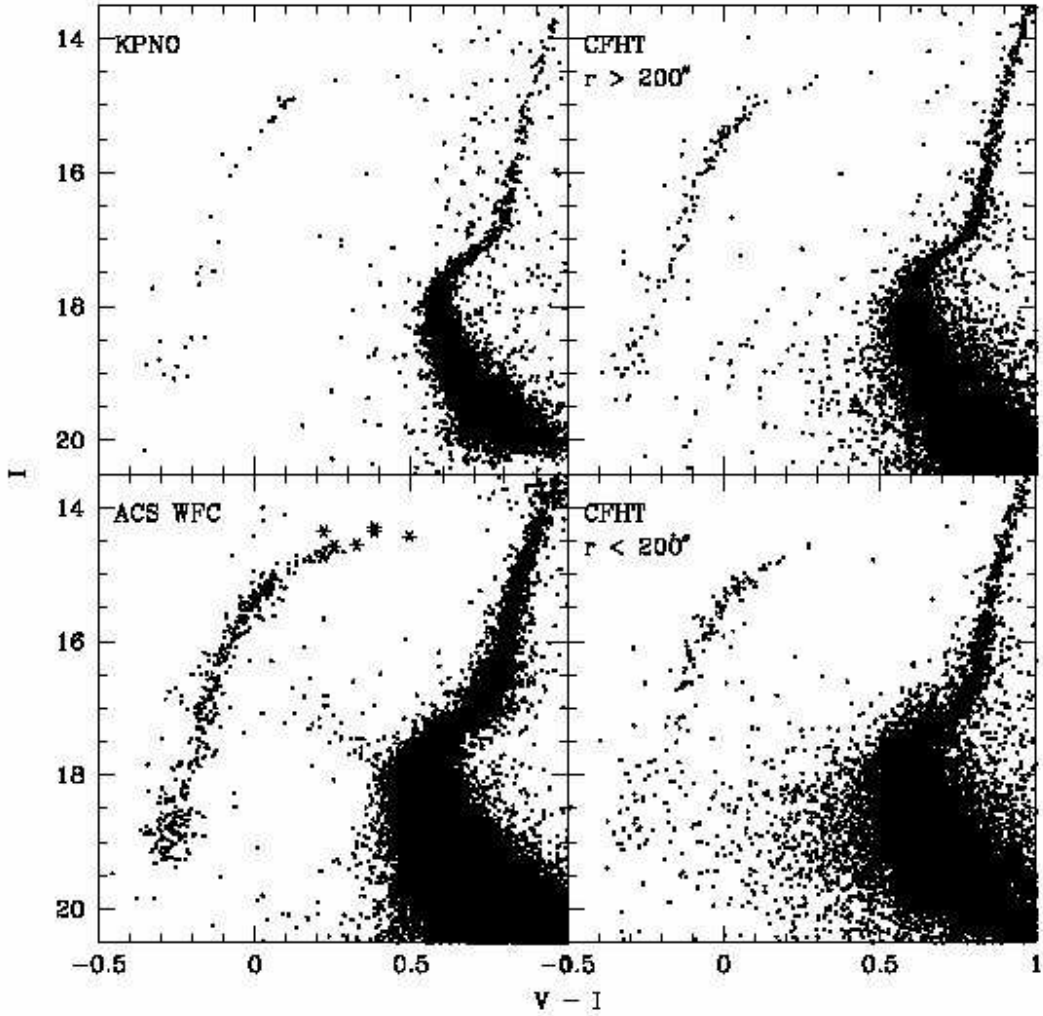}
\caption{($I,V-I$) color-magnitude diagram for HB stars from all photometry
  sources. RR Lyrae stars (shown with $\ast$ symbols) are found within the ACS
  field, and the photometry used is not averaged over the oscillation
  period.\label{vicmd}}  
\end{figure}

\begin{figure}
\epsscale{.90}
\plotone{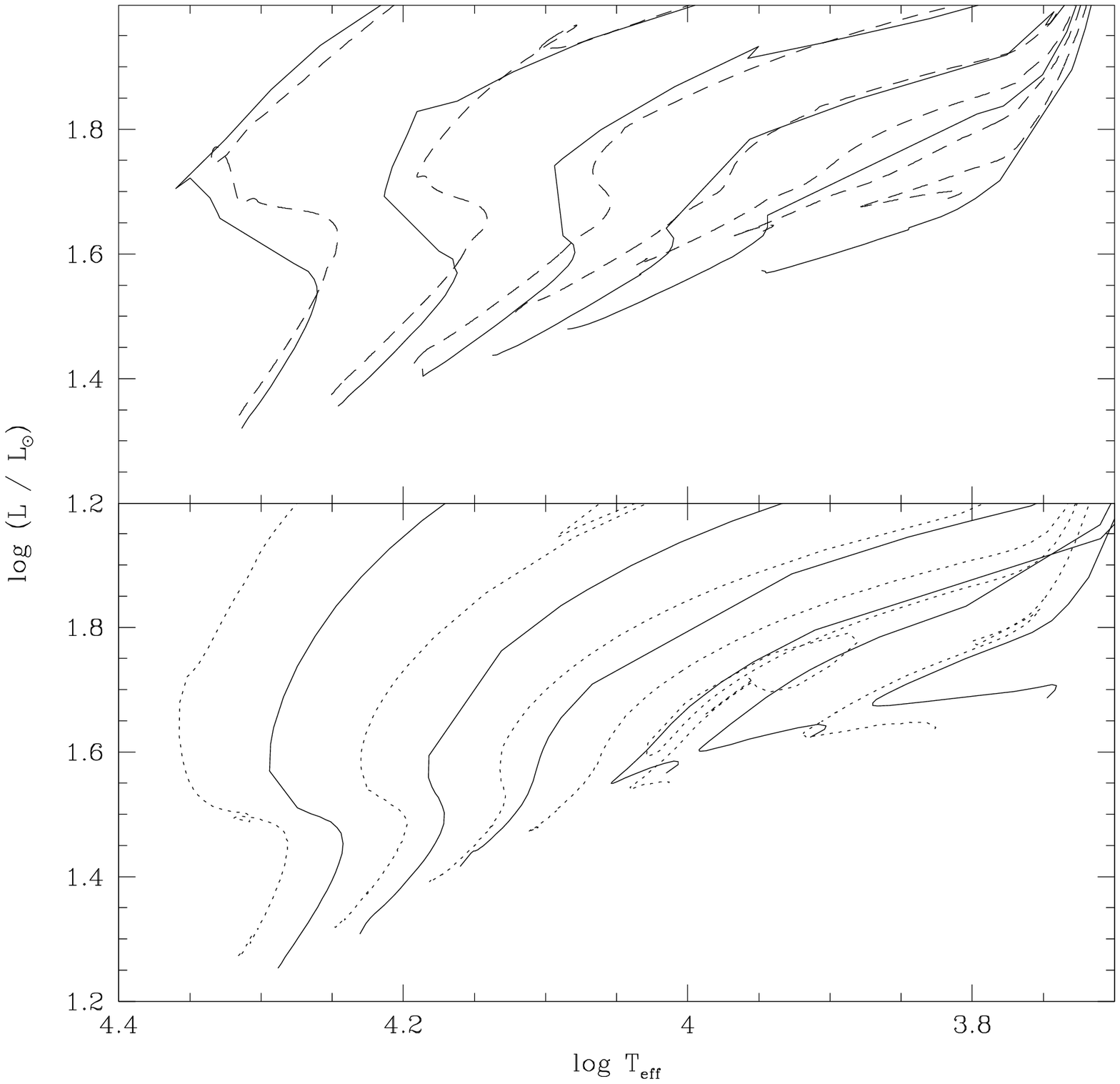}
\caption{HR diagrams for HB stars from theoretical models.
  {\it Top panel:} Models from \citet{dsep} ({\it solid line}) for $Z =
  0.00107$, [$\alpha$/Fe] = +0.4 and from \citet{piet06} ({\it dashed line})
  for $Z = 0.001$ (0.52, 0.54, 0.56, 0.58, 0.6, 0.65 $\msun$ for both sets.)
  {\it Bottom panel:} Models from \citet{dro} for [Fe/H] $= -1.48$, [O/Fe] =
  0.63 ({\it solid line}; 0.52, 0.54, 0.56, 0.59, 0.61, 0.65 $\msun$) and from
  \citet{yi} for $Z = 0.001$ ({\it dotted line}; 0.52, 0.54, 0.56, 0.58, 0.60,
  0.64 $\msun$).\label{hrcomp}}
\end{figure}

\begin{figure}
\epsscale{.90}
\plotone{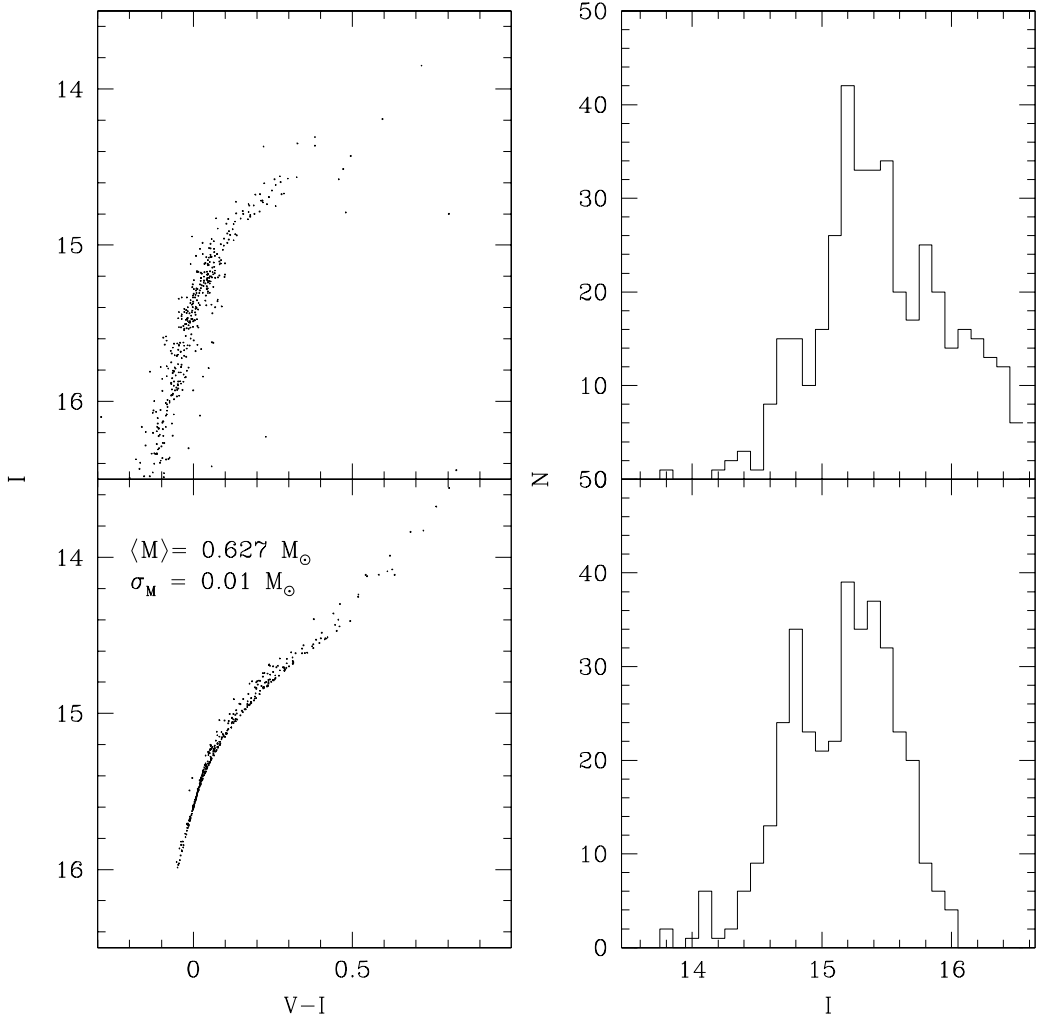}
\caption{Comparison of observations ({\it upper panels}) with DSEP models
  ({\it lower panels}; assuming $(m-M)_I = 14.28$) of the primary peak on
  M13's HB. {\it Left panels:} ($I, V-I$) color-magnitude diagrams. {\it Right
  panels:} $I$-band distributions.
\label{p1shb}}
\end{figure}

\begin{figure}
\epsscale{.80}
\plotone{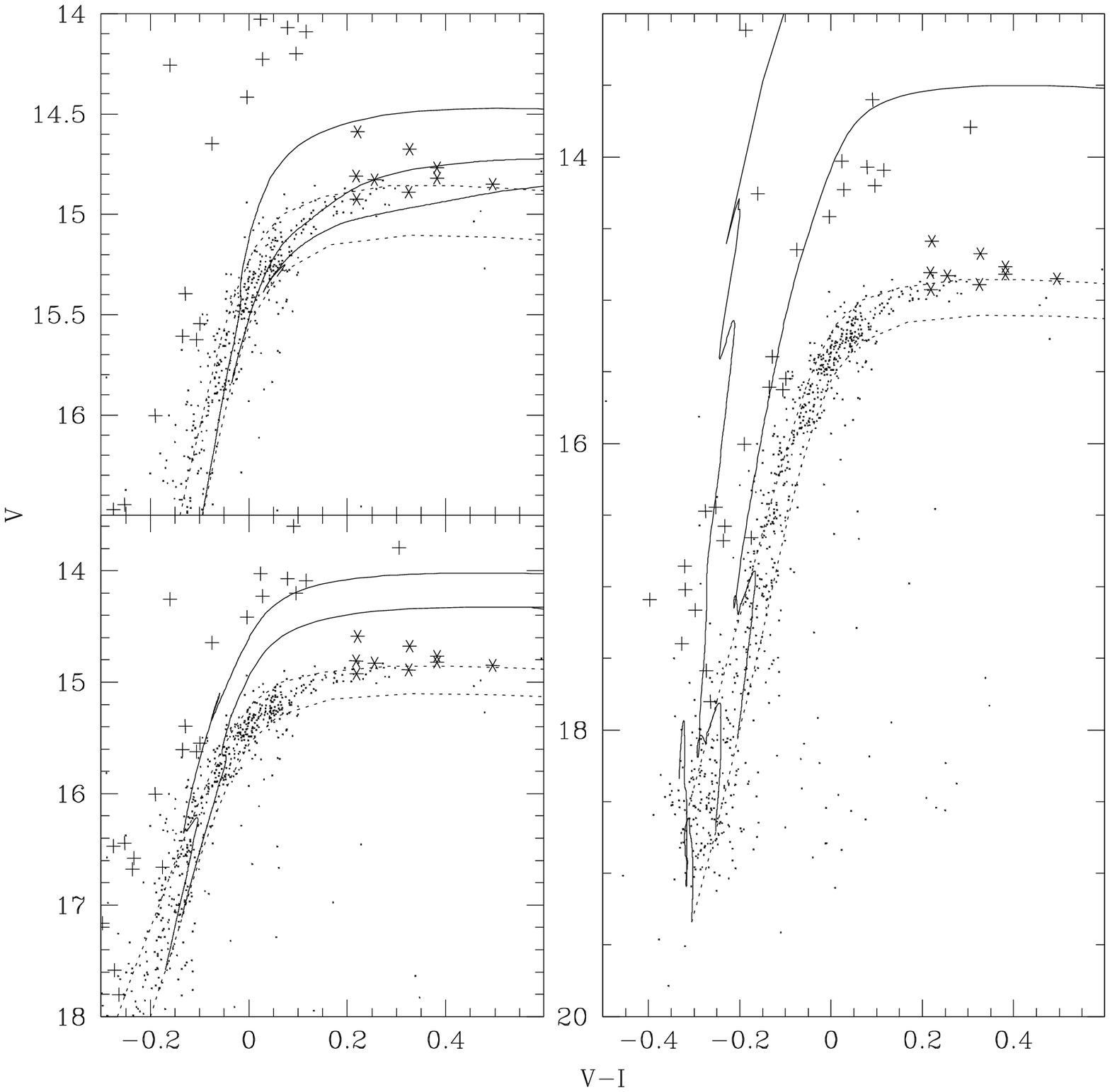}
\caption{Comparison of mastered photometry with Teramo \citep{piet06} HB
  models. In all panels theoretical values are shifted by 14.55 in $V$ and
  0.01 in $V-I$ to fit the envelope of HB stars, and have [Fe/H]$=-1.62$, $Y =
  0.246$, and [$\alpha$/Fe] $\sim +0.4$ ([M/H] $= -1.27$, $Z = 0.001$).  The
  ZAHB and the $Y_c = 0.05$ locuses are shown with dotted lines. Evolution
  tracks are shown for 0.57, 0.60, and 0.62 $\msun$ ({\it upper left panel}),
  0.54 and 0.56 $\msun$ ({\it lower left panel}), and 0.4912, 0.50, and 0.52
  $\msun$ ({\it right panel}). \label{hbtervi}}
\end{figure}

\begin{figure}
\epsscale{.80}
\plotone{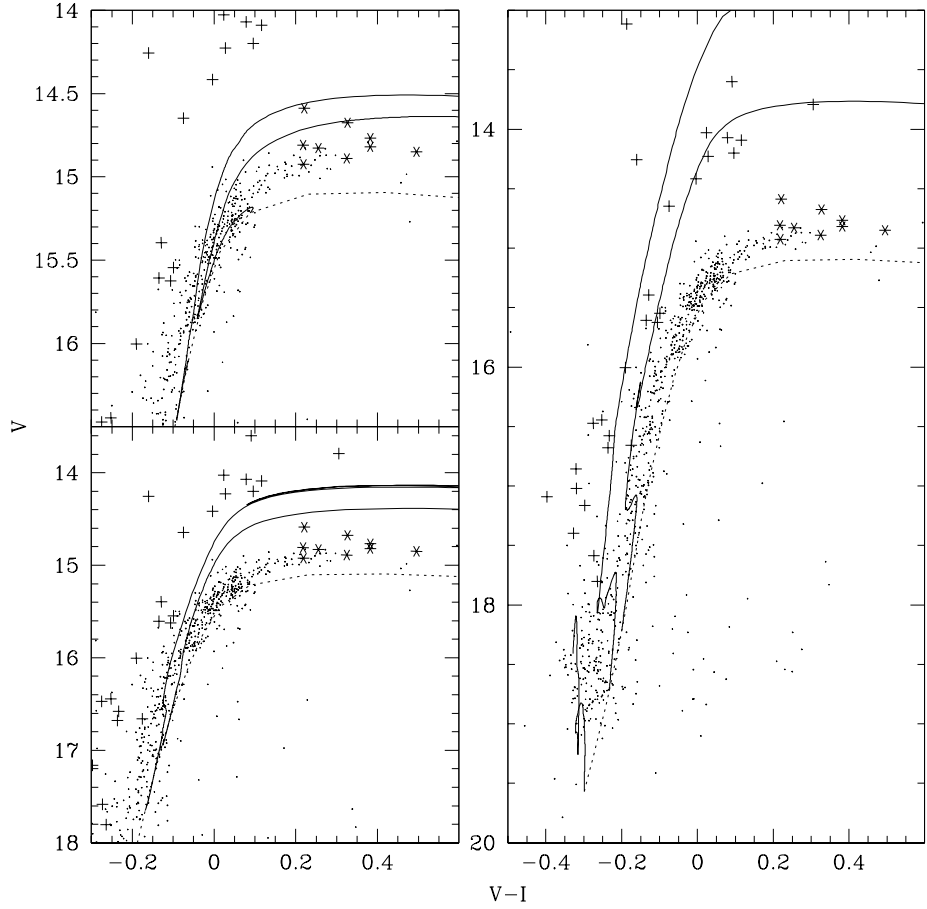}
\caption{Comparison of mastered photometry with Teramo \citep{piet06} helium
  enriched ($Y = 0.30$, $Z = 0.000924$) HB models. In all panels theoretical
  values are shifted by 14.75 in $V$ and 0.01 in $V-I$ to fit the envelope of
  HB stars.  The ZAHB locus is shown with a dotted line.  Evolution tracks are
  shown for 0.57 and 0.61 $\msun$ ({\it upper left panel}), 0.54 and 0.56
  $\msun$ ({\it lower left panel}), and 0.481, 0.50, and 0.52 $\msun$ ({\it
    right panel}).
  \label{hbterviy}}
\end{figure}

Using a combination of ultraviolet and optical CMDs, we attempted to identify
all stars that have evolved away from the hotter parts of the HB. As discussed
in \S \ref{redhb}, theoretical HB models \citep{dsep,piet06} with no helium
enrichment agree that stars originating from the primary peak in the HB
distribution produce AGB stars exclusively --- the evolution of the stars
keeps them within about 0.2 mag of the HB until shortly before central He
exhaustion. Core helium exhaustion occurs over a large portion of the star's
core due to convection during helium fusion, causing a star to adjust its
structure rapidly (in about a Kelvin-Helmholtz timescale) and move to the AGB
clump.  However, during the time leading up to core exhaustion, the core
convection zone is decreasing in size, leaving behind a composition gradient.
In typical AGB stars, almost half of the AGB lifetime is taken up while the
new He fusion shell eats through this composition gradient.  During this time
the star's evolution pauses, producing a fairly well populated AGB clump about
1 $V$ mag brighter than the HB \citep{ferr99}.
Figs. \ref{hbtracks} and \ref{hbter} show illustrative theoretical tracks and
synthetic HB populations from DSEP and Teramo using canonical physics and
reasonable choices for chemical composition (nearly primordial helium
abundance, for example). The top rows in both plots show tracks for stars
which evolve near the HB and have normal AGB phases including an AGB clump.


The middle rows in Figs. \ref{hbtracks} and \ref{hbter} show stars that
produce stars that spend part of their lives as somewhat bluer than average
AGB stars, but do not have an AGB clump phase. For these stars, the evolution
as the He fusion shell consumes the composition gradient take longer (more than a Kelvin-Helmholtz timescale) and occurs at higher
surface temperature. Most of these stars can be put into one of two categories
based on whether they have colors bluer or redder than the ``knee'' of the HB
in optical CMDs ($B-I \approx -0.1$).  One group of stars sitting
approximately 1 mag above what would be the horizontal part of the normal HB
($14 < B < 14.5$ in Fig.  \ref{hbopt}; $V \approx 14.1$) corresponds to
``supra-HB'' stars previously identified by \citet{ss} and \citet{zinn}.  This
group includes ZNG 3 (AGB 75) and ZNG 4 (AGB 8), as well as AGB 27, 32, and 33
(detected in F160BW using HST, and the reddest post-HB star marked in the
upper left panel of Fig.  \ref{hbuv}), 50 (also detected in HST F160BW),
81, and two of the known BL Her pulsating variable stars (V1 and V6;
\citealt{kopacki02}). Type II Cepheids (a group that includes BL Her stars)
are only found among stellar populations having a significant blue HB
component, as was noted by \citet{wall} and \citet{smwhe}.  As reviewed by
\citet{wall02}, the great majority of well-studied Type II Cepheids show no
period change or increasing period, consistent with evolution toward the red
and larger size.

Even hotter post-HB stars are harder to identify in optical CMDs because of the
steepness of the HB in magnitude. Relatively small color errors can cause
normal HB stars to overlap stars that have evolved significantly in luminosity
from hotter on the HB. When ultraviolet observations are used, these stars are
much more easily separated by brightness (see Fig. \ref{hbuv}). So-called AGB
\manq ~ stars fall in this category --- they are stars that have evolved away
from the HB that will not reach the traditional AGB because the star's
envelope does not have enough mass to produce a giant-like structure.
Observationally, this definition is hard to use because it requires
reliable knowledge of the star's future evolution in the CMD. However,
theoretical calculations imply that there is a big change in track morphology
on the blue HB. The bottom rows of Figs. \ref{hbtracks} and
\ref{hbter} show stars that would not have an identifiable AGB phase. 


The main difference in the models with total mass less than about 0.54 $\msun$
at M13's metallicity is that hydrogen shell fusion is unable to provide much
of the luminosity even during the core contraction following helium
exhaustion. It appears that if $L_H$ at maximum does not surpass $L_{He}$
after the H fusion shell has reignited and stabilized, the star will remain at
high temperature during the relatively slow helium shell adjustment.


The supra-HB and AGB \manq ~ stars in M13 identified with crosses generally fall
more than a magnitude above the faint envelope of HB stars. So for a fairly
standard chemical composition, models predict that the only evolution tracks
that produce hot post-HB stars come from bluer than some point between the two
HB peaks in M13.
Observationally, there is an apparent gap between two sets of
``UV-bright'' stars in the UIT CMD in Fig. \ref{hbuv}. While individual
optical CMDs in Fig. \ref{hbopt} may leave the impression of color gaps
between groups of hot post-HB stars, the union of the optical CMDs (see Fig.
\ref{hbtervi}) indicates that there is a thin, fairly uniformly populated band
paralleling the HB from its bluest to its reddest colors with a couple of
post-HB stars even closer to the AGB clump.  There are a few stars in the ACS
field without UV photometry that will partially fill the gap between the two
groups of ``UV bright'' stars in the UIT CMD. As a result, we believe these
stars are tracing out the slowest phase of post-HB evolution for stars that do
not follow traditional AGB tracks. A relatively small change in envelope mass
leads to a drastic change in the morphology of the tracks for these stars.
Because M13's HB is well-populated near the end of the canonical HB (more so
than many other clusters with extended blue HB tails), the cluster is
providing us with a means of observationally ``seeing'' where the post-HB
evolution pauses for a little.

Independent of how believable the gap in the UIT CMD is, the usable portion of
the UIT field does contain the majority of the hot post-HB stars, and does
perhaps allow us to observationally identify the transition between an AGB
\manq ~ track and an post-early AGB track. AGB \manq ~ evolution is expected
to be largely in luminosity, with redward color evolution increasing for
cooler, more massive HB stars. The red edge of the group ($m_{1620} - V
\approx -1.7$) is thus a conservative upper limit for the HB stars producing
AGB \manq stars, and this limit is bluer than the blue end of the primary HB
peak.  The numbers of post-HB stars help to constrain the tracks further (see
\S \ref{r2}).

As can be seen from Fig. \ref{hbterviy}, the shape of post-HB evolution tracks
does not change drastically with a modest increase in helium abundance to $Y =
0.30$. Increasing helium abundance creates a larger luminosity gap between the
blue end of the HB and the AGB \manq ~ stars, greatly increases the number of
AGB \manq ~ stars relative to HB stars, and produces bluer \manq ~ stars that
are more concentrated in color \citep{brownm32}.  Because models of M13's HB
involving helium-enriched stars predict that the bluest HB stars have $Y
\approx 0.38$ \citep{dant08b}, these effects could be tested. Detailed
simulations are beyond the scope of this study, but to first order, the number
of hot post-HB stars is quite small compared to HB stars in the same color
range (17 / 191 = 0.09 for $m_{1620} - V < -1.9$), which argues against large
helium enrichment.

\section{Population Ratios}\label{ratios}

One of the primary reasons for this study was to examine evolutionary
timescales for stars in different life stages. We first recalculate the HB
type $(N_{BHB} - N_{RHB}) / (N_{BHB} + N_{VHB} + N_{RHB})$. Here BHB, VHB, and
RHB stand for stars bluer, within and redder than the RR Lyrae instability
strip.) For M13, $N_{VHB} = 9$, and we have identified five possible RHB
stars. Thus we have HB$_{\mbox{type}} = 0.976\pm0.006$.

\subsection{The Helium-Sensitive $R$ Ratio}\label{rrat}

The $R = N_{HB} / N_{RGB}$ ratio compares the evolutionary lifetimes of HB and
RGB stars, and is a sensitive helium abundance indicator (e.g.
\citealt{cass03}). Because of the claim that M13 stars may have high helium
abundances \citep{jb,cda}, it is particularly important to examine this
population ratio.  Following the definition of \citet{sal04} and
\citet{zocc00}, we used $V_{ZAHB}$ (the magnitude of the zero-age HB) as the
cutoff for the RGB sample.
The determination of the faint limit for the RGB sample is important because
the numbers of RGB stars rises quickly with increasing magnitude. 

To determine $V_{ZAHB}$, we applied two methods. For the first, we
followed the procedure of \citet{reb}, using a
template cluster of comparable metallicity and a well-studied RR Lyrae
population (used to define the reference HB level $\langle V_{RR}
\rangle$) to determine a relative magnitude shift. \citeauthor{reb} used NGC
1904 as the template for clusters near M13's metallicity after first
determining $\langle V_{RR} \rangle$ through a comparison with M3. In their
comparison of photometry for NGC 1904 and M13 from the WFPC2 camera, they
found that M13's sequences are 1.20 mag brighter in $V$ than NGC 1904, with no
relative shift in color.
The corresponding $\langle V_{RR} \rangle$ value for M13 is therefore
predicted to be $14.97 \pm 0.07$, which results in $V_{ZAHB} = 15.06$. 



For the second method, we determined the ZAHB almost exclusively from data on
the reddest blue HB stars, under the assumption that they are relatively
unevolved and unenriched HB stars. (Based on additional arguments in \S
\ref{bimodal}, we believe this is incorrect.) Our photometry does not have
adequate time-coverage for direct determination of $\langle V_{RR} \rangle$.
The value determined by \citet{kopacki02} ($14.83\pm0.02$) was set by
comparison to photometry by \citet{rey} that was in turn calibrated to
\cite{lan92} standards. Zeropoint differences between our dataset and
\citeauthor{kopacki02} should therefore be small. However, nonvariable stars
at the blue end of the instability strip have $V = 14.90$ on average. Using
this as representative of the average HB level, $V_{ZAHB} = 14.99$.

We find $N_{RGB} = 483$ using the first method, giving $R = 795 / 483 = 1.65
\pm 0.09$ (error estimate from Poisson statistics). For the second method,
$N_{RGB} = 465$ and $R = 1.71 \pm 0.10$. In both cases, M13's $R$ value is
about $3 \sigma$ higher than the theoretical calibration presented in
\citet{sal04} for $Y = 0.245$, and in agreement with their measurement of
$R = 1.719 \pm 0.197$ using a considerably smaller sample.

This does not account for the effects of HB morphology, however.  Theoretical
models uniformly predict that blue HB stars have longer evolutionary times
than RR Lyrae variables or redder HB stars, which would produce high $R$
values in clusters with blue HB morphologies. \citet{zocc00}, for example,
conducted an early examination of the effect of HB lifetimes on the $R$ ratio,
and \citet{sal04} found that clusters with HB$_{type} \geq 0.8$
show a larger spread around the mean than clusters with redder morphologies,
which implied that the exact morphology of blue HB clusters might be influencing
the $R$ ratio.  In \citet{sh}, we described a method for correcting $R$ for
variations in HB lifetimes in a way that is based on observable quantities and
is largely independent of chemical composition. The reader should see that
article for more details, but we briefly summarize it below.

\begin{figure}
\epsscale{.90}
\plotone{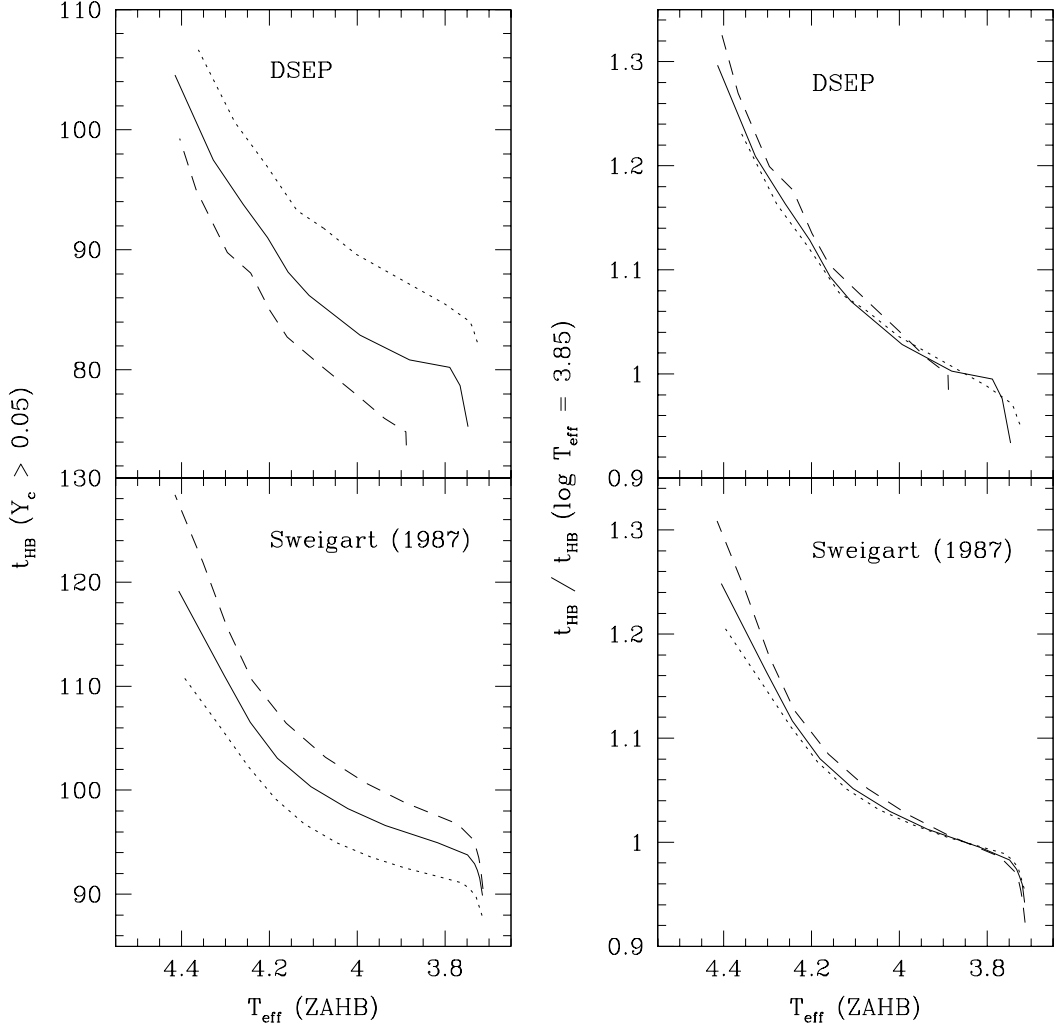}
\caption{HB lifetimes as a function of $T_{eff}$ from DSEP \citep{dsep} models for (from
  top to bottom) [Fe/H] $= -1.0, -1.5$, and $-2.0$ ([$\alpha$/Fe] = 0.2), and
  Sweigart models $Y = 0.30, 0.25$, and 0.20 ($Z = 0.001$).{\it Left panels:}
  lifetimes in Myr. {\it Right panels:} lifetimes normalized to the stars
  reaching the zero-age HB at $\log T_{eff} = 3.85$. (For the DSEP model with
  [Fe/H] $= -2.0$, the most massive models do not reach that temperature, and
  were normalized at $\log T_{eff} = 3.89$.)
\label{hblife}}
\end{figure}

According to theoretical models, HB lifetimes vary similarly as a function of
stellar mass and effective temperature, although the absolute value of the
lifetime does depend on composition (see the lefthand panels of Fig.
\ref{hblife}).
Based on this behavior, we defined a weighting factor
for each star in the HB sample:
\[w_i = \frac{t_{HB}(\log T_{eff} = 3.85)}{t_{HB}(\log T_{eff})} \]
The largest portion of the variation in HB lifetimes occurs on the blue tail
for $T_{eff} \ga 10^4$ K for stars with low-mass hydrogen envelopes. By
choosing $\log T_{eff} = 3.85$ (near the blue edge of the instability strip)
as the normalization point, we can correct the blue HB lifetimes back to
values representative of the more common variable and red HB stars, and the
$R$ value can be realistically compared to values from large studies of
clusters with redder HBs \citep{sal04,zocc00}. The variation of the weighting
factors with $T_{eff}$ is shown in the righthand panels of Fig. \ref{hblife}.
Although there is some variation in the weighting factors with composition,
they clearly describe the lifetime variation to first order, so that residual
uncertainties are at the level of a few percent. Further, the weightings {\it
  only} remove the effects of position on the HB, and do not reference the
absolute value of the HB lifetime, which depends on composition and physics
inputs to the stellar evolution codes (e.g., the
$^{12}$C($\alpha,\gamma)^{16}$O reaction rate; \citealt{cass03}).

Although color-$T_{eff}$ relationships remain imperfect, it is much more
reliable to derive $T_{eff}$ from photometry than it is to derive the stellar
mass. For example, the Grundahl $u$-jump has been found in many clusters with
blue HB stars with $T_{eff} \ga 11500$ K \citep{ujumps}. In addition, the
canonical HB also seems to have a reasonably well-defined termination near
30000 K. For the purposes of this study of M13, we have used the $T_{eff}$
determinations from \citet{moe03} using Stromgren photometry.
\citeauthor{moe03} also measured $T_{eff}$ spectroscopically, and while the
spectroscopic measurements are in good agreement with the photometric
measurements, the spectroscopic measurements are subject to significant model
uncertainties at the faint end and appear to deviate systematically from
photometric determinations at the red end of the HB. These measurements for
M13 do not go all the way to the faint end of the HB, so we have assigned
$T_{eff} = 31000$ K to stars at the cutoff in the distribution ($I = 19.5$)
based on measurements of NGC 6752 \citep{moe00}, which has similar metallicity
and HB extent.

As expected from the analysis above, the use of different sets of models has a
small effect on the weighted HB star total. In the case of some model sets
(DSEP, \citealt{swei87}), we needed to extrapolate the weighting corrections
in the range $4.4 \le \log T_{eff} \le 4.5$ for an extremely small number of
stars. Overall the models agree that the HB sample should be corrected
downward by approximately 10\%, with \citet{cass04} ($Y = 0.23, Z = 0.0006$)
and Sweigart ($Y = 0.25, Z = 0.001$) models giving $N_{HB}^{\prime} = 716$,
and DSEP ($Y = 0.248$, [Fe/H] = $-1.5$) and Sweigart ($Y = 0.30, Z = 0.001$)
models producing a lower $N_{HB}^{\prime} = 703$. Thus, the corrected ratio is
$R \approx 1.47 \pm 0.09$ and $1.51 \pm 0.10$ for our two methods of
determining $V_{ZAHB}$.  (The uncertainty introduced by the choice of models
is significantly smaller than the Poisson uncertainties.) These values deviate
from theoretical models for $Y = 0.245$ by about $1 \sigma$.  Based on these
arguments, the $R$ ratio appears to rule out a global helium enrichment of
$\Delta Y = 0.04$ at a $3\sigma$ level (since $dR/dY \approx 10$;
\citealt{cass03}).

The value of $R$ produced by this analysis rests on the accuracy of
the weightings used to correct the HB star total and on the
determination of the faint cutoff for the RGB sample. If incorrect
weightings were hiding a helium enrichment of $\Delta Y = 0.04$ among
M13 HB stars, they would need to have been overestimated by about
28\%. We remind the reader that the weightings are corrections {\it
  relative} to HB stars near the instability strip and are independent
of the absolute value of lifetimes, so we believe that an overestimate
of this magnitude would be difficult to produce. In addition, because
we used weightings derived from models with $0.23 < Y < 0.25$, this
would {\it overestimate} the weights for the bluest HB stars if those
stars are actually helium-enriched, not underestimate them. If
anything, our $R$ measurement is biased slightly too high.

To hide a helium enrichment through a systematic error in the RGB
sample, it would be necessary to overestimate the number of stars, which would
require a faint limit that was too faint. To produce a measured helium
enrichment of $\Delta Y = 0.04$, our faint limit would need to have
been at $V = 14.79$, or 0.20 mag brighter than our brightest
estimate. Only a gross error in judging the level of the HB could
produce an error this big, but the majority of the RR Lyrae stars in
the cluster dispute that, having $\langle V \rangle \approx 14.85$
\citep{kopacki02}. Even if the RR Lyrae stars are somewhat evolved,
they are still {\it fainter} than the ZAHB level needed to produce a
result of $\Delta Y = 0.04$.  (The RR Lyrae stars will be discussed
more below.) From other indications (such as $\Delta V_{TO}^{HB}$),
the HB of M13 is likely to be brighter than average for globular
clusters if it is deviant at all. As a result, we don't see a clear
reason to dispute the helium abundance implied by the $R$ ratio.

To avoid issues with systematic effects, it is worth doing a {\it relative}
comparison between M13 and M3. We assembled photometry for M3 from
\citet{ferrm3} and \citet{roodm3}, along with averaged photometry for the
large population of RR Lyrae variables from \citet{benko} and \citet{ccm3}.
This sample completely covers a $7\arcmin \times 7\arcmin$ area roughly
centered on the cluster core. This contains 602 HB stars, which we
subsequently corrected for lifetime effects. Because M3's HB morphology is
much more horizontal than M13's, we have used the $V-I$ color to determine
lifetime corrections, in part due to questions about the calibration of the
$B$ data in this cluster \citep{valm3}. For stars that had no $I$ measurement,
we determined an empirical transformation from $B-V$ to $V-I$ using
nonvariable HB stars from the \citet{ferrm3} data. For a handful of RR Lyrae
variables that did not have average brightness information, we assigned the
star a color equal to the average color of variables of the same type (RRab or
RRc). The details are fairly unimportant because the normalization for the
lifetimes is based on stars near the blue edge of the instability strip (near
the middle of M3's HB) and because lifetimes are theoretically predicted to
vary little except on the blue tail. For M3, the total lifetime correction
amounted to just over 1\%, with a corrected value of 594 stars.

The cutoff magnitude for the RGB sample ($\langle V_{RR} \rangle$) for M3 can
be determined quite accurately, so that we only need to worry about possible
zeropoint differences between the variable and non-variable star photometry.
We used $V_{ZAHB} = 15.70 \pm 0.03$ from \citet{ferrm3}, which was also
the source of the zeropoint for the HST observations of \citet{roodm3}. The
RGB sample brighter than $V_{ZAHB}$ is 444 stars, which gives $R = 1.34 \pm
0.08$. According to this analysis, the M3 sample is consistent with having
lower average helium abundance than M13, but the values differ by slightly
more than the error bars and the implied difference in $Y$ is only about
0.015. This is not enough to explain morphological differences in the CMD on
the subgiant branch or HB.

This is an interesting and somewhat surprising result that contradicts a
number of previous studies of M3 and M13 that indicated that M13 is enriched
in helium compared to M3. For example, in their discussion of globular
clusters with multiple stellar populations, \citet{dant08} identified M13 as a
cluster composed almost entirely of stars enriched in helium by about $\Delta
Y = 0.04$. In a later simulation, \citet{dant08b} also model M13's HB
population exclusively with helium enriched stars, but in this case with 70\%
of stars having ranging from $0.27 < Y < 0.35$ and 30\% having $Y = 0.38$.
For comparison, their interpretation of M3's populations involves 50\% of the
stars having a near-canonical value of $Y = 0.24$ and 50\% having $0.26 < Y <
0.28$.  Even if M13's $R$ value is not corrected for lifetime variations, it
still falls short of the value expected for the relatively small helium
enrichment of $\Delta Y = 0.04$. We will discuss the cluster helium abundance
in \S \ref{disc}.

\subsection{The $R_2$ Population Ratio and Post-HB Evolution}\label{r2}

As the morphology of the HB becomes more blue, the evolutionary tracks
of stars after the HB phase start to shift from traditional AGB tracks
(starting in a clump at the faint end of the AGB and subsequently
following a track paralleling the RGB) to abbreviated AGB tracks
(having an evolutionary pause separate from the traditional clump,
touching on the traditional AGB at moderate luminosities, and peeling
away before reaching the tip) to \manq ~ tracks (retaining a surface
temperature thousands of degrees higher than any part of the
traditional AGB). M13 contains HB stars that clearly span this range
according to the number of traditional AGB, supra-HB, and AGB \manq ~
stars we identified in \S 5. In view of the continuing difficulties in
explaining the blue HBs of clusters like M13, we consider whether the
AGB stars can reveal anything about the evolution or structure of the
progenitor HB stars.

At the simplest level, the $R_2=N_{AGB}/N_{HB}$ ratio compares the relative
lifetimes in the AGB and HB evolutionary phases.  \citet{vargas} found that
for a sample of clusters with large samples of evolved stars (more than 200 HB
stars) that the population ratio $R_2$ dropped well below the theoretically
predicted value of 0.12 \citep{cass04} for clusters with a HB type $\ga
0.8$ (mostly bluer than the instability strip).  When we include supra-HB and
AGB \manq ~ stars in the AGB population, 
we derive a value $R_2 = 90 / 795 = 0.113\pm0.012$.  For the 4 clusters with
the reddest morphologies (HB$_{type} < 0.2$), \citet{sb04} found $\langle R_2
\rangle = 0.106 \pm0.011$.  Clusters with $0.2 < \mbox{HB}_{type} < 0.8$ had
higher values (M5: $0.176 \pm 0.018$, \citealt{sb04}; M55: $0.156\pm0.023$,
\citealt{vargas}). Values for bluer clusters may be underestimated due to the
difficulty of identifying AGB \manq ~ stars using only optical filters.  Had
we not included AGB \manq ~ and supra-HB stars in the AGB star counts, we
would have calculated $R_2 = 61 / 824 = 0.074 \pm 0.009$, completely
consistent with the low values plotted there for M30 and NGC 6752. In the blue
HB cluster NGC 2808, \citet{castel} were able to identify \manq ~ stars, and
found an $R_2$ value similar to ours. With proper identification of the
different types of stars, the $R_2$ values are consistent with observational
values for bluer clusters and with theoretical predictions.


Based on the bimodal distribution of HB stars (approximately 47\% of HB stars
in the brighter peak, and about 28\% in the fainter peak), we might expect to
see evidence of an almost bimodal distribution of AGB stars with the \manq and
supra-HB stars evolving from the bluest HB stars.  The detected AGB \manq and
supra-HB population is $N_{manq}/N_{AGB} = 29 / 90 = 0.32\pm0.06$ of the total
population. If we assume as a first-order approximation that all AGB stars
have equal lifetimes, then this implies that the \manq stars originate in the
bluest 32\% of the HB stars ($I \ga 17.75$; $V \ga 17.55$). This excludes the
bright peak and the majority of the intermediate HB population (see Fig.
\ref{ihist}). Taking into account that bluer stars have longer lives (and so
are over-represented in the HB population), the red boundary for HB stars
producing \manq stars should be further to the red.  When this is accounted
for using the weighting factors discussed in \S \ref{rrat}, the boundary falls
at $I \approx 17.3$ ($V \approx 17.15$; $m_{1620} - V \approx -3$). Both imply
the track morphology switches between the two maxima of the HB distribution,
but closer to the fainter peak.



\section{Discussion}\label{disc}

M13 is distinguished by its horizontal branch, and its horizontal branch is
notable for two reasons: the fact that the stars are almost exclusively bluer
than the instability strip, and the fact that there is a secondary population
of stars grouped near the extreme end of the horizontal branch. Below we
discuss evidence bearing on each of these points and try to put M13 in context
of other globular clusters. However, before we do, we would like to state that
this is not intended as a complete survey of the subject, and we apologize for
our limited ability to reference the many previous studies.

\subsection{The Bimodal Horizontal Branch}\label{bimodal}

Independent of the position of the primary peak of the HB star distribution,
the cause of blue HB tails has not been definitively identified. Some studies
have identified relationships between HB tails and cluster dynamical
parameters such as central density \citep{bluetail} and total luminosity
\citep{rb06}, although neither of these can clearly explain differences
between M3 and M13 because of their structural similarities (see Table 1).
While blue tails can plausibly result from a process (such as mass loss or
chemical self-enrichment) producing a large dispersion in properties, M13 has
a clear secondary peak in its HB distribution and this requires multiple star
populations or the action of multiple physical processes.  For example,
theoretical models predict that varying amounts of mass loss on the RGB from
star to star can produce a spectrum of outcomes from normal HB stars to early
hot flashers (stars which leave the RGB before the helium flash, but ignite
helium shortly afterward) to late hot flashers (stars which ignite helium on
the white dwarf cooling curve) to helium white dwarfs (which never ignite
helium). To produce a secondary HB peak in this way, some mechanism must be
concentrating stars in the CMD, and it is difficult to see how this can be
accomplished without almost complete loss of the hydrogen envelope for many
stars on the RGB. On the other hand, the self-enrichment hypothesis explains
peaks on the HB (not only at the blue end) via discrete populations of stars
with different chemical compositions. A peak at the blue end of the HB can be
accomplished with a population of stars having extreme helium abundances ($Y >
0.35$) because enriched stars leave the main sequence with lower total mass,
and if the mass is low enough they will have almost no envelope by the time
they reach the TRGB. He abundances for these hot stars probably can't answer
the question directly (due to the action of diffusive and mixing processes),
but correlated enrichment of helium and carbon among the hottest stars in
$\omega$ Cen and NGC 2808 \citep{moe04,moeom} is not predicted in the self
enrichment picture.  As yet, there is no clear way of distinguishing between
these scenarios for EHB stars. However, clearer understanding of these stars
may provide new clues, so we compare M13 with other clusters with blue HB
tails below.

Like {\it some} of the most massive clusters [NGC 2419, NGC 2808, $\omega$
  Cen, and NGC 6273 (M19)], M13 has a clear second peak at the blue end of the
HB in optical filter bands. Even so, there is a range in the way the secondary
peak appears. Some of the best comparisons of different clusters (with HBs
aligned in the optical) can be found in \citet{pio}, \citet{dale}, and
\citet{sh}.  Often there is a gap or edge feature in the HB star distributions
as a function of an optical magnitude ($M_V \approx 4.5, M_B \approx 4.2$) and
this feature appears to separate blue hook stars from the EHB. In M13 (and
also NGC 6752; \citealt{sabbi}), the secondary peak is almost entirely
brighter than the position of the gap. The secondary peak in NGC 2419 reaches
its maximum among the blue hook stars, but appears to straddle the gap. The
secondary peak in M19 straddles the position of the gap, but differential
reddening prevents the clear identification of a feature.  $\omega$ Cen shows
a sharp edge in the HB distribution at about the position of the gap in NGC
2419, with a much smaller fraction of stars just brighter than the gap. NGC
2808 on the other hand appears to have almost all of its hottest HB stars
fainter than the gap, but its population is a considerably smaller fraction of
its HB stars.  Many other clusters with long blue HB tails do not show a
secondary maximum at the blue end of the HB, however. M54 \citep{momany} and
M15 appear to have produced blue hook stars, but there is little or no sign of
a secondary maximum. M80 \citep{gaps} and M2 \citep{momany} have HBs with a
similar extent to that of M13, but do not show maxima. So clusters are capable
of producing EHB populations comprising from nearly 0 up to 30\% (for NGC
2419) of the total HB tally, with distinguishing features even among the
very bluest HB stars in different clusters.

As emphasized in models \citep{brown} and observations \citep{momany},
evolution of hot HB stars is mostly in luminosity and this can allow
the evolved stars to masquerade as brighter but less evolved HB stars
in optical photometry due to steep bolometric corrections. To better
characterize the populations at the end of blue end of the HB,
ultraviolet observations are needed. In near UV photometry ($U$,
F336W), the effects of luminosity evolution become better separated
from color distribution. As can be seen in Fig. \ref{hbuv}, far UV
photometry (UIT B5, F160BW) mostly flattens the blue end of the HB in
magnitude, and should allow the optimum separation of luminosity and
color effects. Far UV observations are not very common, and it is even
rarer to connect observations in different UV wavelength bands.

\begin{figure}
\epsscale{.90}
\plotone{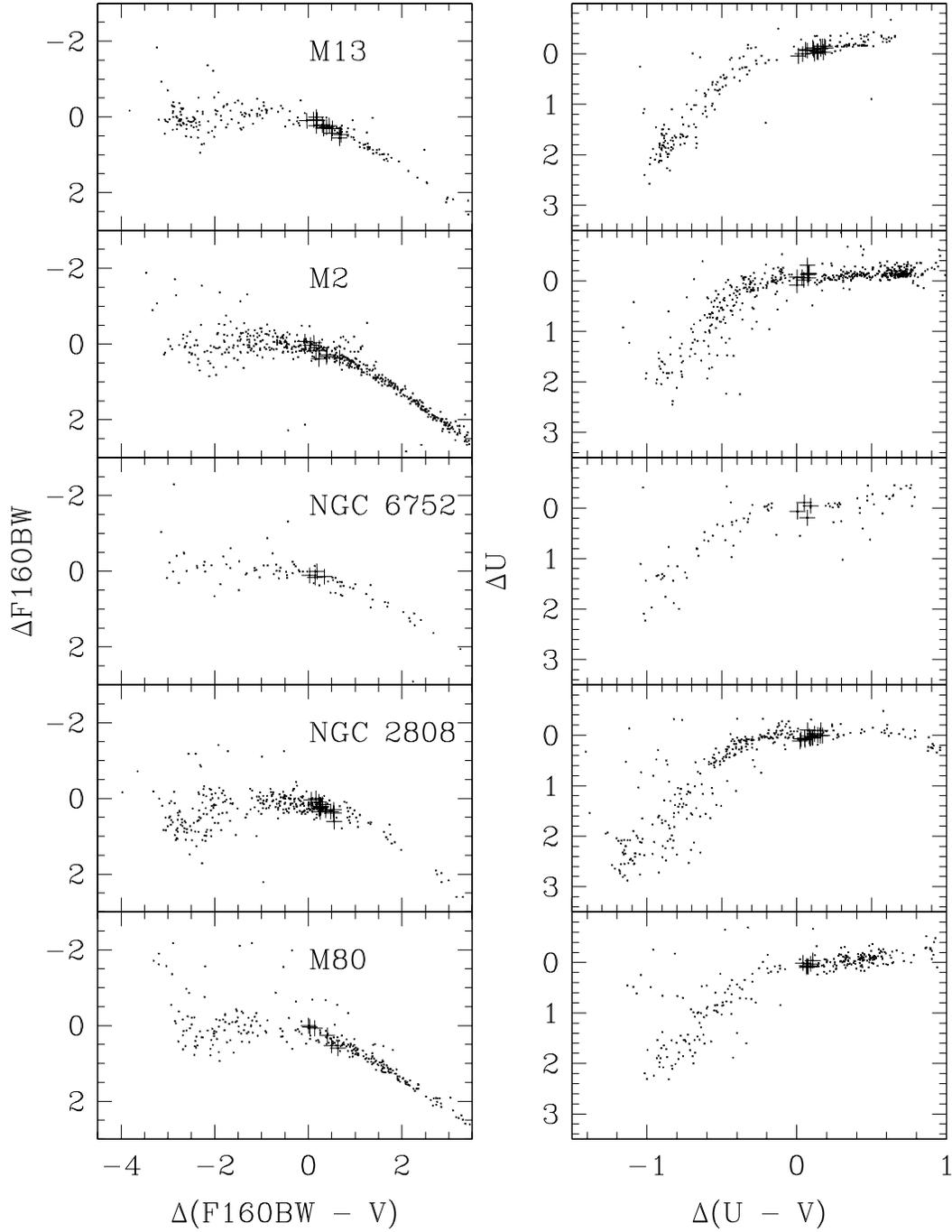}
\caption{CMDs of 5 globular clusters with blue HB tails in the F160BW and
F336W (calibrated to $U$) filters from HST. Crosses show stars identified at
the red end of the Grundahl $u$ jump, and all CMDs have been shifted so that
the bluest of these stars is at (0,0).\label{cmduf1}}
\end{figure}

A small number of clusters have archival data in the near and far UV as well
as optical bands. HST WFPC2 data is available from proposals 5903 (M80; F.
Ferraro PI), 6804 (NGC 2808; F. Fusi Pecci PI), and 8709 (NGC 6752, M2; F.
Ferraro PI). We processed images using HSTPhot in a manner similar to the M13
images. To compare different clusters, we identified stars at the red end of
the Grundahl $u$ jump in $U$ band from HB morphology or with the help of
spectroscopic information (NGC 6752, \citealt{moni}; NGC 2808,
\citealt{moni09}). The $u$ jump appears to have a common $\teff$ in all
clusters \citep{ujumps}, and so we used it to register the CMDs by
temperature. In addition, blue HBs have a long segment at nearly constant $U$
magnitude \citep{ferruv} that includes the $u$ jump, and in the F160BW band
the $u$ jump falls near the red end of a fairly flat segment reaching nearly
to the end of the HB. Thus, this registration provides a convenient means of
comparing the extent of blue HBs in color and magnitude. It also avoids the
need to make large and uncertain corrections for reddening in UV
bands \citep{ccm} or for the time dependence of WFPC2 throughput
in the UV \citep{holtz}. Fig. \ref{cmduf1} shows sets of CMDs shifted so that
the red end of the $u$ jump is at (0,0).

A few conclusions can be drawn from simple comparisons. First, the large
population of blue hook stars in NGC 2808 falls in a portion of the CMD
($\Delta U > 2, \Delta(U-V) < -1$) that is not occupied by more than a few
stars in any other cluster. It is worth remembering that these are not the
bluest stars in the $\Delta$ F160BW, $\Delta(\mbox{F160BW} - V))$ CMD. They
are fainter ($\Delta$ F160BW$ > 0.5$) and slightly redder than the blue end.
This is the origin of the ``blue hook'' moniker, and it is thought to be the
result of a helium-rich atmosphere resulting from flash mixing \citep{brown}.

\begin{figure}
\epsscale{.90}
\plotone{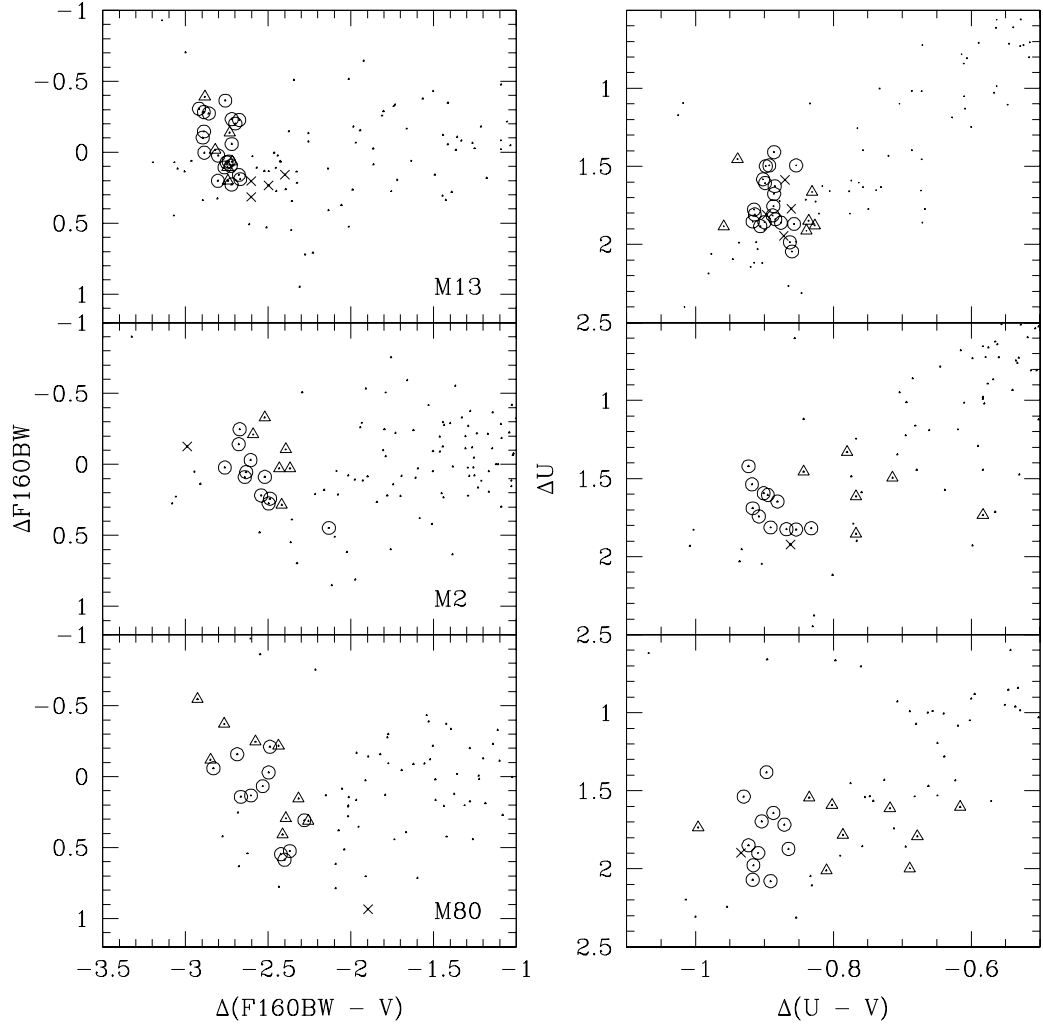}
\caption{Similar to Fig. \ref{cmduf1}, but zoomed on the blue end of the HB.
  $\bigtriangleup$ show are stars that were identified in the jump feature in
  the F160BW filter, $\times$ were ones identified in the F336W ($U$) filter,
  and $\bigcirc$ ones that were identified in both.
\label{uf1zoom}}
\end{figure}

M13, M2, and M80 have similar morphologies at the blue end of the $\Delta$
F160BW, $\Delta(\mbox{F160BW} - V))$ CMD, though the distributions of stars
differ. The HB sequence dips faintward in the range $-2.4 \la
\Delta(\mbox{F160BW} - V)) \la -2$ before rising again up to $(\mbox{F160BW} -
V)) = -3$. Most of the stars in the rising portion are also in the Momany $U$
jump (see Fig. \ref{uf1zoom}). The identification between the upturn in F160BW
and the Momany $U$ jump is clearest in M13 because of the large population of
stars at the end of the HB. Some stars are found to the red in the ($U, U-V$)
CMD for M2 and M80, although these may be blends with main sequence stars in
$V$ band\footnote{Main sequence stars contribute very little light in UV
  bands, and so blends shift almost horizontally to the red in the CMD.
  Because the F160BW$-V$ color has a longer wavelength baseline than $U-V$,
  blends don't shift stars as much.}.

The correlation between stars in the Momany jump and the upturn in F160BW is
not perfect, but it suggests that there is a coherent group of stars to be
found on the extreme blue HB. The group does not fall at the precise end of
the HB though, as we find in each cluster small numbers of stars that are
bluer and fainter in the F160BW and $U$ CMDs. This interpretation differs from
that of \citet{momany}, who stated that the HB sequence appeared to make an
almost discontinuous jump in color ($\sim 0.3$ mag) at constant $U$ magnitude,
and extending faintward in $U$ (by $\sim 0.5 - 0.7$ mag) at nearly constant
color. The color histograms make it clear that the three clusters have very
different distributions, with M2's distribution declining toward the blue end,
M80 with a broad but evenly spread group, and M13 with a well-defined peak and
a large fraction of the stars in the Momany jump.


While no spectroscopic data exists in the literature for M13 stars in
this range, \citet{moe03} examined cooler HB stars (8000 K $\la \teff
\la$ 21000 K) and found evidence of strong helium depletion and iron
enrichment for $\teff \ga$ 12000 K. We can get additional guidance by
contrasting spectroscopic results for NGC 6752 \citep{moni} against
those for NGC 2808 \citep{moe04} and $\omega$ Cen \citep{moeom}. NGC
6752, $\omega$ Cen, and NGC 2808 all have noticeable vertical jump
features in ($U, U-V$) CMDs, but the most obvious jumps occur at
larger $\teff$ in NGC 2808 and $\omega$ Cen. In NGC 6752, there is
consistently helium depletion in the atmospheres of stars hotter than
the Grundahl jump at $\teff \approx 11000$ K, including the hottest
stars observed.  In NGC 2808 and $\omega$ Cen, stars with near-solar
and super-solar helium abundances (and concurrent carbon enrichment)
appear at temperatures above those seen in NGC 6752 ($\teff \approx
31000$ K). Both spectroscopic signatures are consistent with the
predictions of ``late hot flashers'': stars that ignite helium on the
white dwarf cooling curve, and initiating convective mixing of the
envelope.

We are led to the same conclusion reached by \citet{momany}: that the
stars in the second $U$ jump are likely to be ``early hot
flashers''. In this case, the existence of a working hydrogen-fusion
shell is thought to inhibit convective mixing into the outer envelope
during core helium ignition. As a result, no clear spectroscopic
signature is expected and none has been found to date. Because helium
enrichment of the envelope plays a role in producing blue hook stars
\citep{brown} by reducing atmospheric opacity shortward of the Lyman
limit (and reducing redistribution of flux to longer wavelengths), we
examined the data for spectroscopically studied stars in the Momany
jump within NGC 6752 \citep{moni}.  We found no correlation between
measured helium abundances [which covered the range $-3.26 \le \log
  (n_{He} / n_{H}) \le -1.58$] and $U$ magnitude, which seems to
confirm that the envelope is still dominated by hydrogen. The scatter
in helium abundance appears to be interesting, however --- when
\citet{moni09} combined their measurements for M80 and NGC 5986 stars
with those from NGC 6752, they found consistent helium depletion for
stars hotter than the Grundahl jump at $\teff \approx 11000$ K, but
also found stars with larger depletions and generally larger
star-to-star scatter in two temperature ranges (13000 K $\la \teff
\la$ 18000 K and 25000 K $\la \teff \la$ 31000 K).

Both of the helium depletion features appear to be associated with the $U$
jumps, although the details probably differ. For the Grundahl jump, stars that
are hotter than the jump are uniformly brighter in the Stromgren $u$ filter
than a canonical ZAHB. By contrast the stars associated with the second $U$
jump in M13 are brighter than the majority of HB stars of similar (but higher
or lower) temperature, as seen in the $U$ and F160BW filters.
Observationally, the radiative levitation of heavy elements and the downward
diffusion of helium almost certainly must be involved because all stars with
$\teff \la 11000$ K (excepting blue hook stars) are found to have some degree
of surface helium depletion, but the mechanics are still being debated. The
transition at 11000 K is associated with the near-disappearance of the surface
convection zone, but recent models \citep{mich} indicate a need for a low-mass
mixed layer near the surface to moderate the build-up of metals. There may be
reason to attend to the effects of rotation because a discontinuity in
average rotation speeds also appears at around this temperature (with hotter
stars largely being slow rotators; \citealt{pete,behr03}).

Based on the evidence above, we considered whether progenitors of extreme HB
stars such as early and late hot flashers can be identified on the RGB. Early
hot flashers are theoretically expected to only leave the RGB near the tip,
while late hot flashers could leave at lower luminosities. While M13 appears
able to produce both kinds of hot flashers, the second $U$ jump
stars (our early hot flasher candidates) are much more abundant. Thus, we
would expect that clusters with similar HB morphologies (like NGC 6752, M2,
and M80) to show symptoms of this near the TRGB if anywhere.

Strangely, \citet{sned} found that M13's known ``super O-poor'' stars ([O/Fe]
$< -0.4$) have high Na abundances and most appear very close to the RGB tip
($M_V^0 < -2.3$), indicating the possible exposure of heavily processed
gas at the surface. For perspective, \citet{carr06} summarize literature data
for the well-known anticorrelation between [O/Fe] and [Na/Fe] for stars in 20
clusters, and \citet{carbrag} present new spectroscopic data for 19 clusters.
The extent of the O-Na anticorrelation varies from cluster to cluster, but it
probably requires nuclear processing under conditions that cannot be produced
in gas that can be mixed to the surface in low-mass stars. This has led to the
supposition that it results from pollution by previous generations of more
massive stars. Even so, super O-poor stars are quite rare and, to the best of
our knowledge, have not been seen among relatively unevolved turnoff and
subgiant branch stars \citep[see][]{carr06}, although this could be due to
difficulties in detecting low oxygen abundances among faint stars. 

In addition, the cluster NGC 2808 has a large population of blue hook stars
and seemingly few or no early hot flasher candidates, and contains super
O-poor stars as well \citep{carr06} but these are fairly uniformly spread down
the RGB (although those authors did not observe stars at the tip of the RGB).
$\omega$ Cen also has a large population of blue hook stars, and appears to
have a population of super O-poor stars \citep{nda}. In the cases of NGC 2808
and $\omega$ Cen though, there is clear evidence of both self-enrichment and
multiple populations, which may complicate the interpretation.

Examination of bright giants in NGC 6752 \citep{yong03,carbrag} have
revealed no super O-poor stars. In addition, Mg isotope ratios in M13
\citep{sned} and NGC 6752 are correlated with the O depletions among bright
giants, but the conditions necessary to process Mg are not expected in RGB
stars based on current nuclear reaction rates. The sample of \citet{carbrag}
hints at the possibility of smaller populations of super O-poor stars in
clusters like NGC 3201 and M5, neither of which have extreme HB stars.

Though there is strong evidence of mass loss effects on the extreme BHB, there
is still little to connect this directly to the properties of stars still on
the RGB. One outstanding question that deserves theoretical attention involves
the position (redder than the extreme end of the HB) and clumping of the early
hot flasher candidates in the CMD.  Even if helium is enriched overall in the
bluest HB stars, some non-standard physics is probably required to produce
their observed characteristics because helium-enriched EHB stars should be
fainter than ones with primordial helium abundance. We can suggest two types
of observations that might help to clarify the situation. First, in clusters
with stars concentrated near the blue end of the HB, stars at the TRGB may
have chemical peculiarities similar to those in M13 if they are preparing to
have early hot helium flashes. If found, this could result from excessive mass
loss or from highly helium-enriched stars, but it would connect RGB and
extreme HB stars. Second, in clusters with predominantly red HBs, spectra for
blue HB stars in a limited range of $T_{eff}$ near the instability strip may
reveal their initial helium abundances \citep{villa}. This would be a direct
observational test of whether helium enrichment has produced {\it any} blue HB
stars. While there is circumstantial evidence of helium enrichment in clusters
like NGC 6388 and NGC 6441 \citep{busso}, spectroscopic proof should be
possible.

\subsection{``Second Parameter'' Effects on the HB}

Because the $R$ ratio seems to rule out previously suggestions of
helium enrichment in M13, we discuss the literature on M13 to see if the
observations can be reconciled.  The main evidence pointing toward a helium
enrichment falls into several categories: 1) morphology of the HB, 2)
luminosity of the HB and the RR Lyrae variables, and 3) morphology of the
subgiant branch.

\subsubsection{HB Morphology}

M13 and M3 have long been a ``second parameter pair'' based on the
distribution of stars on their HBs --- the small difference in [Fe/H]
(the ``first parameter'') between the two clusters is unable to
explain the much bluer HB stars in M13. Before discussing further, we
emphasize that we are mainly considering the shift in the CMD position
of the ``most representative'' HB stars in the two clusters, and not
the differences in the shape of the HB distributions. Our goal is not
to model the HB morphology in detail. M13 has an unexplained
bimodal HB distribution as well as a significant number of stars
between the two peaks, but arguably the brighter peak ($15 < V <
15.5$; slightly bluer than the instability strip) is the most
representative based on star numbers.  By contrast, M3 has a unimodal
distribution with the peak of the distribution falling within the
instability strip \citep{valm3}.

Earlier studies \citep{fusi93,bluetail} have used $(B-V)_{peak}$ (the
dereddened color of the peak of the HB star distribution) as a way of
describing the most representative HB stars in clusters, although this
indicator starts to lose sensitivity on the steep blue tail. We used
our earlier results from histogramming the $I$ magnitudes of HB stars,
and then translated the resulting position to the $(B-V)$ color using
a fiducial line.  We find $I_{peak} \approx 15.25$ and $(B-V)_{peak}
\approx -0.05$. For comparison, \citet{bluetail} give $(B-V)_{peak} =
0.30$ for M3. If the clusters have nearly identical chemical
compositions, this corresponds to a difference of about $0.06 \msun$
in HB star mass (Teramo models).  Unfortunately, this does not
transform to a similar mass difference at the cluster turnoff if
common red giant mass loss formulas are applied because they are
nonlinear with star mass. If M13 is helium enriched compared to M3 by
$\Delta Y = 0.05$, the mass difference is virtually the same. Helium
enrichment in M13 does not remove the need for a difference in mean
mass between its HB stars and those in M3.

The mass difference can be explained with reasonable assumptions in
both the $\Delta t$ and $\Delta Y$ hypotheses, and we refer the reader
to \S 1 for a brief summary of recent attempts to model M13's HB.  In
the $\Delta t$ hypothesis, M13's stars are older than M3's. In the
$\Delta Y$ hypothesis, helium enrichment allows less massive stars to
leave the main sequence earlier, {\it but} a large amounts of mass
loss is still necessary for each star. While this may be correct, the
picture is not fully motivated by physical reasoning.  As discussed in
\S \ref{redhb}, the sharp red edge of the bright peak and the population
of redward-evolving HB stars may provide new constraints on fits in
the $\Delta Y$ hypothesis. We encourage synthetic HB studies that pay
closer attention to these features.

\subsubsection{The Luminosity of the HB}\label{lhb}

The luminosity of the HB is most strongly affected by metallicity and helium
abundance ($\delta V_{HB} / \delta Y \approx -4$ mag), and is insensitive to
age. However, without accurate and independent measurements of distance and
extinction, the luminosity must be determined relative to other cluster
stars. Most commonly, the reference landmarks are found among MS stars (such
as the turnoff or a fainter point identified with the help of the turnoff) or
RGB stars (like the RGB bump).  These methods have generally indicated that
M13's HB is luminous compared to other clusters, but the reference points
usually have the disadvantage of having their own dependencies on
composition and age that complicate the interpretation. In the discussion
below, we will focus on the $\Delta Y = 0.04$ \citep{jb} and $\Delta t = 1.7$
Gyr \citep{rey} hypotheses for explaining differences between M3 and M13, but
we must consider the evolution hypothesis (see \S 1) at the same time.

When the cluster turnoff is used as a reference, it introduces a dependence on
age ($\delta V_{TO} / \delta t \approx 0.075$ mag / Gyr; \citealt{cda}) and
increases the dependence on helium ($\delta V_{TO} / \delta Y \approx 1.5$
mag; \citealt{cda}).  We determined $V_{TO} = 18.59\pm0.05$, which gives
$\Delta V_{TO}^{HB} = 3.69 \pm 0.07$ if we apply our value for $V_{HB}$ at the
blue edge of the instability strip. This value is larger than any previously
quoted (see \citealt{jb} for a summary), and if we use the \citet{kopacki02}
value for $\langle V_{RR}\rangle$, the value would be higher still (3.76 mag).
The larger values of $\Delta V_{TO}^{HB}$ depend entirely on
whether the RR Lyraes and ``reddest blue HB'' stars near the instability strip
(see \S \ref{redhb}) are evolved or not. Both \citet{jb} and \citet{rey}
observed much smaller samples in M13, and an examination of their CMDs shows
that together they only observed two ``reddest BHB'' stars among the 55 we
identified --- understandable considering that most of these stars are in the
core of the cluster. As a result, \citeauthor{jb} quote a value for $V_{HB}$
that is 0.19 mag fainter than ours.  We also note that like the studies of
\citeauthor{jb} and \citeauthor{rey}, our photometry for the HB and turnoff
regions are on a consistent zeropoint.  We see no reason to dispute earlier
values of $\Delta V_{TO}^{HB}$ for M3, given that M3's HB is well-populated on
either side of the instability strip.  So if M13's reddest HB stars are not
evolved, it has an extreme value for $\Delta V_{TO}^{HB}$ (almost 0.2 mag
larger than that of M3), and this is consistent with both the $\Delta t$ and
$\Delta Y$ hypotheses.  To find the cause (the HB, TO, or both), we need
other reference points.

\citet{ferruv} conducted a similar comparison between M3 and M13 using HST $U$
observations. In $U$, the subgiant branch and a section of the blue HB become
flat, making them excellent magnitude references. The authors found no
significant difference between the $\Delta U_{SGB}^{HB}$ derived for the two clusters.
While this does not directly bear on the luminosity of the HB near the
instability strip, it is an indication that the bluer HB of M13 has a normal
luminosity compared to M3.

The red giant bump can also be used as a reference point, dependent on helium
abundance ($\delta V_{bump} / \delta Y \approx 1.5 - 2 $ mag;
\citealt{riello,cda}) and age ($\delta V_{bump} / \delta t \approx 0.035$ mag
/ Gyr; \citealt{cda}).  The observed value for the difference $\Delta
V^{bump}_{HB}$ can be put in context using the catalogs of \citet{ferr99} and
\citet{dice}. To ensure a reliable comparison, we calculated the correction
from $V_{HB}$ to $V_{ZAHB} = 14.96$ according to the \citeauthor{ferr99}
prescription.  We therefore derive $\Delta V_{HB}^{bump} = -0.21$, which falls
right among those for clusters of similar metallicity (for example, M3 has
$-0.23 \pm 0.07$).  However, their tabulated value for $V_{ZAHB}$ is 0.14 mag
fainter than ours --- this is partly a reflection that the photometry they
used \citep{palt} lacked stars near the instability strip, and partly that
they used synthetic HB calculations to derive the level of the ZAHB relative
to the HB stars they did observe. Using the fainter ZAHB, M13's value for
$\Delta V_{HB}^{bump}$ becomes significantly different from M3's value, and
falls at the low end of the distribution for similar clusters. So if the RR
Lyraes are showing us the true ZAHB level, neither age nor helium changes are
necessary but can't be ruled out because the expected changes are $1 \sigma$
level. If the ZAHB has a fainter level, rather large differences are needed
($\Delta Y = 0.06 - 0.09$, or $\Delta t = 3.5$ Gyr). The tabulation of
\citet{dice} measured values for both M3 and M13 using homogeneous photometry
and a template-fitting method for determining $V_{ZAHB}$ for troublesome
clusters like M13 that have very blue HBs. Their $\Delta V_{HB}^{bump}$ values
are in agreement to within the measurement errors.\footnote{\citet{cda} also
  used the position of the RGB bump relative to the turnoff ($\Delta
  V_{TO}^{bump}$) as a potential helium indicator, with larger magnitude
  differences implying higher helium. In their comparison, they found that M13
  had a magnitude difference that was $0.14 \pm 0.09$ larger than that of M3,
  which they admitted had relatively low significance.  Because our photometry
  of the bump and main sequence constitutes a homogeneous dataset with a much
  larger number of stars, we checked their quoted bump and turnoff magnitudes
  (derived from the photometry of \citealt{palt}). We find there is a rather
  large difference in color (0.08 mag) between the \citeauthor{palt}
  photometry and ours, but the $V$ magnitudes are almost identical to ours.
On their own, the RGB bumps of the M13-M3 pair do not give a strong constraint
on helium abundance, although \citeauthor{cda} showed that clusters with
predominantly blue HBs had larger $\Delta V_{TO}^{bump}$ than clusters that
have strong populations in the instability strip.}

The brightness of the AGB clump is fairly independent of metallicity
\citep{ccp}, but when the magnitude difference with the HB is formed, the
dependences on helium abundance and RGB progenitor mass almost completely
disappear \citep{pulone,cass01}. Although M13 has a sparsely populated AGB
clump, there is a fairly clear grouping of 12 stars at $V = 14.19$. Using the
bright ZAHB level from the previous paragraph, we have $\Delta V_{HB}^{AGB} = -0.77$.
This is slightly smaller than the value for M3 ($-0.88 \pm 0.10$) from
\citet{ferr99}, but only about $1 \sigma$ different. If the fainter ZAHB level
is applied, the M3 and M13 values agree to within about 0.03 mag. The results
might also be affected by likely lower mean HB mass in M13 compared to M3,
which affects how the clump is populated and would tend to reduce the
luminosity of the clump. We consider this weak evidence in favor of the
fainter ZAHB level.

\begin{figure}
\plotone{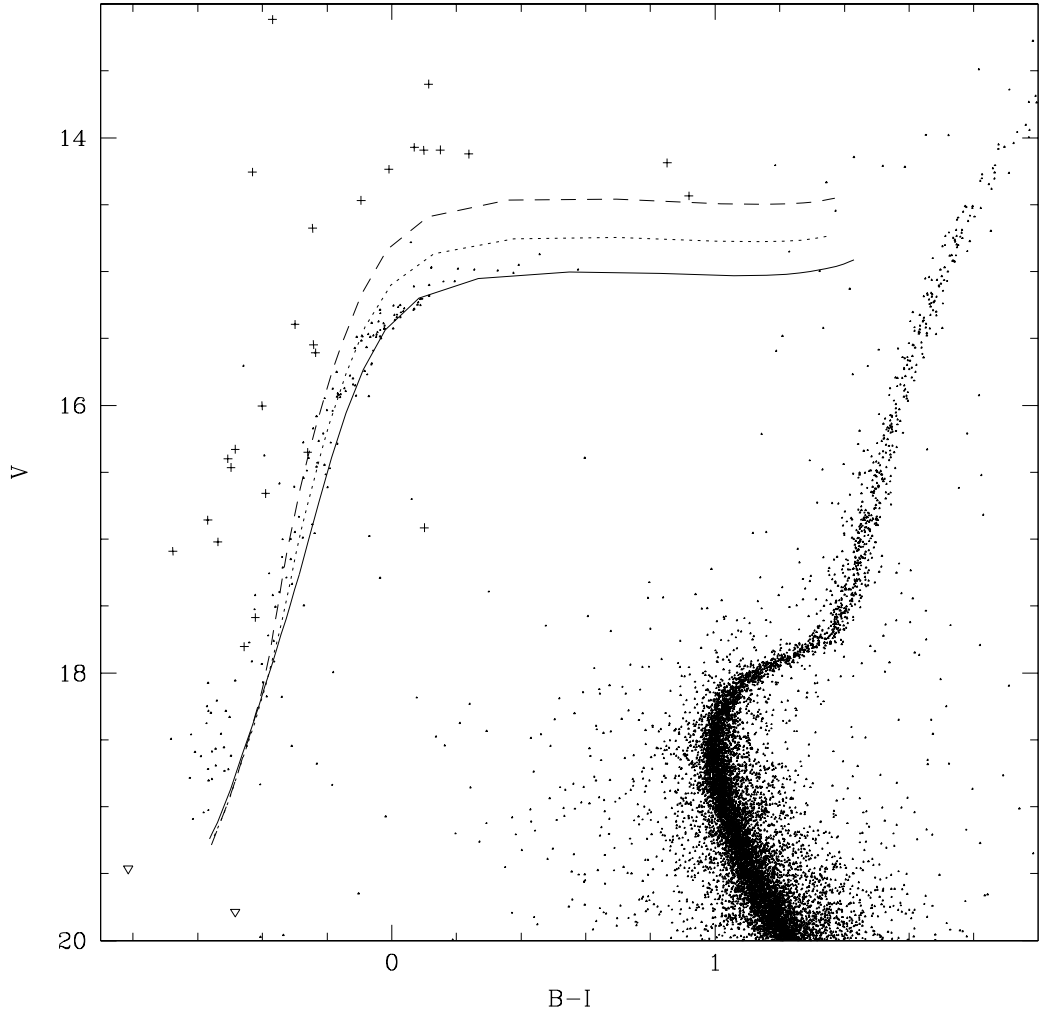}
\caption{Comparison of ground-based CFHT and KPNO photometry ($r >
  200\arcsec$) with Teramo \citep{piet06} ZAHB models for $Y = 0.246$ ({\it
    solid line}), 0.30 ({\it dotted line}), and 0.35 ({\it dashed line}). All
  ZAHB curves have been shifted by 0.02 in $B-I$, but by 14.45, 14.40, and
  14.30 mag respectively in $V$. Blue hook candidates are shown with
  $\bigtriangledown$ symbols.\label{hbends}}
\end{figure}

The HB luminosity can be constrained using the end of canonical HB as a
reference in clusters with long blue tails.  Like NGC 6752, M13's HB appears
to terminate essentially at the end of the canonical HB, and we identified
blue hook candidates (in \S \ref{notable}) and second $U$ jump stars (in \S
\ref{bimodal}) that mark points in the HB near the end.  Models show that the
difference in magnitude between the termination of the blue HB and the blue
end of the instability strip grows substantially larger with increasing helium
abundance, mostly because the horizontal part of the HB gets brighter. The
faint end of the HB is fairly insensitive to helium abundance because the
hydrogen-rich envelopes of the stars are almost gone.  If the end of
theoretical ZAHBs are shifted in magnitude to fit the end of M13's HB, the
bright end is consistent with primordial helium values (see Fig.
\ref{hbends}). While the fit is understandably uncertain due to the gap
between the HB stars and the blue hook stars in optical filters, the Teramo
evolutionary tracks do go nearly parallel to the faint envelope of HB stars
(see Fig. \ref{hbtervi}).  The fit seems to rule against helium enrichment
among stars within and redder than the primary peak.  Evolutionary effects
among the reddest HB stars would tend to bias toward an indication of helium
enrichment, so this result seems fairly robust. Spectroscopic measurements of
stars near the red end of the HB in M13 and NGC 6752 also support the idea
that the outer envelopes have {\it not} been significantly processed through
the CNO cycle, and (in NGC 6752) still have primordial helium abundance
\citep{villa}.

Alternately, if the ZAHBs are fit to the red end of the HB, helium-enriched
ZAHBs are far too red at the faint end, lending additional credence to models
with canonical composition. However, current models of the bluest HB stars do
not fully incorporate important physics such as diffusion and radiative
levitation, and these effects seem to be responsible for the $u$ jump and the
intra-peak HB stars that are brighter than canonical models. As a result, 
fits to the HB should be treated somewhat skeptically.

Finally, we consider the distance moduli implied by the HB (specifically the
RR Lyrae stars) and by the TRGB. Both of these features are in common use as
standard candles, and a comparison of the resulting values might also give us
clues on whether the HB is unusually bright. The absolute magnitude of the
TRGB in $I$ is almost independent of age and chemical composition. From the
calibration of \citet{bella}, we calculate $M_I^{TRGB} = -4.07$, which gives
$(m-M)_I = 14.35$. The distance modulus could be smaller than this if by
chance M13 does not contain RGB stars very near the flash stage. To evaluate
this possibility, we can use a binomial distribution to calculate the
probability that at least one star is within a certain magnitude range of the
TRGB. In our cluster-wide sample, there were 271 stars brighter than the RGB
bump ($I < 13.6$).  Using theoretical RGB luminosity functions that fit M13's
population (see \S \ref{rgb}), stars brighter than that level have an
approximately 0.4\% chance of being within 0.01 mag of the TRGB, so that a
calculation using the binomial distribution predicts a 70\% chance of having
at least one of the 271 stars within that interval. The probability rises to
about 90\% that at least one star is within 0.05 mag of the TRGB, so this is
probably a minor source of error. 

All of the brightest RGB stars are known to be semi-regular variables
including the brightest (V11), having an amplitude of about 0.13 mag in $V$ and
a timescale for variability of about 30 d \citep{kopacki02}. As a result, the
variability of the brightest giants is a significant source of error in our
determination of $I_{TRGB}$. Though \citeauthor{kopacki02} made observations
over 70 d (more than two cycles for most giant variables), it is possible they
did not sample the full amplitude of variation.  Our CFHT $V$ measurement of
V11 (which was used above) was 0.15 mag brighter than the average given by
\citeauthor{kopacki02}, while our KPNO observation was only about 0.06 mag
brighter (barring zeropoint differences). If we can assume that the $I$
observations are off by similar amounts, this gives $(m-M)_I \approx 14.5$.
The range of variation for other bright RGB stars (V17, V24, V38, and V42)
overlap that of V11, and all but one give corrected distance moduli within
0.03 mag of the V11 value. V17 gives a significantly brighter value (14.37),
but it also has the largest amplitude (0.38 mag) of the red giant variables
tabulated by \citeauthor{kopacki02}.
We believe the agreement among 4 of the 5 variables is sufficient to
identify the TRGB, which in turn implies $(m-M)_0 = 14.48$ with a very
conservative uncertainty of 0.10 mag.

Although there is still considerable discussion of the RR Lyrae
$M_V-$[Fe/H] relation, recent consensus views \citep[e.g.][]{catrr}
imply a mean RR Lyrae value $M_V^{RR} = 0.62$ for M13's metallicity.
With $\langle V_{RR}\rangle$ from the \citet{kopacki02} observations
of M13 RR Lyraes, this gives $(m-M)_V = 14.21$ . The disagreement with
the TRGB distance modulus can again be relieved if the RR Lyraes in
M13 were significantly evolved or were helium enriched ($\Delta Y
\approx 0.05$). If we use the TRGB distance modulus and $M_V^{RR}$ to
calculate the brightness of the HB, we find $V_{HB} = 15.07$. This is
fainter than all of the ``reddest BHB'' stars but approximately the
same brightness as the blue end of the primary peak.

Once again, a comparison with M3 is useful, thanks to that cluster's
huge RR Lyrae population. \citet{jurc} identified 4 groups of RR
Lyraes with different brightness and light curve parameters, with the
most abundant group of 50 stars having $\langle V\rangle =
15.67$. Their most evolved group (32 stars), which the authors
hypothesized were evolving redward, had $\langle V\rangle =
15.53$. The difference between $\langle V\rangle$ for the most evolved
RR Lyraes in M3 and that of M13's RR Lyraes is most consistent with
magnitude differences derived from comparing points like the RGB bump
and AGB clump \citep{ferr99}.

Evolution is also expected to affect the pulsational properties of stars
within the instability strip because the mean density of evolved stars is
lower than for unevolved stars, resulting in larger periods. We could expect
to discern evolutionary effects in several ways: based on observations derived
solely from the light curves (such as the Bailey period-amplitude diagram),
period-luminosity relationships (with luminous RR Lyraes having longer
periods), and color distributions (with evolved stars likely to be spread more
evenly through the instability strip due to their short evolutionary timescale).

\begin{figure}
\epsscale{.80}
\plotone{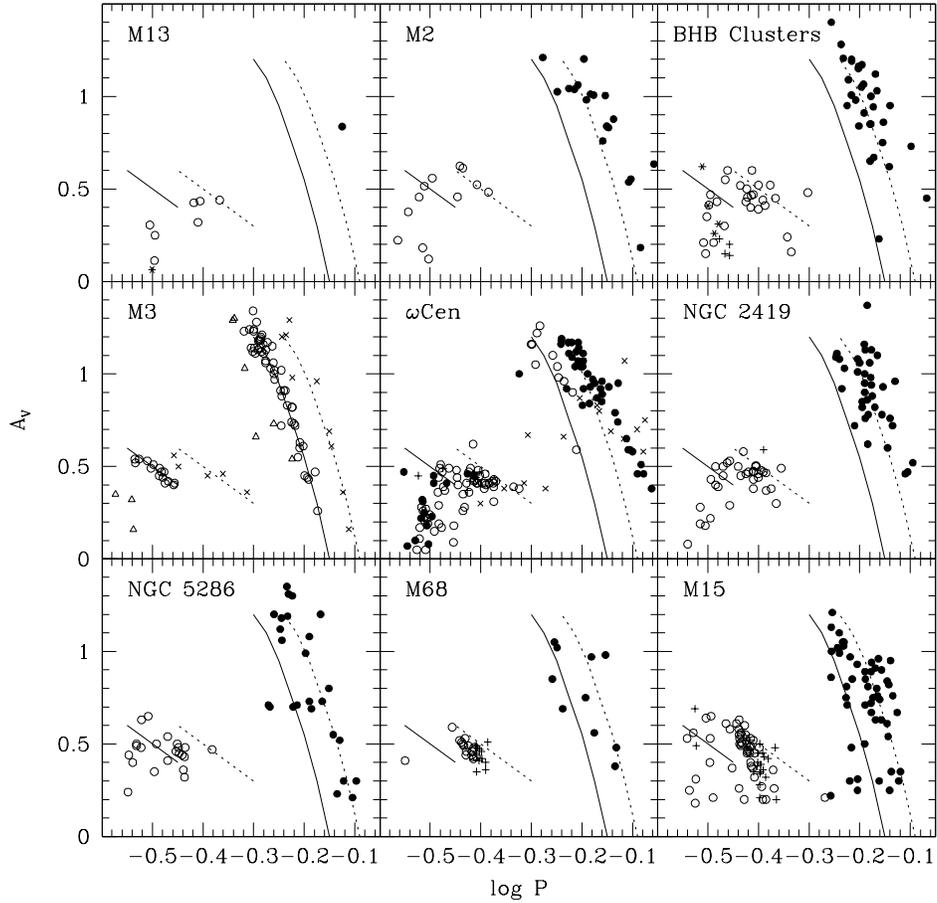}
\caption{Comparison of period-amplitude values for RR Lyrae stars in M3
  \citep{caccm3}, M13 \citep{kopacki02}, and other Oosterhoff II group
  clusters. For M3, stars identified as ``evolved'' (long period / high
  amplitude) are shown with $\times$, and small amplitude variables are shown
  with $\bigtriangleup$ (among the RRab stars, these are suspected Blazhko
  variables). For other clusters, fundamental mode pulsators are shown with
  $\bullet$, first harmonic pulsators have as $\bigcirc$, double-mode
  pulsators have $+$ (and are plotted with the dominant first overtone
  period), and non-radial pulsators have $\ast$. Solid lines are mean
  relations for regular M3 variables, and dotted lines are mean relations for
  ``evolved M3 variables'' \citep{caccm3}. In $\omega$ Cen, stars are
  segregated by magnitude: for $\langle V \rangle > 14.6$, RRab stars have
  $\bigcirc$ and RRc stars have $\bullet$; for $14.4 < \langle V \rangle <
  14.6$, RRab stars have $\bullet$ and RRc stars have $\bigcirc$; and for
  $\langle V \rangle < 14.4$, all RRab stars have $\times$.
\label{avp}}
\end{figure}

If variability information (such as period and variation amplitude) can give
unambiguous evolutionary information, it should be preferred because such
measurements can generally be made to higher precision than can determinations
of average magnitude or color. Fig. \ref{avp} shows $V$ amplitude and period
data for M13 RR Lyrae stars from \citet{kopacki02} in the Bailey diagram. The
single known RRab cluster variable (V8) has an even longer period than the
``well-evolved'' stars in M3 identified by \citep{caccm3}. Because V8 is also
the reddest RR Lyrae and evolutionary tracks in this part of the CMD are
expected to slope brightward toward the AGB, it would not be surprising if it
showed an evolution signature. In M13, the three BL Her stars (Population II
Cepheid variables) fall approximately in the part of the CMD where stars that
have evolved from the blue HB tail are predicted to pass, and their periods
(all greater than a day) also support evolved status. However, V8's $\langle V
\rangle$ is fainter than most of the RRc variables. Given the amount of
scatter seen in larger samples of RR Lyraes (see below), this can't be
considered evidence.

RRc variables are more abundant in M13, so an evolutionary signature among
these stars would be more compelling. Four of the RRc variables have periods
consistent with the ``well-evolved'' RRc stars in M3, but with somewhat lower
amplitude (one exception falls on the M3 relation). The remaining 4 (V7, V31,
V35, V36) have periods consistent with the regular variables in M3 but again
with low amplitudes. There is no evidence of separation in average $V$
magnitude between the two groups however, weakening the idea that evolutionary
effects can be identified solely from the periods of the RRc variables.

To put M13 in better context, we assembled literature data for
other Oosterhoff group II clusters (hereafter, OoII). For a few
in this group, the instability strip is heavily populated, leading to
an expectation (based on evolution timescales) that a large
proportion of the RR Lyraes should be near the ZAHB. Clusters in this
category include $\omega$ Cen \citep{kom,oom}, NGC 2419 \citep{rip}, 
NGC 5286 \citep{zoro}, M68 \citep{walkm68,clemm68}, and M15
\citep{corm15,ssm15}. These clusters have relatively low HB$_{type}$ values
($\le 0.8$), and NGC 2419 and M68 have significant red HB populations as
well. Both of these factors imply that there are likely to be unevolved stars
in the instability strip. This group of OoII clusters, however, is outnumbered
by clusters like M13 that have almost exclusively blue HB stars. For example,
M2 \citep{lazm2,lcm2} is a massive cluster that has more RR Lyrae stars than
M13, but still has a very small fraction of its stars in the instability
strip. For comparison, we compiled data for clusters with only a few RR
Lyraes, under the hypothesis that these stars are more likely to have evolved
from the blue HB\footnote{References for RR Lyraes in ``BHB Clusters'' are: M9
  \citep{csm9};
M30 \citep{pkm30}; M55
  \citep{olechm55}; M80 \citep{wbhm80}; M92 \citep{kopm92,cm92}; NGC
  288 \citep{kkn288}; NGC 5897 \citep{wn5897,crn5897}; NGC 5986
  \citep{abon5986}. 
}.  The Bailey diagrams ($V$ amplitude versus
period) for the different samples are shown in Fig. \ref{avp}.

M13 RRc variables fall in parts of the Bailey diagram that are commonly
occupied by variables in other OoII clusters.  In the BHB cluster group and
most of the other clusters, a significant fraction of the RRc variables fall
near $(A_V, \log P) = (0.45, -0.41)$.  This group appears largely independent
of cluster metallicity.  The main exceptions are NGC 5286, which is more metal
rich than M3 and has an analogous group that is much closer in period to M3
RRc variables, and M2, which doesn't have a clear concentration of RRc
variables in the Bailey diagram.  The second identifiable group of M13 RRc
variables overlaps with groups in the BHB Cluster sample and in $\omega$ Cen,
and has $\log P \approx -0.5$ and $A_V \le 0.5$. From these diagrams, we
conclude that there are not clear evolutionary signatures that be inferred
from the Bailey diagram alone.

Strangely, there are proportionately few short period, low amplitude RRc
variables in the ``protoypical'' OoII clusters M15 and M68. In addition, both
of these clusters have populations of double-mode RRd variables that largely
reside in a small range of color between the RRab and RRc variables
\citep{ssm15,walkm68}.  Such a distribution is unlikely if most of the stars
are evolved BHB stars --- evolution accelerates as a star moves redward toward
the AGB.  Conversely, we find that the shortest period ($\log P \la -0.46$)
RRc stars are the bluest RR Lyrae stars in OoII clusters. Many of the known non-radial pulsators in OoII clusters
\citep{kopacki02,olechm55,kopm92} are also found within this group.

At best, we find that Bailey diagrams for OoII RRc stars provide some
information about average colors. This is somewhat useful for
examining the effects of evolution within individual clusters because
some color distributions (especially ones biased toward the red) are
incompatible with model predictions. However, the RRc don't appear to
follow patterns implied by M3, in which ``well evolved'' stars appear 
shifted to greater period.

In summary, the majority of the evidence from photometric indicators implies
that the M13 RR Lyraes (and nonvariable stars of similar color) are
significantly brighter than the ZAHB level. The most notable exception
involves the comparison to the RGB bump, while the pulsation properties of the
RR Lyraes themselves 
are somewhat ambiguous. If
the reddest HB stars are discounted as significantly evolved, then M13's blue
HB appears to fall at approximately the same luminosity level as M3's and the
evidence for helium enrichment or age differences from this is removed.

\subsubsection{Subgiant Branch Morphology}

As mentioned earlier, one of the more unusual aspects of M13's CMD is the
steeper slope of its subgiant branch in comparison to M3's. This was
interpreted as resulting either from helium enrichment \citep{jb} or greater
age \citep{rey}. Both of the cited papers used the so-called ``horizontal
method'' to compare the ages of clusters with other clusters or isochrones.
Higher helium abundance or age tends to shorten the length of the subgiant
branch, so that if the turnoffs are aligned in color and fainter main
sequence points (usually 0.05 mag redder than the turnoff) are subsequently aligned in
magnitude, the relative position of the giant branch reveals the difference.
It has been verified repeatedly that there is a difference between M3 and M13,
but the cause is unclear.

We can attempt to look at the problem from a different angle by realizing that
age and helium enrichment affect the absolute colors of the turnoff and red
giant branch in different ways. An increase in helium
abundance reduces the opacity of the stellar envelope, making surface
temperature higher at both the turnoff and on the red giant branch. On the
other hand, increased age makes both the turnoff and red giant branch redder,
although the effect on the giant branch is very small.

Because reddening and metal abundance differences have large effects, reliable
comparisons of absolute colors can only be done when these are
well-determined --- M3 and M13 are probably the best example of such a pair.
\citet{stet98} discusses them in this respect, but we expand on the
arguments here.  The reddenings for the two clusters appear to be small and
very similar (M13 probably with the larger reddening by $\Delta E(B-V) <
0.01$; \citealt{schlegel}) and their metallicities appear to agree to within
0.1 dex (with M13 being the more metal poor; \citealt{sned}). One additional
benefit of using the M3/M13 pair is that the effects of the reddening and
metallicity differences on the turnoff should partly cancel, with the
metallicity differences leading to an expectation that M13's turnoff should be
bluer than M3's by about 0.01 in $B-I$, and less in other optical colors.

\begin{figure}
\epsscale{.90}
\plotone{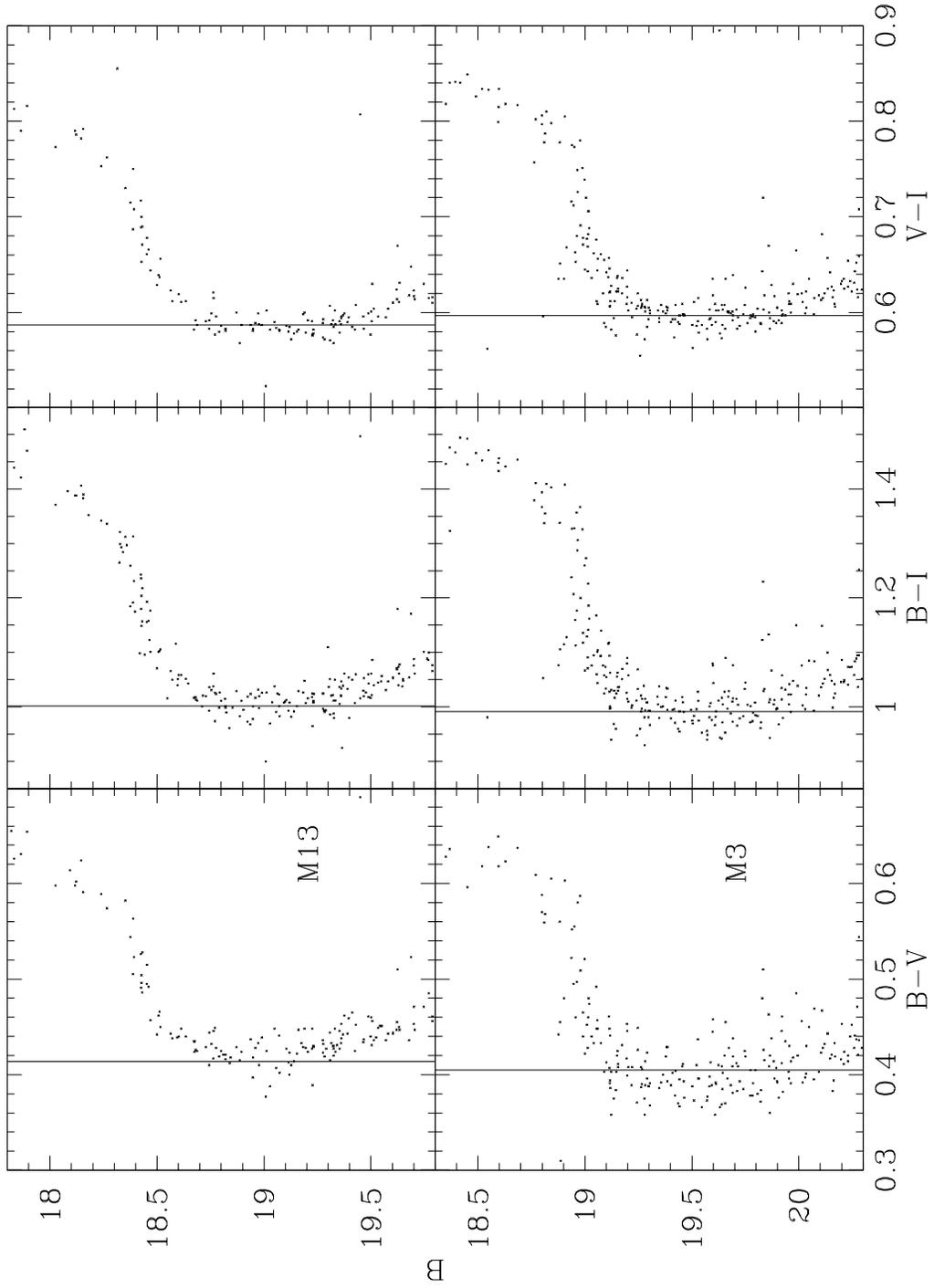}
\caption{Stetson standard star values for M13 and M3 at the cluster turnoff in
  different optical colors. The vertical lines are the median color values of
stars within about 0.2 mag of the cluster turnoff in brightness.
\label{tos}}
\end{figure}

A comparison in absolute colors also requires datasets for which the
calibration can be done uniformly for both clusters. While we are
unable to do this because our deep exposures of M3 and M13 were not
taken on the same photometric night, it has been done in the most
recent studies, although using different colors. \citet{jb} used $V-I$
and found that M13 was bluer at the turnoff and giant branch by
similar amounts ($0.01 - 0.02$). This implies a tiny helium enrichment
($\Delta Y \approx 0.01$) at most. On the other hand, \citet{rey} used
the $B-V$ color and found that the turnoff colors were virtually
identical, although M13's red giant branch was still significantly
bluer. We examined the standard star photometry of \citet{stet} for
the two clusters (8 November 2007 update), and found that in the $V-I$
color M13 is bluer than M3 on the giant branch and at the turnoff (in
agreement with \citealt{jb}), while in $B-V$ or $B-I$ colors M13's
turnoff is significantly {\it redder} (even accounting for small
differences in reddening) and the giant branches have nearly identical
colors (see Fig. \ref{tos})\footnote{While the Stetson standard stars
  in M3 and M13 are rigorously tied to the system of Landolt standard
  stars, the observations for the two clusters were taken under
  varied conditions. M3 and M13 can be observed on
  the same night using the same equipment, and so it should be
  possible to get good relative photometry for the pair. We note that
  \citet{stet98} reported that M13 had a bluer turnoff than M3 by
  0.014 mag in $B-I$ (about 0.04 mag when reddening was accounted for)
  for images of the two clusters taken on the same night under
  photometric conditions using the same equipment.}.  Based only on
the Stetson standard stars in $B-V$ and $B-I$ colors, an age
difference appears tenable.  With the $V-I$ color included, neither
helium nor age differences seem capable of explaining the observations
because neither is expected to affect one color differently than
others.
We conclude that current photometric datasets do not paint a consistent
picture, and may still be influenced by subtle systematic effects. To make
this a stringent test, effort should go into deep and carefully calibrated
photometry using filters with a wide wavelength baseline. The $B-I$ color still
appears to be a good choice.




\section{Conclusions}

To our minds, some of the most important questions regarding M13 remain in
dispute. One question that we have reopened here is whether the reddest of the
blue HB stars in M13 are significantly evolved (and whether they are therefore
good representatives of the brightness of the horizontal branch). The weight
of the observational and theoretical evidence leans toward the idea
that they are significantly evolved, and that the red edge of the primary HB
population is a decent indicator of the true HB level. 

The distribution of stars on the horizontal branch in M13 is complex,
and the most notable questions regard 1) how the color of the primary
peak in M13 could have been shifted so far relative to M3's when the
gross composition of clusters appear nearly identical, and 2) how a
large fraction of M13's RGB stars become blue stars near the end of
the canonical HB.  Our examination of the luminosity function shows
little sign that a large fraction of stars leave the bright RGB before
having a core flash, in agreement with the massive cluster NGC 2419
\citep{sh} but not with NGC 2808 \citep{sm}.  We do not find any clear
evidence of helium enrichment among the stars of the dominant (redder)
HB population, and in fact, the helium abundance indicator $R$ and the
relative brightness of the HB argue against significant
enrichment. The HB and RGB stars (and different subsets of these) do
not show significant signs of radial segregation within the
cluster. The small color difference between the main sequence turnoff
and the giant branch of M13 (in comparison to M3) remains unexplained,
but careful examination of the absolute colors of both clusters would
provide a new test.

Our thorough search of M13's HB population has revealed second $U$
jump and blue hook stars that imply that many of these stars have very
low-mass hydrogen-rich envelopes. Far UV observations show that many
of the stars in the second $U$ jump are more luminous than stars with
similar colors at the end of the HB. The reason is unclear,
however. Spectroscopic data on similar stars in NGC 6752 \citep{moni}
indicate that the excess brightness is not related to enhanced atmospheric
helium abundance, so further study is required.

Spectroscopic measurements may help to clarify our understanding of
extreme HB stars in a number of ways. We particularly encourage
studies of: stars near the red end of the HB in M13 where helium
abundances can be accurately determined \citep{villa}; stars in the
extreme HB to look for signs of unusual Mg abundances (a species that
appears to be minimally affected by diffusion) that could connect them
to giant stars; O, Na, and Mg for stars at the red giant tip of other
clusters to determine whether they are super O-poor; relatively
unevolved turnoff and subgiant stars in M13 and NGC 2808 to search for
large O depletions ([O/Fe]$ < -0.4$) and check whether this is the
result of external pollution or not.

\acknowledgements We would like to thank D. Pollard for contributions to the
paper, S. Cassisi for graciously providing tabulations of CMD positions of HB
stars for various degrees of central helium depletion, and the anonymous
referee for a careful reading. This work has been funded through AST grants
00-98696 and 05-07785 from the National Science Foundation to E.L.S. and
M.B. M.G. was partially supported as part of a Research Experiences for
Undergraduates (REU) program at San Diego State University under grant AST
04-53609 from the National Science Foundation.

This research used the facilities of the Canadian
Astronomy Data Centre operated by the National Research Council of
Canada with the support of the Canadian Space Agency. Some of the data
presented in this paper were obtained from the Multimission Archive at
the Sapce Telescope Science Institute (MAST). STScI is operated by the
Association of Universities for Research in Astronomy, Inc., under
NASA contract NAS5-26555. Support for MAST for non-HST data is
provided by the NASA Office of Space Science via grant NAG5-7584 and
by other grants and contracts.

\clearpage

\clearpage

\clearpage

\clearpage

\clearpage

\clearpage

\clearpage

\clearpage

\clearpage
\begin{deluxetable}{lcc}
\tablewidth{0pt} 
\tablecaption{Characteristics of M13 and M3}
\tablehead{\colhead{} & \colhead{M13} & \colhead{M3}}
\startdata
[Fe/H]$_{CG97}$ \tablenotemark{a} & $-1.39$ & $-1.34$ \\
$R_a$ (kpc) \tablenotemark{b} & $21.5 \pm 4.7$ & $13.4\pm0.8$ \\
$R_p$ (kpc) \tablenotemark{b} & $5.0 \pm 0.5$ & $5.5\pm0.8$ \\
$e$ \tablenotemark{b}& $0.62\pm0.06$ & $0.42 \pm 0.07$ \\
$\Psi$ (deg)\tablenotemark{b} & $54\pm5$ & $55\pm2$ \\ 
$L_Z$ (kpc km s$^{-1}$)\tablenotemark{b} & $-376\pm145$ & $705\pm123$\\ 
$E_{tot}$ ($10^5$ km$^2$ s$^{-2}$)\tablenotemark{b} & $-0.476\pm0.088$ & $-0.649\pm0.022$\\ 
$M_{V_t}$\tablenotemark{c} & $-8.66$ & $-8.65$ \\ 
$\log \rho_0$ ($L_{V_\odot} \mbox{pc}^{-3}$) & 3.33 & 3.51 \\
$c$\tablenotemark{c} & 2.11 & 1.56 \\
$r_c$ (pc)\tablenotemark{c} & 1.7 & 1.3 \\
$r_c$ ($\arcsec$) & 46 & 26 \\
$r_h$ (pc)\tablenotemark{c} & 3.5 & 3.7 \\
$r_h$ ($\arcsec$)\tablenotemark{c} & 94 & 73 \\
$d$ (kpc)\tablenotemark{c} & 7.7 & 10.4 \\
E($B-V$)\tablenotemark{d} & 0.017 & 0.013\\
\enddata
\label{m13m3}
\tablenotetext{a}{\citet{cg97}}
\tablenotetext{b}{\citet{dines}}
\tablenotetext{c}{\citet{mclaugh}; power-law models were used for $M_{V,t}$ values}
\tablenotetext{d}{\citet{schlegel}}
\end{deluxetable}

\begin{table}[ht]
\caption{Description of HST data}
\centering
\begin{tabular}{rll}
\hline\hline
Proposal ID & Principal Investigators & Filters\\
\hline
5903 & Ferraro & F160BW,F255W,F336W,F439W,F555W\\
8174 & van Altena & F555W,F785LP\\
8278 & Bailyn & F555W,F814W\\
10775 & Sarajedini & F606W, F814W\\
\hline
\end{tabular}
\label{tbl-1}
\end{table}

\begin{deluxetable}{rrrrrrrl}
\rotate
\tablewidth{0pt}
\tablecaption{Star Cross-Identifications}
\tablehead{\colhead{ID} & \colhead{L ID} & \colhead{K ID} & \colhead{B ID} & 
  \colhead{Flag\tablenotemark{a}} & \colhead{Other} & \colhead{Notes}}
\startdata
 V5 & 806a &  0 &   0 &  0 & 0 &  & RR1; blend in KPNO\\
 V7 &  344 & 432 &   0 &  0 & 0 & & RR1 \\
 V8 &  206 & 389 &   0 &  0 & 0 & & RR0; reddest RRLyr\\
 V9 &  806b&   0 &   0 &  0 & 0 & & RR1\\
V25 &    0 &   0 &   0 &  0 & 0 & & RR1; evolved? bad sampling?; SIMBAD L630 ID is wrong (nearby RGB)\\
V31 &  807 &   0 &1043 &  0 & 0 & & RR1; av affected by blending; BARN201 (blend)\\
V34 &  918 &   0 &   0 &  0 & 0 & & RR1\\
V35 &  571 &   0 &   0 &  0 & 0 & & RR1; gap in ACS\\
V36 &    0 &   0 &   0 &  0 & 0 & & RR2\\
\enddata
\label{idtab}
\tablecomments{Table \ref{idtab} is published in its entirety in the
  electronic edition of the Astronomical Journal.  A portion is shown here for
  guidance regarding its form and content. Sources for IDs: L:
  \citet{lud}; K: \citet{kad}; A: \citet{arp}; B:
  \citet{barn31}} \tablenotetext{a}{UIT Photometry Flag. 0: No UIT detection;
  1: Optimal photometry with no likely blending effects; 2: Photometry
  probably minimally affected by neighbors; 3: Photometry contamination of
  uncertain magnitude; 4: Photometric contamination certain; 5: Blending of
  two UIT sources.}
\end{deluxetable}

\clearpage
\begin{deluxetable}{rrrrrrrrrrrrrrrrr}
\rotate
\tabletypesize{\footnotesize}
\tablewidth{0pt}
\tablecaption{HB Star Photometry}
\tablehead{\colhead{ID} & \colhead{$\Delta \alpha$} & \colhead{$\Delta
    \delta$} & \colhead{$B$} & \colhead{$V$} & \colhead{$I$} &
  \colhead{Source} & \colhead{$m_{1620}$} & \colhead{$\sigma_{1620}$} &
  \colhead{Flag} & \colhead{F160BW} & \colhead{$\sigma_{160}$} & \colhead{F255W} &
  \colhead{$\sigma_{255}$} & \colhead{F336W} & \colhead{$\sigma_{336}$} &
  \colhead{$P_{PM}$}}
\startdata
 V5 &  57.711  &  1.414 & 15.252 & 14.690 & 14.308 & ACS &   &  & 0 &  &  & 16.180 & 0.042 & 15.487 & 0.004 &  \\
 V7 & $-$52.404 & $-$67.743 & 15.283 & 14.944 & 14.726 & ACS &   &  & 0 & &  & 16.196 & 0.181 & 15.352 & 0.005 & 87 \\
 V8 &$-$106.089 &  25.345 & 99.999 & 14.922 & 14.428 & ACS &    &  & 0 & & & & &  & & 99 \\
 V9 &  58.994 &   3.471 & 15.003 & 14.745 & 14.363 & ACS &   &  & 0 & 18.010 & 0.118 & 16.231 & 0.071 & 15.297 & 0.003 & \\
 V25 &  14.847 &  $-$5.506 & 14.753 & 14.589 & 14.368 & ACS &  &  & 0 & 17.240 & 0.100 & 15.590 & 0.042 & 14.954 & 0.003 & \\
 V31 &  57.728 &  78.754 & 99.999 & 14.890 & 14.565 & ACS &   &  & 0 & & & & & & & 99 \\
 V35 &  91.867 & $-$28.731 & 15.070 & 14.833 & 14.579 & ACS &  &  & 0 & & & 16.487 & 0.191 & 15.143 & 0.003 & \\
 V36 &  $-$0.226 &  10.760 & 14.834 & 14.676 & 14.349 & CFHT &  &  & 0 & 17.209 & 0.083 & 15.999 & 0.060 & 15.257 & 0.004 & \\
\enddata
\label{hbtab}
\tablecomments{Table \ref{hbtab} is published in its entirety in the
  electronic edition of the Astronomical Journal.  A portion is shown here for
  guidance regarding its form and content. }
\end{deluxetable}

\begin{deluxetable}{rrrrrrrrrrrrrrrrr}
\rotate
\tabletypesize{\footnotesize}
\tablewidth{0pt}
\tablecaption{AGB Star Photometry}
\tablehead{\colhead{ID} & \colhead{$\Delta \alpha$} & \colhead{$\Delta
    \delta$} & \colhead{$B$} & \colhead{$V$} & \colhead{$I$} &
  \colhead{Source} & \colhead{$m_{1620}$} & \colhead{$\sigma_{1620}$} &
  \colhead{Flag} & \colhead{F160BW} & \colhead{$\sigma_{160}$} & \colhead{F255W} &
  \colhead{$\sigma_{255}$} & \colhead{F336W} & \colhead{$\sigma_{336}$} &
  \colhead{$P_{PM}$}}
\startdata
  1 & $-$612.315 & 557.213 & 17.108 & 17.398 & 17.725 & KPNO & 12.83 & 0.03 & 1 & & & & & & & \\
  2 & $-$403.576 & 148.486 & 15.225 & 15.395 & 15.524 & CFHT & 12.84 & 0.03 & 1 & & & & & & & 99\\
  3 & $-$233.816 & 112.551 & 16.444 & 16.659 & 16.834 & CFHT & 13.42 & 0.04 & 1 & & & & & & & \\
  4 & $-$222.956 &  29.858 & 14.766 & 14.144 & 13.336 & CFHT &  &  & 0 & &  & & & & & 99\\
  5 & $-$184.247 & 134.084 & 15.793 & 16.004 & 16.194 & CFHT & 12.96 & 0.02 & 1 & & & & & & & \\
\enddata
\label{agbtab}
\tablecomments{Table \ref{agbtab} is published in its entirety in the
  electronic edition of the Astronomical Journal.  A portion is shown here for
  guidance regarding its form and content. }
\end{deluxetable}

\begin{deluxetable}{rrrrrrrrrrrr}
\tabletypesize{\footnotesize}
\rotate
\tablewidth{0pt}
\tablecaption{RGB Star Photometry}
\tablehead{\colhead{ID} & \colhead{$\Delta \alpha$} & \colhead{$\Delta
    \delta$} & \colhead{$B$} & \colhead{$V$} & \colhead{$I$} &
  \colhead{Source} & \colhead{F255W} &
  \colhead{$\sigma_{255}$} & \colhead{F336W} & \colhead{$\sigma_{336}$} &
  \colhead{$P_{PM}$}}
\startdata
  1 & $-$707.261 &   7.295 & 14.373 & 13.365 & 12.236 & KPNO & & & & & 99\\
  2 & $-$556.761 & 489.164 & 15.380 & 14.515 & 13.533 & KPNO & & & & & 99\\
  3 & $-$543.549 &$-$318.220 & 15.149 & 14.304 & 13.285 & CFHT & & & & & 99\\
  4 & $-$468.740 &$-$80.657 & 15.734 & 14.968 & 14.024 & CFHT & & & & & 99\\
  5 & $-$439.069 &  44.185 & 15.543 & 14.739 & 13.785 & CFHT & & & & & 99\\
\enddata
\label{rgbtab}
\tablecomments{Table \ref{rgbtab} is published in its entirety in the
  electronic edition of the Astronomical Journal.  A portion is shown here for
  guidance regarding its form and content. }
\end{deluxetable}

\begin{deluxetable}{rrrrrrrr}
\tabletypesize{\footnotesize}
\rotate
\tablewidth{0pt}
\tablecaption{Bright Nonmember Star Photometry}
\tablehead{\colhead{ID} & \colhead{$\Delta \alpha$} & \colhead{$\Delta
    \delta$} & \colhead{$B$} & \colhead{$V$} & \colhead{$I$} &
  \colhead{Source} & \colhead{$P_{PM}$}}
\startdata
K178 & $-$785.649 & $-$294.677 & 14.410 & 13.740 & & PM &    0 \\
K180 & $-$763.792 & $-$269.343 & 12.280 & 11.070 & & PM &    0 \\
K185 & $-$725.361 & $-$382.757 & 15.380 & 14.700 & & PM &    0 \\
K190 & $-$702.285 &  186.036 & 15.784 & 14.926 & 14.078 & KPNO &  0 \\
K196 & $-$664.843 &  556.797 & 15.468 & 14.764 & 14.019 & KPNO & 12\\
\enddata
\label{nmtab}
\tablecomments{Table \ref{nmtab} is published in its entirety in the
  electronic edition of the Astronomical Journal.  A portion is shown here for
  guidance regarding its form and content. ID Notes: L: \citet{lud}; K: \citet{kad}; CM: \citet{cmpm}; SA, SB: \citet{sav}; F: this paper.}
\end{deluxetable}

\clearpage
\begin{deluxetable}{ccccc}
\tablewidth{0pt}
\tablecaption{HB Star Distribution}
\tablehead{\colhead{Sample} & \colhead{$I < 16.25$} & \colhead{$16.25 < I <
    18$} & \colhead{$I > 18$} & \\ & \colhead{$f_{P1}$} & \colhead{$f_I$} &
  \colhead{$f_{P2}$} & \colhead{$f_{P1} - f_{P2}$}}
\startdata
Total & 365 & 197 & 222 & \\
& $0.47\pm0.02$ & $0.25\pm0.02$ & $0.28\pm0.02$ & $0.18\pm0.02$\\
$r < r_c/2$ & 25 & 10 & 21 & \\
& $0.45\pm0.07$ & $0.18\pm0.05$ & $0.38\pm0.06$ & $0.07\pm0.09$\\
$r < r_c$ & 82 & 36 & 61 & \\
& $0.46\pm0.04$ & $0.20\pm0.03$ & $0.34\pm0.04$ & $0.12\pm0.05$\\
$r_c/2 < r < r_h$ & 145 & 80 & 99 & \\
& $0.45\pm0.03$ & $0.25\pm0.02$ & $0.30\pm0.03$ & $0.14\pm0.04$\\
$r_h < r < 3.5 r_h$ & 157 & 94 & 87 & \\
& $0.46\pm0.03$ & $0.28\pm0.02$ & $0.26\pm0.02$ & $0.21\pm0.04$\\
$r > 3.5 r_h$ & 38 & 13 & 15 & \\
& $0.58\pm0.06$ & $0.20\pm0.05$ & $0.23\pm0.05$ & $0.35\pm0.08$\\
\enddata
\label{tbl-hb}
\end{deluxetable}

\clearpage
\begin{deluxetable}{cccc|cccc}
\tablewidth{0pt}
\tablecaption{Kolmogorov-Smirnov Test Results for Comparisons with
  Victoria-Regina RGB Models}
\tablehead{\multicolumn{4}{c}{M13 Sample} & \multicolumn{4}{c}{Combined M13 +
    M5 Sample}\\ \colhead{$I$ Cutoff} & \colhead{$D$} & \colhead{$P$} & 
\colhead{$N$} & \colhead{$I - I_{TRGB}$ Cutoff} & \colhead{$D$} &
\colhead{$P$} & \colhead{$N$}}
\startdata
13.4 & 0.080 & 0.10 & 232 & 3.3 & 0.036 & 0.46 & 545\\
13.2 & 0.078 & 0.19 & 193 & 3.1 & 0.040 & 0.44 & 460\\
13.0 & 0.068 & 0.45 & 159 & 2.9 & 0.052 & 0.23 & 397\\
12.8 & 0.098 & 0.16 & 130 & 2.7 & 0.039 & 0.69 & 329\\
12.6 & 0.110 & 0.13 & 109 & 2.5 & 0.048 & 0.56 & 273\\
12.4 & 0.102 & 0.25 & 97  & 2.3 & 0.046 & 0.69 & 234\\
12.2 & 0.093 & 0.43 & 86  & 2.1 & 0.049 & 0.72 & 200\\
12.0 & 0.126 & 0.19 & 73  & 1.9 & 0.073 & 0.30 & 171\\
11.8 & 0.118 & 0.40 & 55  & 1.7 & 0.097 & 0.12 & 148\\
11.6 & 0.139 & 0.34 & 44  & 1.5 & 0.083 & 0.40 & 114\\
11.4 & 0.130 & 0.66 & 30  & 1.3 & 0.087 & 0.48 & 91 \\
11.2 & 0.161 & 0.58 & 22  & 1.1 & 0.091 & 0.59 & 70\\
\enddata
\label{tbl-2}
\end{deluxetable}


\begin{thebibliography}{}
\bibitem[Ambika et al.(2004)]{ambika} Ambika, S., Parthasarathy, M., Aoki, W.,
Fujii, T., Nakada, Y., Ita, Y., \& Izumiura, H.\ 2004, \aap, 417, 293
\bibitem[Alves et al.(2001)]{abon5986} Alves, D.~R., Bond, 
H.~E., \& Onken, C.\ 2001, \aj, 121, 318
\bibitem[Arellano Ferro et al.(2008)]{arn5466} Arellano Ferro, 
A., Rojas L{\'o}pez, V., Giridhar, S., \& Bramich, D.~M.\ 2008, \mnras, 384, 1444 
\bibitem[Arp(1955)]{arp} Arp, H.~C.\ 1955, \aj, 60, 317
\bibitem[Barnard(1909)]{barn09}Barnard, E. E. 1909, \apj, 29, 72
\bibitem[Barnard(1914)]{barn14}Barnard, E. E. 1914, \apj, 40, 173
\bibitem[Barnard(1931)]{barn31} Barnard, E.~E.\ 1931, 
Publications of the Yerkes Observatory, 6, 1
\bibitem[Behr et al.(2000)]{behrm13} Behr, B.~B., Djorgovski, 
S.~G., Cohen, J.~G., McCarthy, J.~K., C{\^o}t{\'e}, P., Piotto, G., 
\& Zoccali, M.\ 2000, \apj, 528, 849
\bibitem[Behr(2003)]{behr03} Behr, B.~B.\ 2003, \apjs, 149, 67
\bibitem[Bedin et al.(2000)]{heap} Bedin, L.~R., Piotto, G., 
Zoccali, M., Stetson, P.~B., Saviane, I., Cassisi, S., \& Bono, G.\ 2000, 
\aap, 363, 159
\bibitem[Bedin et al.(2004)]{bed04} Bedin, L.~R., Piotto, G., 
Anderson, J., Cassisi, S., King, I.~R., Momany, Y., 
\& Carraro, G.\ 2004, \apjl, 605, L125
\bibitem[Bellazzini et al.(2004)]{bella} Bellazzini, M., Ferraro, F.~R.,
  Sollima, A., Pancino, E., \& Origlia, L.\ 2004, \aap, 424, 199
\bibitem[Benk{\H o} et al.(2006)]{benko} Benk{\H o}, J.~M., 
Bakos, G.~{\'A}., \& Nuspl, J.\ 2006, \mnras, 372, 1657
\bibitem[Bergbusch(1993)]{berg} Bergbusch, P.~A.\ 1993, \aj, 106, 1024
\bibitem[Bjork \& Chaboyer(2006)]{bjork} Bjork, S.~R., \&
  Chaboyer, B.\ 2006, \apj, 641, 1102
\bibitem[Bono et al.(1995)]{bono95} Bono, G., Caputo, F., 
\& Marconi, M.\ 1995, \aj, 110, 2365
\bibitem[Bono et al.(1997)]{bono97} Bono, G., Caputo, F., Castellani, V., \& Marconi, M.\ 1997, \aaps, 121, 327
\bibitem[Bono et al.(2001)]{bono} Bono, G., Cassisi, S., 
Zoccali, M., \& Piotto, G.\ 2001, \apjl, 546, L109
\bibitem[Briley et al.(2004)]{briley} Briley, M.~M., Cohen, 
J.~G., \& Stetson, P.~B.\ 2004, \aj, 127, 1579
\bibitem[Brown et al.(2001)]{brown} Brown, T.~M., Sweigart, 
A.~V., Lanz, T., Landsman, W.~B., \& Hubeny, I.\ 2001, \apj, 562, 368
\bibitem[Brown et al.(2008)]{brownm32} Brown, T.~M., Smith, E., 
Ferguson, H.~C., Sweigart, A.~V., Kimble, R.~A., 
\& Bowers, C.~W.\ 2008, \apj, 682, 319
\bibitem[Buonanno et al.(1994)]{buon94} Buonanno, R., Corsi, 
C.~E., Buzzoni, A., Cacciari, C., Ferraro, F.~R., \& Fusi Pecci, F.\ 1994, 
\aap, 290, 69 
\bibitem[Buonanno et al.(1997)]{bluetail} Buonanno, R., Corsi, 
C., Bellazzini, M., Ferraro, F.~R., \& Pecci, F.~F.\ 1997, \aj, 113, 706
\bibitem[Busso et al.(2007)]{busso} Busso, G., et al.\ 2007, \aap, 474, 105 
\bibitem[Cacciari et al.(2005)]{caccm3} Cacciari, C., Corwin, 
T.~M., \& Carney, B.~W.\ 2005, \aj, 129, 267
\bibitem[Caloi \& D'Antona(2005)]{cda} Caloi, V., \& 
D'Antona, F.\ 2005, \aap, 435, 987
\bibitem[Cardelli et al.(1989)]{ccm} Cardelli, J.~A., 
Clayton, G.~C., \& Mathis, J.~S.\ 1989, \apj, 345, 245
\bibitem[Carretta et al.(2008)]{carbrag} Carretta, E., 
Bragaglia, A., Gratton, R.~G., \& Lucatello, S.\ 2008, arXiv:0811.3591
\bibitem[Carretta et al.(2006)]{carr06} Carretta, E., Bragaglia, A., Gratton,
  R.~G., Leone, F., Recio-Blanco, A., \& Lucatello, S.\ 2006, \aap, 450, 523
\bibitem[Carretta \& Gratton(1997)]{cg97} Carretta, E., \& Gratton, R.~G.\
  1997, \aaps, 121, 95
\bibitem[Carretta et al.(2000)]{car} Carretta, E., Gratton, 
R.~G., Clementini, G., \& Fusi Pecci, F.\ 2000, \apj, 533, 215
\bibitem[Carretta et al.(2005)]{car05} Carretta, E., Gratton, R.~G.,
  Lucatello, S., Bragaglia, A., \& Bonifacio, P. 2005, \aap, 433, 597
\bibitem[Carretta et al.(2007)]{carrb} Carretta, E., 
Recio-Blanco, A., Gratton, R.~G., Piotto, G., 
\& Bragaglia, A.\ 2007, \apjl, 671, L125
\bibitem[Cassisi et al.(2001)]{cass01} Cassisi, S., Castellani, V., Degl'Innocenti,
S., Piotto, G., \& Salaris, M.\ 2001, \aap, 366, 578
\bibitem[Cassisi \& Salaris(1997)]{casa} Cassisi, S., \& 
Salaris, M.\ 1997, \mnras, 285, 593
\bibitem[Cassisi et al.(2004)]{ter} Cassisi, S., Salaris, 
M., Castelli, F., \& Pietrinferni, A.\ 2004, \apj, 616, 498
\bibitem[Cassisi et al.(2003)]{cass03} Cassisi, S., Salaris, 
M., \& Irwin, A.~W.\ 2003, \apj, 588, 862
\bibitem[Cassisi et al.(2004)]{cass04} Cassisi, S., Castellani, M., Caputo,
  F., \& Castellani,   V. 2004, \aap, 426, 641
\bibitem[Castellani \& Castellani(1993)]{cast93} Castellani, M., \&
  Castellani, V.\ 1993, \apj, 407, 649
\bibitem[Castellani et al.(1991)]{ccp} Castellani, V., 
Chieffi, A., \& Pulone, L.\ 1991, \apjs, 76, 911
\bibitem[Castellani et al.(2006a)]{cast06} Castellani, M., Castellani, V., \&
  Prada Moroni, P.~G.\ 2006, \aap, 457, 569
\bibitem[Castellani et al.(2006b)]{castel} Castellani, V.,
  Iannicola, G., Bono, G., Zoccali, M., Cassisi, S., \& Buonanno, R.\ 2006,
  \aap, 446, 569
\bibitem[Catelan et al.(1996)]{cat96} Catelan, M., de Freitas 
Pacheco, J.~A., \& Horvath, J.~E.\ 1996, \apj, 461, 231
\bibitem[Catelan(1999)]{cat98} Catelan, M. 1998, \apjl, 495, L81
\bibitem[Catelan et al.(2001)]{catm3} Catelan, M., Ferraro, 
F.~R., \& Rood, R.~T.\ 2001, \apj, 560, 970
\bibitem[Catelan et al.(2004)]{catrr} Catelan, M., Pritzl, 
B.~J., \& Smith, H.~A.\ 2004, \apjs, 154, 633
\bibitem[Catelan(2009)]{catrev} Catelan, M.\ 2009, \apss, 18
\bibitem[Cho et al.(2005)]{cho} Cho, D.-H., Lee, S.-G., 
Jeon, Y.-B., \& Sim, K.~J.\ 2005, \aj, 129, 1922
\bibitem[Clement et al.(1993)]{clemm68} Clement, C.~M., 
Ferance, S., \& Simon, N.~R.\ 1993, \apj, 412, 183
\bibitem[Clement \& Rowe(2000)]{cromega} Clement, C.~M., \& Rowe, J.\ 2000, \aj, 120, 2579
\bibitem[Clement \& Rowe(2001)]{crn5897} Clement, C.~M., \& Rowe, J.~F.\ 2001, \aj, 122, 1464
\bibitem[Clement \& Shelton(1999)]{cs}Clement, C. M. \& Shelton,
  I. 1999, \apjl, 515, L85
\bibitem[Clement \& Shelton(1999)]{csm9} Clement, C.~M., \& Shelton, I.\ 1999,
\aj, 118, 453
\bibitem[Cohen \& Matthews(1992)]{cm92} Cohen, J.~G., \& Matthews, K.\ 1992,
\pasp, 104, 1205
\bibitem[Cohen \& Mel{\'e}ndez(2005)]{cm} Cohen, J.~G., \& Mel{\'e}ndez, J.\
2005, \aj, 129, 303
\bibitem[Cohen et al.(1997)]{cohm13} Cohen, R.~L., 
Guhathakurta, P., Yanny, B., Schneider, D.~P., \& Bahcall, J.~N.\ 1997, 
\aj, 113, 669
\bibitem[Conlon et al.(1994)]{con94} Conlon, E.~S., Dufton, P.~L., \& Keenan,
F.~P.\ 1994, \aap, 290, 897
\bibitem[Corwin et al.(2008)]{corm15} Corwin, T.~M., 
Borissova, J., Stetson, P.~B., Catelan, M., Smith, H.~A., Kurtev, R., 
\& Stephens, A.~W.\ 2008, \aj, 135, 1459
\bibitem[Corwin \& Carney(2001)]{ccm3} Corwin, T.~M., \&
  Carney, B.~W.\ 2001, \aj, 122, 3183
\bibitem[Corwin et al.(1999)]{ccn5466} Corwin, T.~M., Carney, 
B.~W., \& Nifong, B.~G.\ 1999, \aj, 118, 2875
\bibitem[Cudworth(1979)]{cud} Cudworth, K.\ 1979, \aj, 84, 1005
\bibitem[Cudworth \& Monet(1979)]{cmpm} Cudworth, K.~M., \& 
Monet, D.~G.\ 1979, \aj, 84, 774
\bibitem[Dalessandro et al.(2008)]{dale} Dalessandro, E., 
Lanzoni, B., Ferraro, F.~R., Vespe, F., Bellazzini, M., 
\& Rood, R.~T.\ 2008, \apj, 681, 311
\bibitem[D'Antona \& Caloi(2008)]{dant08} D'Antona, F., \& Caloi, V.\ 2008,
in Proc. of IAU Symp. 246, Dynamical Evolution of Dense Stellar Systems,
ed. E. Vesperini, M. Gierzs, \& A. Sills (Dordrecht: Kluwer), 156
\bibitem[D'Antona \& Caloi(2008b)]{dant08b} D'Antona, F., \& Caloi, V.\ 2008,
  \mnras, 390, 693 
\bibitem[D'Cruz et al.(1996)]{dcruzbhk} D'Cruz, N.~L., Dorman, 
B., Rood, R.~T., \& O'Connell, R.~W.\ 1996, \apj, 466, 359
\bibitem[D'Cruz et al.(2000)]{dcruzomega} D'Cruz, N.~L., et al.\ 
2000, \apj, 530, 352
\bibitem[Denisenkov \& Denisenkova(1989)]{dandd} Denisenkov, P.~A., \& Denisenkova, S.~N.\ 1989, Astronomicheskij Tsirkulyar, 1538, 11
\bibitem[Di Cecco et al.(2010)]{dice} Di Cecco, A., et al.\ 
2010, \apj, 712, 527
\bibitem[Dieball et al.(2005)]{dien2808} Dieball, A., Knigge, 
C., Zurek, D.~R., Shara, M.~M., \& Long, K.~S.\ 2005, \apj, 625, 156
\bibitem[Dieball et al.(2007)]{diem15} Dieball, A., Knigge, 
C., Zurek, D.~R., Shara, M.~M., Long, K.~S., Charles, P.~A., 
\& Hannikainen, D.\ 2007, \apj, 670, 379 
\bibitem[Dinescu et al.(1999)]{dines} Dinescu, D.~I., Girard, 
T.~M., \& van Altena, W.~F.\ 1999, \aj, 117, 1792
\bibitem[Dolphin(2000a)]{hstphot} Dolphin, A.~E.\ 2000, \pasp, 112, 1383
\bibitem[Dolphin(2000b)]{hstcal} Dolphin, A.~E.\ 2000, \pasp, 112, 1391
\bibitem[Dorman(1992)]{dorman} Dorman, B.\ 1992, \apjs, 81, 
221
\bibitem[Dorman et al.(1993)]{dro} Dorman, B., Rood, R.~T., 
\& O'Connell, R.~W.\ 1993, \apj, 419, 596
\bibitem[Dotter et al.(2007)]{dsep} Dotter, A., Chaboyer, 
B., Jevremovi{\'c}, D., Baron, E., Ferguson, J.~W., Sarajedini, A., \& 
Anderson, J.\ 2007, \aj, 134, 376
\bibitem[Edmonds \& Gilliland(1996)]{edm} Edmonds, P.~D., 
\& Gilliland, R.~L.\ 1996, \apjl, 464, L157
\bibitem[Fabbian et al.(2005)]{fabb05} Fabbian, D., 
Recio-Blanco, A., Gratton, R.~G., \& Piotto, G.\ 2005, \aap, 434, 235
\bibitem[Fekadu et al.(2007)]{fek} Fekadu, N., Sandquist, 
E.~L., \& Bolte, M.\ 2007, \apj, 663, 277
\bibitem[Ferraro et al.(1997)]{ferrm3} Ferraro, F.~R., Carretta, E., Corsi,
  C.~E., Fusi Pecci, F., Cacciari, C., Buonanno, R., Paltrinieri, B., \&
  Hamilton, D.\ 1997, \aap, 320, 757 
\bibitem[Ferraro et al.(1997)]{faintuv} Ferraro, F.~R., 
Paltrinieri, B., Fusi Pecci, F., Rood, R.~T., \& Dorman, B.\ 1997, \mnras,
292, L45
\bibitem[Ferraro et al.(1997)]{ferruv} Ferraro, F.~R., 
Paltrinieri, B., Fusi Pecci, F., Cacciari, C., Dorman, B., 
\& Rood, R.~T.\ 1997, \apjl, 484, L145
\bibitem[Ferraro et al.(1998)]{gaps} Ferraro, F.~R., 
Paltrinieri, B., Fusi Pecci, F., Rood, R.~T., \& Dorman, B.\ 1998, \apj, 500, 
311 
\bibitem[Ferraro et al.(1999)]{ferr99} Ferraro, F.~R., 
Messineo, M., Fusi Pecci, F., de Palo, M.~A., Straniero, O., Chieffi, A., 
\& Limongi, M.\ 1999, \aj, 118, 1738
\bibitem[Fusi Pecci et al.(1993)]{fusi93} Fusi Pecci, F., 
Ferraro, F.~R., Bellazzini, M., Djorgovski, S., Piotto, G., 
\& Buonanno, R.\ 1993, \aj, 105, 1145
\bibitem[Fusi Pecci et al.(1992)]{fusi} Fusi Pecci, F., 
Ferraro, F.~R., Corsi, C.~E., Cacciari, C., 
\& Buonanno, R.\ 1992, \aj, 104, 1831
\bibitem[Grundahl et al.(1998)]{ujumpm13} Grundahl, F., 
Vandenberg, D.~A., \& Andersen, M.~I.\ 1998, \apjl, 500, L179
\bibitem[Grundahl et al.(1999)]{ujumps} Grundahl, F., Catelan, 
M., Landsman, W.~B., Stetson, P.~B., \& Andersen, M.~I.\ 1999, \apj, 524, 
242
\bibitem[Haft et al.(1994)]{hrw} Haft, M., Raffelt, G., 
\& Weiss, A.\ 1994, \apj, 425, 222
\bibitem[Hargis et al.(2004)]{har} Hargis, J.~R., 
Sandquist, E.~L., \& Bolte, M.\ 2004, \apj, 608, 243
\bibitem[Harris(1996)]{harris} Harris, W.~E.\ 1996, \aj, 112, 1487 
\bibitem[Holtzman et al.(1995)]{holtz} Holtzman, J.~A., 
Burrows, C.~J., Casertano, S., Hester, J.~J., Trauger, J.~T., Watson, 
A.~M., \& Worthey, G.\ 1995, \pasp, 107, 1065
\bibitem[Itoh et al.(1996)]{itoh} Itoh, N., Hayashi, H., 
Nishikawa, A., \& Kohyama, Y.\ 1996, \apjs, 102, 411
\bibitem[Johnson et al.(2005)]{jkp} Johnson, C.~I., Kraft, 
R.~P., Pilachowski, C.~A., Sneden, C., Ivans, I.~I., 
\& Benman, G.\ 2005, \pasp, 117, 1308
\bibitem[Johnson \& Bolte(1998)]{jb} Johnson, J.~A., \& 
Bolte, M.\ 1998, \aj, 115, 693
\bibitem[Jurcsik et al.(2003)]{jurc} Jurcsik, J., Benk{\H o}, J.~M., Bakos,
  G.~{\'A}., Szeidl, B., \& Szab{\'o}, R.\ 2003, \apjl, 597, L49
\bibitem[Kadla(1966)]{kad} Kadla, E.~I.\ 1966, Izvestiya 
Glavnoj Astronomicheskoj Observatorii v Pulkove, 181, 93
\bibitem[Kalirai et al.(2001)]{kali} Kalirai, J.~S., et al.\ 
2001, \aj, 122, 257
\bibitem[Kaluzny et al.(2004)]{kom} Kaluzny, J., Olech, A., Thompson, I.~B., Pych, W., Krzemi{\'n}ski, W., \& Schwarzenberg-Czerny, A.\ 2004, \aap, 424, 1101 
\bibitem[Kaluzny et al.(1997)]{kkn288} Kaluzny, J., Krzeminski,
  W., \& Nalezyty, M.\ 1997, \aaps, 125, 337
\bibitem[Kinman et al.(2000)]{kinfield} Kinman, T., Castelli, F., Cacciari,
  C., Bragaglia, A., Harmer, D., \& Valdes, F.\ 2000, \aap, 364, 102 
\bibitem[Kopacki(2001)]{kopm92} Kopacki, G.\ 2001, \aap, 369, 862
\bibitem[Kopacki et al.(2003)]{kopacki02} Kopacki, G., Kolaczkowski, Z., \&
  Pigulski, A.  2003, \aap, 398, 541
\bibitem[Landolt(1992)]{lan92}  Landolt, A.~U. 1992, \aj, 104, 340
\bibitem[Langer et al.(1993)]{lang} Langer, G.~E., Hoffman, 
R., \& Sneden, C.\ 1993, \pasp, 105, 301 
\bibitem[L{\'a}zaro et al.(2006)]{lazm2} L{\'a}zaro, C., 
Ferro, A.~A., Ar{\'e}valo, M.~J., Bramich, D.~M., Giridhar, S., 
\& Poretti, E.\ 2006, \mnras, 372, 69
\bibitem[Lee \& Carney(1999)]{lcm2} Lee, J.-W., \& Carney, B.~W.\ 1999, \aj,
117, 2868
\bibitem[Lee et al.(1994)]{ldz} Lee, Y.-W., Demarque, P., 
\& Zinn, R.\ 1994, \apj, 423, 248
\bibitem[Ludendorff(1905)]{lud} Ludendorff, H.\ 1905, 
Publikationen des Astrophysikalischen Observatoriums zu Potsdam, 50, 1
\bibitem[McLaughlin \& van der Marel(2005)]{mclaugh} 
McLaughlin, D.~E., \& van der Marel, R.~P.\ 2005, \apjs, 161, 304
\bibitem[Michaud et al.(2007)]{mich07} Michaud, G., Richer, 
J., \& Richard, O.\ 2007, \apj, 670, 1178
\bibitem[Michaud et al.(2008)]{mich} Michaud, G., Richer, 
J., \& Richard, O.\ 2008, \apj, 675, 1223
\bibitem[Miller Bertolami et al.(2008)]{millb} Miller Bertolami, M.~M.,
  Althaus, L.~G., Unglaub, K., \& Weiss, A.\ 2008, \aap, 491, 253
\bibitem[Moehler et al.(2007)]{moeom} Moehler, S., Dreizler, S., Lanz,
  T., Bono, G., Sweigart, A. V., Calamida, A., Monelli, M., \& Nonino,
  M. 2007, \aap, 475, L5
\bibitem[Moehler et al.(1998)]{moe98} Moehler, S., Heber, U., Lemke, M., \&
Napiwotzki, R.\ 1998, \aap, 339, 537 
\bibitem[Moehler et al.(2003)]{moe03} Moehler, S., Landsman, 
W.~B., Sweigart, A.~V., \& Grundahl, F.\ 2003, \aap, 405, 135
\bibitem[Moehler et al.(2000)]{moe00} Moehler, S., Sweigart, 
A.~V., Landsman, W.~B., \& Heber, U.\ 2000, \aap, 360, 120
\bibitem[Moehler et al.(2004)]{moe04} Moehler, S., Sweigart, 
A.~V., Landsman, W.~B., Hammer, N. J., \& Dreizler, S.\ 2004, \aap, 415, 313
\bibitem[Momany et al.(2002)]{momany02} Momany, Y., Piotto, G., 
Recio-Blanco, A., Bedin, L.~R., Cassisi, S., \& Bono, G.\ 2002, \apjl, 576, 
L65
\bibitem[Momany et al.(2004)]{momany} Momany, Y., Bedin, 
L.~R., Cassisi, S., Piotto, G., Ortolani, S., Recio-Blanco, A., De Angeli, 
F., \& Castelli, F.\ 2004, \aap, 420, 605 
\bibitem[Moni Bidin et al.(2006)]{moni06} Moni Bidin, C., Moehler, S., Piotto,
  G., Recio-Blanco, A., Momany, Y., \& M{\'e}ndez, R.~A.\ 2006, \aap, 451, 499
\bibitem[Moni Bidin et al.(2007)]{moni} Moni Bedin, C., Moehler, S., Piotto, 
G., Momany, Y., \& Recio-Blanco, A. 2007, \aap, 474, 505
\bibitem[Moni Bidin et al.(2009)]{moni09} Moni Bidin, C., Moehler, S., Piotto,
  G., Momany, Y., \& Recio-Blanco, A.\ 2009, \aap, 498, 737
\bibitem[Norris \& Da Costa(1995)]{nda} Norris, J.~E., \& Da Costa, G.~S.\
  1995, \apj, 447, 680 
\bibitem[Nemec(2004)]{nem5053} Nemec, J.~M.\ 2004, \aj, 127, 
2185
\bibitem[Olech et al.(1999)]{olechm55} Olech, A., Kaluzny, J., 
Thompson, I.~B., Pych, W., Krzeminski, W., 
\& Shwarzenberg-Czerny, A.\ 1999, \aj, 118, 442
\bibitem[Olech \& Moskalik(2009)]{oom} Olech, A., \& Moskalik, P.\ 2009, \aap,
  494, L17
\bibitem[Pace et al.(2006)]{pace07} Pace, G., Recio-Blanco, 
A., Piotto, G., \& Momany, Y.\ 2006, \aap, 452, 493
\bibitem[Paltrinieri et al.(1998)]{palt} Paltrinieri, B., 
Ferraro, F.~R., Carretta, E., \& Fusi Pecci, F.\ 1998, \mnras, 293, 434
\bibitem[Papadakis et al.(2000)]{pap6426} Papadakis, I., 
Hatzidimitriou, D., Croke, B.~F.~W., 
\& Papamastorakis, I.\ 2000, \aj, 119, 851
\bibitem[Parise et al.(1994)]{par} Parise, R.~A., et al.\  1994, \apj, 423, 305
\bibitem[Peterson(1983)]{p83} Peterson, R.~C.\ 1983, \apj, 275, 737
\bibitem[Peterson et al.(1995)]{pete} Peterson, R.~C., Rood, 
R.~T., \& Crocker, D.~A.\ 1995, \apj, 453, 214
\bibitem[Pietrinferni et al.(2006)]{piet06} Pietrinferni, A., 
Cassisi, S., Salaris, M., \& Castelli, F.\ 2006, \apj, 642, 797
\bibitem[Pietrinferni et al.(2009)]{piet} Pietrinferni, A., Cassisi, S.,
  Salaris, M., Percival, S., \& Ferguson, J.~W.\ 2009, \apj, 697, 275
\bibitem[Pietrukowicz \& Kaluzny(2004)]{pkm30} Pietrukowicz, P., \& Kaluzny, J.\ 2004, Acta Astronomica, 54, 19
\bibitem[Pilachowski et al.(1996)]{pskl} Pilachowski, C.~A., 
Sneden, C., Kraft, R.~P., \& Langer, G.~E.\ 1996, \aj, 112, 545
\bibitem[Piotto et al.(1999)]{pio} Piotto, G., Zoccali, M., 
King, I.~R., Djorgovski, S.~G., Sosin, C., Rich, R.~M., 
\& Meylan, G.\ 1999, \aj, 118, 1727
\bibitem[Piotto et al.(2005)]{pio05} Piotto, G., et al.\ 
2005, \apj, 621, 777
\bibitem[Piotto et al.(2007)]{pio07} Piotto, G., et al.\ 
2007, \apjl, 661, L53
\bibitem[Pollard et al.(2005)]{pol} Pollard, D.~L., 
Sandquist, E.~L., Hargis, J.~R., \& Bolte, M.\ 2005, \apj, 628, 729
\bibitem[Pulone(1992)]{pulone} Pulone, L.\ 1992, Memorie della 
Societa Astronomica Italiana, 63, 485
\bibitem[Rey et al.(2001)]{rey} Rey, S.-C., Yoon, S.-J., 
Lee, Y.-W., Chaboyer, B., \& Sarajedini, A.\ 2001, \aj, 122, 3219
\bibitem[Recio-Blanco et al.(2005)]{reb} Recio-Blanco, A., 
et al.\ 2005, \aap, 432, 851
\bibitem[Recio-Blanco et al.(2006)]{rb06} Recio-Blanco, A., Aparicio, A.,
  Piotto, G., de Angeli, F., \& Djorgovski, S.~G.\ 2006, \aap, 452, 875
\bibitem[Riello et al.(2003)]{riello} Riello, M., et al.\ 
2003, \aap, 410, 553
\bibitem[Ripepi et al.(2007)]{rip} Ripepi, V., et al.\ 
2007, \apjl, 667, L61
\bibitem[Rood et al.(1999)]{roodm3} Rood, R.~T., et al.\ 1999, 
\apj, 523, 752
\bibitem[Sabbi et al.(2004)]{sabbi} Sabbi, E., Ferraro, 
F.~R., Sills, A., \& Rood, R.~T.\ 2004, \apj, 617, 1296
\bibitem[Salaris et al.(2004)]{sal04} Salaris, M., Riello, 
M., Cassisi, S., \& Piotto, G.\ 2004, \aap, 420, 911
\bibitem[Sandage(1981)]{sand81a} Sandage, A.\ 1981, \apjl, 244, 
L23
\bibitem[Sandquist \& Bolte(2004)]{sb04} Sandquist, E.~L., 
\& Bolte, M.\ 2004, \apj, 611, 323
\bibitem[Sandquist \& Hess(2008)]{sh} Sandquist, E. L. \& Hess, J. R., 2008,
  \aj, 136, 2259
\bibitem[Sandquist \& Martel(2007)]{sm} Sandquist, E.~L., 
\& Martel, A.~R.\ 2007, \apjl, 654, L65
\bibitem[Sarajedini et al.(2007)]{sara} Sarajedini, A., et 
al.\ 2007, \aj, 133, 1658
\bibitem[Savedoff(1956)]{sav} Savedoff, M.~P.\ 1956, \aj, 
61, 254
\bibitem[Schlegel et al.(1998)]{schlegel} Schlegel, D.~J., 
Finkbeiner, D.~P., \& Davis, M.\ 1998, \apj, 500, 525
\bibitem[Silbermann \& Smith(1995)]{ssm15} Silbermann, N.~A., \& Smith, H.~A.\
  1995, \aj, 110, 704
\bibitem[Simon \& Clement(1993)]{sc93} Simon, N.~R., \& Clement, C.~M.\ 1993,
  \apj, 410, 526
\bibitem[Sirianni et al.(2005)]{sir} Sirianni, M., et al.\ 2005, \pasp, 117,
  1049
\bibitem[Smith(2005)]{smith}Smith, G. H.\ 2005, The Observatory, 125, 244
\bibitem[Smith \& Briley(2006)]{sbcn} Smith, G.~H., \& Briley, M.~M.\ 2006,
  \pasp, 118, 740
\bibitem[Smith et al.(2005)]{sbh} Smith, G.~H., Briley, 
M.~M., \& Harbeck, D.\ 2005, \aj, 129, 1589
\bibitem[Smith et al.(1996)]{smith96}Smith, G. H., Shetrone, M. D.,
  Bell, R. A., Churchill, C. W., \& Briley, M. M. 1996, \aj, 112, 1511
\bibitem[Smith \& Wehlau(1985)]{smwhe} Smith, H.~A., \& Wehlau, A.\ 1985, \apj,
298, 572 
\bibitem[Sneden et al.(2004)]{sned}Sneden, C., Kraft, R. P.,
  Guhathakurta, P., Peterson, R. C., \& Fulbright, J. P.\ 2004, \aj, 127, 2162
\bibitem[Stecher et al.(1997)]{stech} Stecher, T.~P., et al.\ 
1997, \pasp, 109, 584
\bibitem[Stetson(1987)]{daophot} Stetson, P.~B. 1987, \pasp, 99, 191
\bibitem[Stetson(1992)]{stet92} Stetson, P.~B.\ 1992, \jrasc, 86, 71 
\bibitem[Stetson(1998)]{stet98} Stetson, P.~B.\ 1998, Bulletin 
d'information du telescope Canada-France-Hawaii, 38, 1
\bibitem[Stetson(2000)]{stet} Stetson, P.~B.\ 2000, \pasp, 112, 925
\bibitem[Straniero et al.(1997)]{scl97} Straniero, O., 
Chieffi, A., \& Limongi, M.\ 1997, \apj, 490, 425
\bibitem[Strom et al.(1970)]{ss} Strom, S.~E., Strom, K.~M., Rood, R.~T., \&
Iben, I., Jr.\ 1970, \aap, 8, 243
\bibitem[Sweigart(1987)]{swei87} Sweigart, A.~V.\ 1987, \apjs, 65, 95
\bibitem[Thompson et al.(2007)]{thom07} Thompson, H.~M.~A., 
Keenan, F.~P., Dufton, P.~L., Ryans, R.~S.~I., Smoker, J.~V., Lambert, 
D.~L., \& Zijlstra, A.~A.\ 2007, \mnras, 378, 1619
\bibitem[Valcarce \& Catelan(2008)]{valm3} Valcarce, A.~A.~R.,
  \& Catelan, M.\ 2008, \aap, 487, 185
\bibitem[VandenBerg et al.(2006)]{vr} VandenBerg, D.~A., 
Bergbusch, P.~A., \& Dowler, P.~D.\ 2006, \apjs, 162, 375
\bibitem[Vargas {\'A}lvarez \& Sandquist(2007)]{vargas} Vargas 
{\'A}lvarez, C.~A., \& Sandquist, E.~L.\ 2007, \aj, 134, 825
\bibitem[Villanova et al.(2009)]{villa} Villanova, S., 
Piotto, G., \& Gratton, R.~G.\ 2009, \aap, 499, 755
\bibitem[Walker(1994)]{walkm68} Walker, A.~R.\ 1994, \aj, 108, 
555
\bibitem[Wallerstein(1970)]{wall} Wallerstein, G.\ 1970, 
\apj, 160, 345 
\bibitem[Wallerstein(2002)]{wall02} Wallerstein, G.\ 2002, 
\pasp, 114, 689
\bibitem[Wehlau(1990)]{wn5897} Wehlau, A.\ 1990, \aj, 99, 250
\bibitem[Wehlau \& Butterworth(1990)]{wbm28} Wehlau, A., \& Butterworth,
  S.\ 1990, \aj, 100, 686
\bibitem[Wehlau et al.(1990)]{wbhm80} Wehlau, A., Butterworth, 
S., \& Hogg, H.~S.\ 1990, \aj, 99, 1159
\bibitem[Wehlau \& Hogg(1984)]{whm28} Wehlau, A., \& Hogg, H.~S.\ 1984, \aj,
  89, 1005
\bibitem[Whitney et al.(1995)]{whit} Whitney, J.~H., et al.\ 
1995, \aj, 110, 1722
\bibitem[Yi et al.(1997)]{yi} Yi, S., Demarque, P., \& Kim, Y.-C. 1997, \apj,
  482, 677
\bibitem[Yong et al.(2006)]{yong} Yong, D., Aoki, W., 
\& Lambert, D.~L.\ 2006, \apj, 638, 1018
\bibitem[Yong et al.(2009)]{yong09}Yong, D., Grundahl, F., D'Antona,
  F., Karakas, A. I., Lattanzio, J. C., \& Norris, J. E. 2009, \apjl,
  695, L62
\bibitem[Yong et al.(2003)]{yong03} Yong, D., Grundahl, F., Lambert, D.~L.,
Nissen, P.~E., \& Shetrone, M.~D.\ 2003, \aap, 402, 985
\bibitem[Zinn et al.(1972)]{zng} Zinn, R.~J., Newell, E.~B., \& Gibson, J.~B.\
1972, \aap, 18, 390
\bibitem[Zinn(1974)]{zinn} Zinn, R.\ 1974, \apj, 193, 593
\bibitem[Zoccali et al.(2000)]{zocc00} Zoccali, M., Cassisi, 
S., Bono, G., Piotto, G., Rich, R.~M., \& Djorgovski, S.~G.\ 2000, \apj, 
538, 289
\bibitem[Zoccali et al.(1999)]{zocc99} Zoccali, M., Cassisi, 
S., Piotto, G., Bono, G., \& Salaris, M.\ 1999, \apjl, 518, L49
\bibitem[Zorotovic et al.(2010)]{zoro} Zorotovic, M., et 
al.\ 2010, \aj, 139, 357
\end{thebibliography}
\end{document}